\documentclass[aps, prd, superscriptaddress, twocolumn, 10pt]{revtex4-2}
\usepackage{amsmath}	% AMS Math Package
\usepackage{amsthm}		% Theorem Formatting
\usepackage{amssymb}	% Math symbols such as \mathbb
\usepackage{array}
\usepackage{datetime}
\usepackage{graphicx}
\usepackage{color}
\usepackage{verbatim}
\usepackage{bm}
\usepackage{braket}
\usepackage{natbib}
\usepackage{cancel}
\usepackage{xcolor}
\usepackage{slashed}
\usepackage{braket}
\usepackage{caption}
\usepackage{subcaption}
\usepackage{amsfonts} 
\usepackage{float}
\usepackage{dsfont}
\smallskipamount
\usepackage{feynmp}
\usepackage{letltxmacro} % closed roots
\usepackage[colorlinks=true, citecolor=midblue, linkcolor=midblue, urlcolor=midblue]{hyperref}

\usepackage{setspace}

\usepackage{caption}
\usepackage{tikz,xcolor}
\definecolor{lime}{HTML}{A6CE39}
\DeclareRobustCommand{\orcidicon}{
	\begin{tikzpicture}
	\draw[lime, fill=lime] (0,0) 
	circle [radius=0.16] 
	node[white] {{\fontfamily{qag}\selectfont \tiny ID}};
	\draw[white, fill=white] (-0.0625,0.095) 
	circle [radius=0.007];
	\end{tikzpicture}
	\hspace{-2mm}
}
\foreach \x in {A, ..., Z}{
	\expandafter\xdef\csname orcid\x\endcsname{\noexpand\href{https://orcid.org/\csname orcidauthor\x\endcsname}{\noexpand\orcidicon}}
}

\settimeformat{ampmtime}

\definecolor{grey}{rgb}{0.4,0.4,0.4}
\definecolor{dullmagenta}{rgb}{0.4,0,0.4}
\definecolor{darkblue}{rgb}{0,0,0.4}
\definecolor{midblue}{rgb}{0,0,0.5}
\definecolor{midred}{rgb}{0.5,0,0}
\definecolor{orange}{rgb}{1,0.5,0}
\definecolor{lightbrown}{rgb}{0.75,0.5,0.25}
\definecolor{tan}{cmyk}{0.14,0.42,0.56,0}
\definecolor{djunglegreen}{cmyk}{0.99,0,0.52,0}
\definecolor{lightgreen}{rgb}{0,1,0}
\definecolor{olivegreen}{cmyk}{0.64,0,0.95,0.40}
\definecolor{midgreen}{rgb}{0.0,0.675,0.0}
\definecolor{darkgreen}{rgb}{0,0.5,0}

\newcommand{\q}{\quad}

\newcommand{\vs}{\vspace}

\renewcommand{\.}{\hspace{0.5mm}}

\newcommand{\Grm}{\ensuremath{\mathrm{G}}}
\newcommand{\Hrm}{\ensuremath{\mathrm{H}}}

\newcommand{\Lrm}{\ensuremath{\mathrm{L}}}

\newcommand{\Rrm}{\ensuremath{\mathrm{R}}}

\newcommand{\crm}{\ensuremath{\mathrm{c}}}

\newcommand{\mrm}{\ensuremath{\mathrm{m}}}

\newcommand{\Hcal}{\ensuremath{\mathcal{H}}}

\newcommand{\Kcal}{\ensuremath{\mathcal{K}}}

\newcommand{\Mcal}{\ensuremath{\mathcal{M}}}

\newcommand{\Pcal}{\ensuremath{\mathcal{P}}}

\newcommand{\kbm}{\ensuremath{\bm{k}}}

\newcommand{\qbm}{\ensuremath{\bm{q}}}

\renewcommand{\d}{\ensuremath{\mathrm{d}}}
\newcommand{\eg}{e.g.}

\settimeformat{ampmtime}

\usepackage{float}

\makeatletter
\let\oldr@@t\r@@t
\def\r@@t#1#2{%
\setbox0=\hbox{$\oldr@@t#1{#2\,}$}\dimen0=\ht0
\advance\dimen0-0.2\ht0
\setbox2=\hbox{\vrule height\ht0 depth -\dimen0}%
{\box0\lower0.4pt\box2}}
\LetLtxMacro{\oldsqrt}{\sqrt}
\renewcommand*{\sqrt}[2][\ ]{\oldsqrt[#1]{#2}}
\makeatother

\newcommand{\FourthAffiliation}{\affiliation{
	Physics Faculty, Tomsk State University, Lenin ave. 36, Tomsk 634050, Russia}}

\newcommand{\FirstAffiliation}{\affiliation{
	Arnold Sommerfeld Center,
	Ludwig-Maximilians-Universit{\"a}t,
	Theresienstra{\ss}e 37,
	80333 M{\"u}nchen,
	Germany}}
 
\newcommand{\SecondAffiliation}{\affiliation{
	Max-Planck-Institut f{\"u}r Physik,
	Boltzmannstr.~8, 
    85748 Garching,
	Germany}}

\newcommand{\ThirdAffiliation}{\affiliation{
    Service de Physique Th{\'e}orique,
    C.P. 225, 
    Universit{\'e} Libre de Bruxelles,
    Boulevard du Triomphe, B-1050 Brussels,
    Belgium}}

\begin{document}

%%%%%%%%%%%%%%%%%%%%%%%%%%%%%%%%%%%%%%%
\title{Reconstructing Primordial Black Hole Power Spectra from Gravitational Waves}

\author{Daniel Frolovsky \!\orcidF{}}
\FourthAffiliation

\author{Florian K{\"u}hnel\!\orcidD{}}
\FirstAffiliation
\SecondAffiliation

\author{Ioanna Stamou\!\orcidE{}}
\ThirdAffiliation

%%%%%%%%%%%%%%%%%%%%%%%%%%%%%%%%%%%%%%%
%\date{\formatdate{\day }{ \month }{ \year}, \currenttime}

%%%%%%%%%%%%%%%%%%%%%%%%%%%%%%%%%%%%%%%
\begin{abstract}
A novel methodology for analysing the relation between the energy density in gravitational waves and primordial power spectra is developed. Focusing on scalar-induced gravitational radiation, this methodology is applied to a number of scenarios for the primordial black hole formation. Being differed from conventional Bayesian approaches, its advantages include directness and computational efficiency, which are crucial for handling the complex data characteristic of gravitational wave research. %\textcolor{red}{As an important application, it is demonstrated that it allows to systematically identify all scenarios consistent with current and future pulsar-timing-array data.}
{As an important application, it is demonstrated that this methodology allows for the systematic reconstruction of power spectra across all scenarios using current pulsar timing array data, providing a clear example of its potential in gravitational wave analysis.}
\end{abstract}

\maketitle

%%%%%%%%%%%%%%%%%%%%%%%%%%%%%%%%%%%%%%%
\section{Introduction}
\label{sec:Introduction}

The groundbreaking observations of gravitational waves originating from binary black hole mergers by the LIGO/Virgo collaborations \cite{Abbott:2016blz, Abbott:2017vtc, Abbott:2017gyy, Abbott:2017oio, Abbott:2016nmj} has significantly increased interest in both experimental and theoretical investigations of a possible stochastic gravitational-wave background. Importantly, the recent identification of the latter at nanohertz frequencies by pulsar timing array (PTA) collaborations \cite{NANOGrav:2023gor, NANOGrav:2023hde, NANOGrav:2023hvm, EPTA:2023fyk, EPTA:2023sfo, EPTA:2023akd, EPTA:2023xxk, Miles:2024rjc} has generated an avalanche of interest in acquiring understanding of the signal sources~\cite{NANOGrav:2023hvm}. Moreover, the development of future space-based gravitational-wave interferometers such as LISA, BBO and DECIGO \cite{LISA:2017pwj, Yagi:2011wg}, alongside ground-based detectors, like Einstein Telescope~\cite{Maggiore:2019uih}, promises to enable a more comprehensive detection of gravitational waves across a large frequency range.
 
A variety of theoretical models have been posited as potential sources for the signals detected by PTAs, encompassing phenomena such as scalar-induced %\textcolor{blue}{and first-order}  
gravitational waves ~\cite{Kohri:2018awv, Acquaviva:2002ud,Vagnozzi:2023lwo}, first-order phase transitions~\cite{Fujikura:2023lkn, Addazi:2023jvg, Megias:2023kiy, Zu:2023olm, Winkler:2024olr}, cosmic strings~\cite{Ellis:2023tsl, Kitajima:2023vre, Lazarides:2023ksx, Eichhorn:2023gat}, domain walls~\cite{Kitajima:2023cek, Guo:2023hyp, Blasi:2023sej, Gouttenoire:2023ftk, Babichev_2023} or primordial black holes (PBHs)~\cite{Vaskonen:2020lbd, Franciolini:2023pbf, Inomata:2023zup, Cai:2023dls, Wang:2023ost}. A recent comprehensive investigation has explored these theoretical frameworks in depth, offering new insights and perspectives on the origin of PTA signals~\cite{NANOGrav:2023hvm}, and numerous of papers have investigated the source of these (see \eg~Ref.~\cite{Ellis:2023oxs, Figueroa:2023zhu, Iovino:2024tyg,Vagnozzi:2023lwo}).

A large variety of constraints exist on the abundance of PBHs, as extensively reviewed in recent literature (see Ref.~\cite{Carr:2016drx, Carr:2020gox, Carr:2021bzv} for reviews). Apart from constraints, there are numerous astrophysical detections which might plausibly be allocated to PBHs (see Ref.~\cite{Carr:2023tpt}). One of the most promising detection methods, accessible with current technology, involves identifying subsolar-mass black holes in compact binary coalescences. 
%\textcolor{red}{Interestingly, recent gravitational-wave observations have reported several subsolar-mass candidate events.} 
{  Interestingly, recent gravitational-wave observations have suggested the possibility of subsolar-mass candidates~\cite{LIGOScientific:2021job, Phukon:2021cus, LIGOScientific:2022hai, Morras:2023jvb}. It is important to note that none of these findings currently have sufficient statistical evidence to be considered confirmed. However,} these {indications} 
%\textcolor{red}{findings}
are far reaching to our understanding of dark matter and early-Universe physics. In particular, they offer insights into the primordial fluctuations that seeded their formation and the nature of dark matter~\cite{Prunier:2023cyv, Miller:2024fpo}.

The present work provides a comprehensive study aimed at studying primordial power spectra from observed gravitational-wave data. Focusing on scalar-induced gravitational waves, and combining theoretical models with simulated observational data, we formulate a novel methodology for the inverse problem in gravitational-wave analysis. Our methodology is distinct from Bayesian statistical approaches, emphasising direct, objective analysis based on empirical data. In order to demonstrate its potential, we explore four specific classes of power spectra: {\it Broken Power Law}, {\it Log-Normal Peak} with its variants having {\it Oscillatory Patterns} or an {\it UV Cut-Off} \textcolor{black}{or with a {\it Plateau} at CMB scales and beyond the peak}. These scenarios have been used to generate datasets of gravitational-wave energy densities, aimed at aligning with the PTA NANOGrav results~\cite{Vaskonen:2020lbd, Ellis:2023oxs}.

Using the corresponding datasets, we reconstruct the scalar power spectra, including the incorporation of noise. Subsequent, we focus on assessing the allowed abundances of PBHs, recalculating the PBH formation thresholds in each case. Finally, we investigate which of these cases effectively reconciles the observed gravitational-wave signals with PBH production.
%\textcolor{red}{ Our analysis suggests that the broken-power-law model is a strong candidate, successfully accounting for gravitational-waves while yielding a substantial fraction of PBH dark matter. While the log-normal scenario and its UV cut-off variant tend to overproduce PBHs, oscillatory patterns can efficiently prevent this.} 
{   Our analysis indicates that all scenarios face the issue of PBH overproduction when accounting for the PTA gravitational-wave signal. The case of log-normal power spectrum with oscillatory pattern reduces the PBH abundance, given that the critical threshold gains larger values. However, in the case of future gravitational-wave signals at higher frequencies, this overproduction problem may not persist. We should stress, however, that this article is about a novel methodology, which works regardless of the specific scenario it is applied to.}

The structure of this paper is as follow. In Section~\ref{sec:Methodology-of-Approach-for-Inverse-Problem} we propose a numerical approach to find the power spectrum from the gravitational-wave energy density using fitting techniques. In Section~\ref{sec:Application-to-Specific-Scalar-Power-Spectra} we apply our analysis to some proposed scenarios having power spectra described by either a broken power law, a log-normal peak, or a log-normal peak with a oscillatory pattern or with a UV cut-off { or with a plateau}. Section ~\ref{sec:Gravitational--Wave-Background-and-Primordial-Black-Holes} is devoted to the abundance of PBHs. This is followed by a conclusion and outlook in Section~\ref{sec:Conclusion-and-Outlook}. The Appendix supplements technical details of the evaluation of the gravitational-wave energy density of the scalar power spectrum.

%%%%%%%%%%%%%%%%%%%%%%%%%%%%%%%%%%%%%%%
\section{Methodology of Approach for Inverse Problem}
\label{sec:Methodology-of-Approach-for-Inverse-Problem}

This section outlines the methodology used for addressing the inverse problem in the context of scalar-induced gravitational waves. Our approach integrates theoretical power-spectrum models with artificial observational datasets, facilitating the extraction of significant model parameters. Through numerical integration and optimisation methods, we reconcile these theoretical predictions with controlled, simulated data. We detail the steps for generating and analysing artificial observational data, formulating theoretical models, and applying optimisation techniques to refine our understanding of the gravitational-wave signals.

Our primary objective is to design a methodology for optimal fitting of observational data~\cite{RoperPol:2022iel}. Contrary to the Bayesian statistical approach extensively explored in the literature~(see e.g.~Refs.~\cite{Ellis:2023oxs, NANOGrav:2023hvm, Vagnozzi:2023lwo, Yi:2023mbm, Domenech:2024rks, Chen:2024fir, InternationalPulsarTimingArray:2023mzf, Mitridate:2023oar, You:2023rmn, Handley:2020hdp, Ashok:2024fts, Breschi:2024qlc, Aurrekoetxea:2023vtp, Gupta:2023jrn, Hubner:2019sly, Matas:2020roi, Benetti:2021uea, Lamb:2023jls}, our strategy is anchored in optimisation techniques. This approach emphasises direct and objective analyses based solely on data, without the influence of subjective prior distributions. While Bayesian methods often integrate existing data with prior knowledge, our focus is on a straightforward approach that not only facilitates parameter estimation for theoretical models but can also be extended to more complex fitting techniques. We note that studies with Bayesian analyses using current data are already available (see e.g.~Refs.~\cite{NANOGrav:2023hvm, Mitridate:2023oar}).

Building upon our approach, we present a methodology that stands out for its minimal computational demands, enabling a more efficient exploration of gravitational-wave data, focusing on deriving insights directly from observations without the convolution of subjective priors. This direct, data-centric methodology allows for a more accessible analysis, potentially leveraging advanced computational techniques such as machine learning to unravel the gravitational-wave signals. Through analysing gravitational-wave data, we seek to unveil the origin of these signals, highlighting the significance of primordial black holes.
\vs{5mm}

%%%%%%%%%%%%%%%%%%%%%%%%%%%%%%%%%%%%%%%
\subsection{Observational Data}
\label{sec:Observational-Data}

For our analysis, we consider a dataset of $n$ wavenumbers, $k_{i} = \{ k_{1}, k_{2}, \ldots, k_{n} \}$, with each wavenumber $k_{i}$ corresponding to an observational data point $w_{i}^{\text{exp}}$, representing the observed value of $w^{\text{exp}}$ at $k_{i}$. Consequently, we formulate a comprehensive dataset represented by
\begin{equation}
\label{eq:dataset}
    \Omega_{\rm GW}^{\rm exp}( k )
        =
            \big\{
                w_{1}^{\rm exp}, \dots, w_{n}^{\rm exp}
            \big\}
            \, .
\end{equation}
Each observational data point $w_{i}^{\rm exp}$ comes with an associated error $\sigma_{i}$. We remark that, in our present illustrative study, we use simulated data for this process. In our simulation for the gravitational-wave energy density, $\Omega_{\rm GW}$, we use Gau{\ss}ian noise to realistically mimic observational errors of actual data. 
\vs{2mm}

%%%%%%%%%%%%%%%%%%%%%%%%%%%%%%%%%%%%%%%
\subsection{Theoretical Data}
\label{sec:Theoretical-Data}

In the theoretical framework, we consider a power spectrum model, denoted by $\Pcal_{R}^{\text{th}}(a; k)$, where $a$ represents a set of parameters that characterises the model. For instance, in a Gau{\ss}ian case, these parameters would be $a = (\mu, \sigma)$, corresponding to mean ($\mu$) and standard deviation ($\sigma$). For each wavenumber $k_{i}$ within our observational range, we compute the theoretical value $w_{i}^{\text{th}}( a )$, which corresponds to the gravitational-wave energy density $\Omega_{\rm GW}( k_{i} )$ predicted by the model. This process yields a series of values $\Omega_{\rm GW}^{\text{th}} = \{ w_{1}^{\text{th}}( a ), \ldots, w_{n}^{\text{th}}( a ) \}$.

The computation of $\Omega_{\rm GW}( k )$ involves numerical integration over the spectrum of primordial perturbations (a detailed analysis is presented in the Appendix). Due to the complexity of the integral and the need for computational efficiency, we apply a numerical integration that approximates the integral as a sum over chosen sample points and their corresponding weights. For our two-dimensional integral, the computation is represented as
\begin{equation}
\begin{split}
    &\Omega_{\rm GW}( k )
        =
            0.387\;
            \Omega_{\Rrm}\.
            \frac{ 1 }{ 6 }\mspace{-3mu}
            \left(
                \frac{ g_{*,s}^{4}\,g_{*}^{-3} }
                { 106.75 }
            \right)^{\!-1/3}
            \times
            \\[1mm]
    &\sum_{i,j =1}^{N}
        w_{i}\.w_{j}\.
        \Pcal_{R}\!
        \left(
            \frac {y_{j} - x_{i} }{ 2 }\.k
        \right)\!
        {\Pcal_{R}}\!
        \left(
            \frac{ x_{i}\.y_{j} }{ 2 }\.k
        \right)
        \overline{I_{\rm RD}^{2}( x_{i}, y_{j} )}
        \, .
\end{split}
\label{eq:omega_num}
\end{equation}
Here, $w_{i}$ and $w_{j}$ are the weights associated with the sampling points as determined by the numerical integration process, specifically using Gau{\ss}ian quadrature based on the Gau{\ss}-–Legendre or Gau{\ss}--Kronrod quadrature formul{\ae}~\cite{Piessens1983-wl, 10.5555/232468}. This approach allows for an accurate estimation of $\Omega_{\rm GW}$, which is essential for a comparison to observational data. We note that a Monte Carlo integration can also give accurate result, but it requires substantially more iterations to obtain similar results. Hence, in order to minimise the computational time, we utilise the aforementioned formul{\ae}.
\vs{2mm}

%%%%%%%%%%%%%%%%%%%%%%%%%%%%%%%%%%%%%%%
\subsection{Evaluation with Minimisation}
\label{Evaluation-with-Minimisation}

Optimising the theoretical model to align with observational data requires a meticulous approach, where we employ the chi-squared $\chi^{2}$-function,
\vs{2mm}
\begin{equation}
    \chi^{2}( a )
        =
            \sum_{i=1}^{n}
            \frac{ [w_{i}^{\text{exp}} - w_{i}^{\text{th}}( a )]^{2} }
            { \sigma_{i}^{2} }
            \, .
            \\[-4mm]
            \notag
\end{equation}
%\newpage

\noindent The $\chi^{2}$-minimisation with respect to parameter vector $a$ is pivotal for extracting the optimal model parameters. We use the Nelder--Mead simplex algorithm~\cite{NeldMead65}, being a robust, derivative-free optimisation method. It excels in handling complex, nonlinear problems, especially when derivatives are unavailable or impractical to compute~\cite{Desesquelles:2009kw, 10.5555/232468}. This algorithm iteratively adjusts a simplex{\,---\,}a geometric figure formed by $N + 1$ points in an $N$-dimensional parameter space{\,---\,}to converge on the function's minimum. Whilst it may not be as rapid as gradient-based methods, particularly for high-dimensional problems, its simplicity and versatility make it highly effective in our case, i.e.~for models with a moderate number of parameters, offering a balance between reliability and computational efficiency.

%%%%%%%%%%%%%%%%%%%%%%%%%%%%%%%%%%%%%%%
\section{Application to Specific Scalar Power Spectra}
\label{sec:Application-to-Specific-Scalar-Power-Spectra}

In this section, we apply our methodology to {five} distinct models of power spectra. Firstly, we explore the {\it Broken Power-Law} power spectrum, renowned for its distinct characteristics in single-field inflationary models~(cf.~Refs.~\cite{Vaskonen:2020lbd, Byrnes:2018txb}). Secondly, we delve into the standard {\it Log-Normal} power spectrum, appreciated for its notable features in the abundances of gravitational waves as, for instance, studied in Ref.~\cite{Ellis:2023oxs}. Thirdly, our focus shifts to a nuanced variant of the {\it Log-Normal} model, enhanced by {\it oscillatory elements}, thus offering enriched insights into gravitational-wave analysis as it was discussed in Ref.~\cite{Fumagalli:2021cel}. Lastly, we study a specialised form of the {\it Log-Normal} power spectrum, which integrates an {\it exponential UV cut-off} (see Ref.~\cite{Vaskonen:2020lbd}), as well as a {\it Log-Normal} spectrum with a plateau (see Refs.~\cite{Balaji_2022, Frolovsky_2023}). %\textcolor{blue}{Lastly, we explore the case of Log-Normal power spectrum with an plateau at small scales capable to explain both the production of GWs at small scales and the observable constraits of CMB at large scales \cite{Frolovsky_2023}. }

%%%%%%%%%%%%%%%%%%%%%%%%%%%%%%%%%%%%%%%
\subsection{Broken Power-Law}
\label{sec:Broken-Power--Law}

\begin{figure}
    \centering
    \vs{-5mm}
    \includegraphics[width=0.49\textwidth]{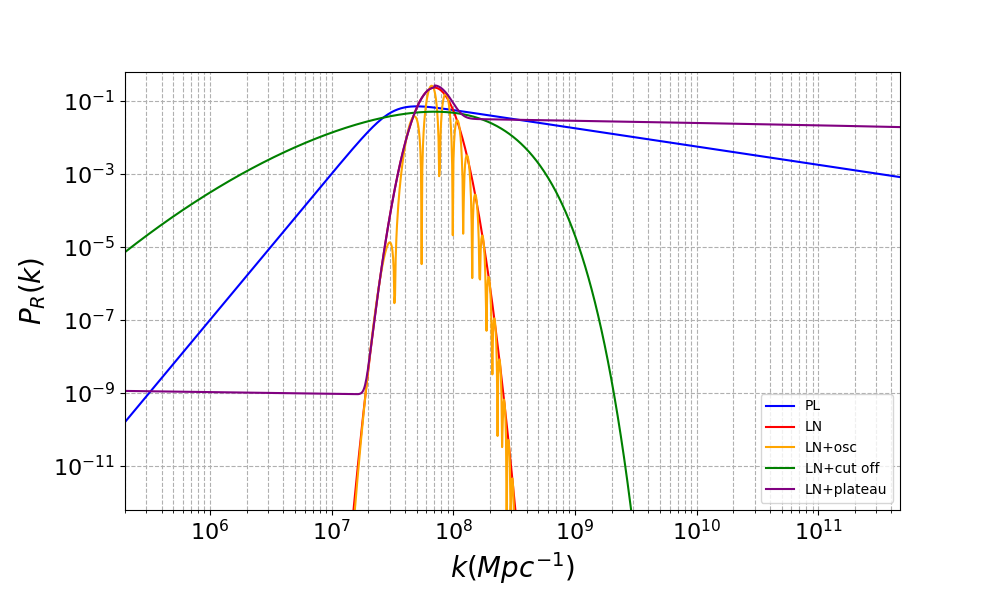}
    \caption{
        Power spectrum with the shapes:
            power law ``PL" (blue line), 
            log-normal ``LN" (red line), 
            log-normal with oscillations ``LN+osc" (orange line) and 
            log-normal with a cut-off ``LN+cut-off"(green line), log-normal with a plateau (purple line).    See main text for parameter specification. 
        }
    \label{fig:power}
\end{figure}

In the context of single-field inflation, peaks are often approximated by a broken-power-law power (PL) spectrum, which is described as
\begin{equation}
    \Pcal_{\rm PL}
        =
            A_{1}\.
            \frac{ \alpha_{1} + \beta_{1} }
            { \beta_{1}\!
            \left(
                k/k_{*}
            \right)^{-\alpha_{1}}
            + \alpha_{1}\!
            \left(
                k/k_{*}
            \right)^{\beta_{1}} }
            \, .
    \label{eq:p_power_Law}
\end{equation}
The parameters $A_{1}$ and $k_{*}$ describe the amplitude and position of the peak, respectively, and $\alpha_{1}$ and $\beta_{1}$ determine the growth and decay around it. The blue line in Fig.~\ref{fig:power} depicts the broken-power-law power spectrum. 

\begin{figure}[ht]
    \centering
    \vs{-6mm}
    \includegraphics[width=0.49\textwidth]{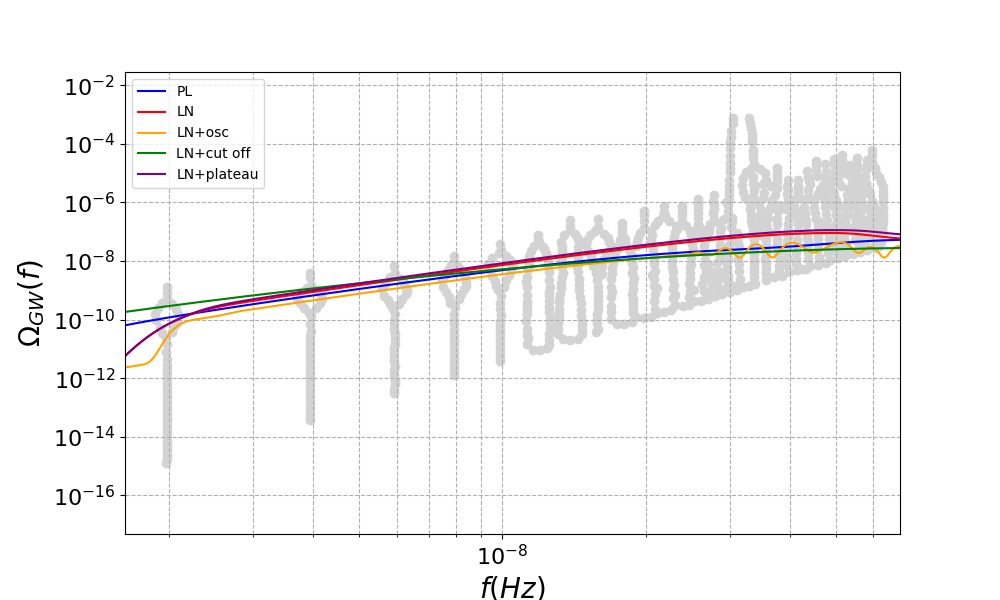}
    \caption{
        Energy density for the cases of power law ``PL" (blue line), 
            log-normal ``LN" (red line), 
            log-normal with oscillations ``LN+osc" (orange line),  
            log-normal with a cut-off ``LN+cut-off"(green line), \textcolor{black}{log-normal with a plateau (purple line). }
        Gray dots corresponds to the NANOGrav signal. 
        \vs{2mm}
        }
    \label{fig:gw_pta}
\end{figure}

\begin{figure}[ht]
    \centering
    \vs{-5mm}
    \includegraphics[width=0.49\textwidth]{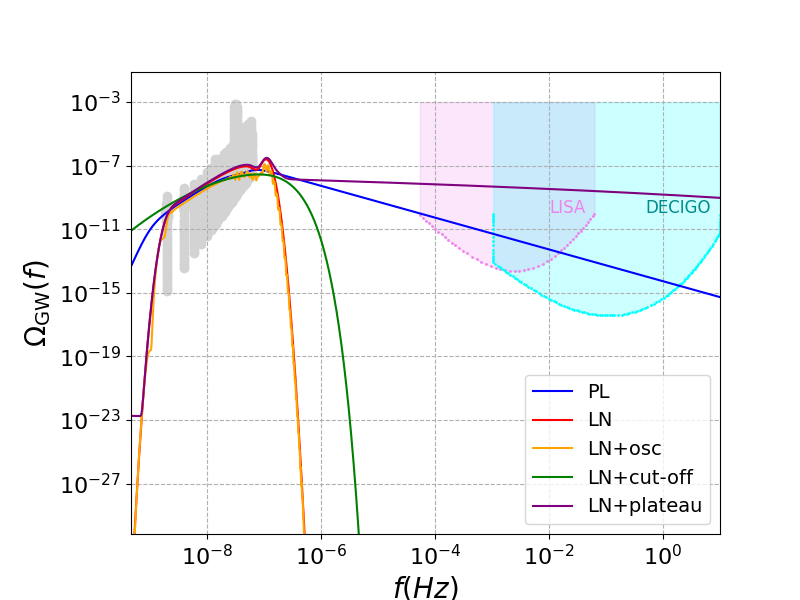}
    \caption{
        Energy density for Eqs.~(\ref{eq:p_power_Law}-\ref{eq:p_Log_normal_cut}) 
        with blue, red, orange, green \textcolor{black}{ and purple} color, respectively, as in Fig.\ref{fig:gw_pta}. Gray dots corresponds to NANOGrav signal. Also included are the prospective sensitivity regions of LISA (violet) and DECIGO (cyan).
        \vs{3mm}
        }
    \label{fig:qw_all}
\end{figure}

\begin{figure*}[ht]
    \centering
    \vs{-5mm}
    \includegraphics[width=0.49\textwidth]{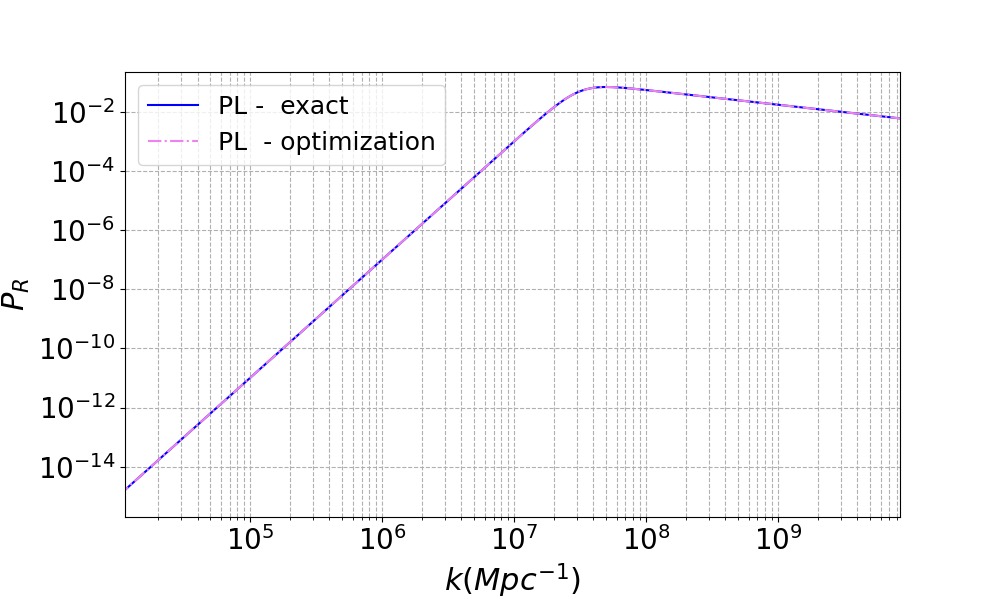}\hspace{2.5mm}
    \includegraphics[width=0.49\textwidth]{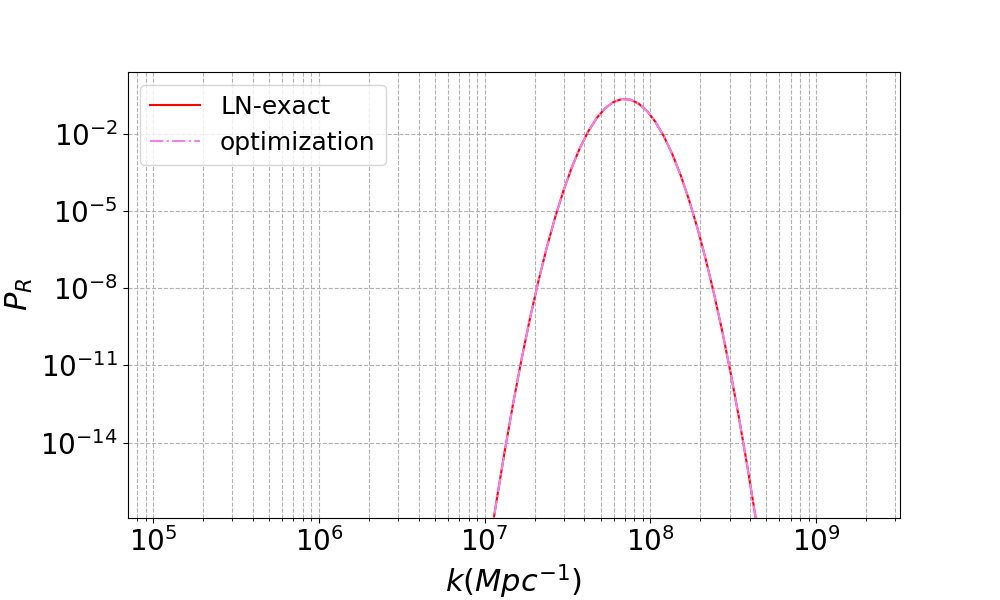}\hspace{3mm}\\[-3.5mm]
    \includegraphics[width=0.49\textwidth]{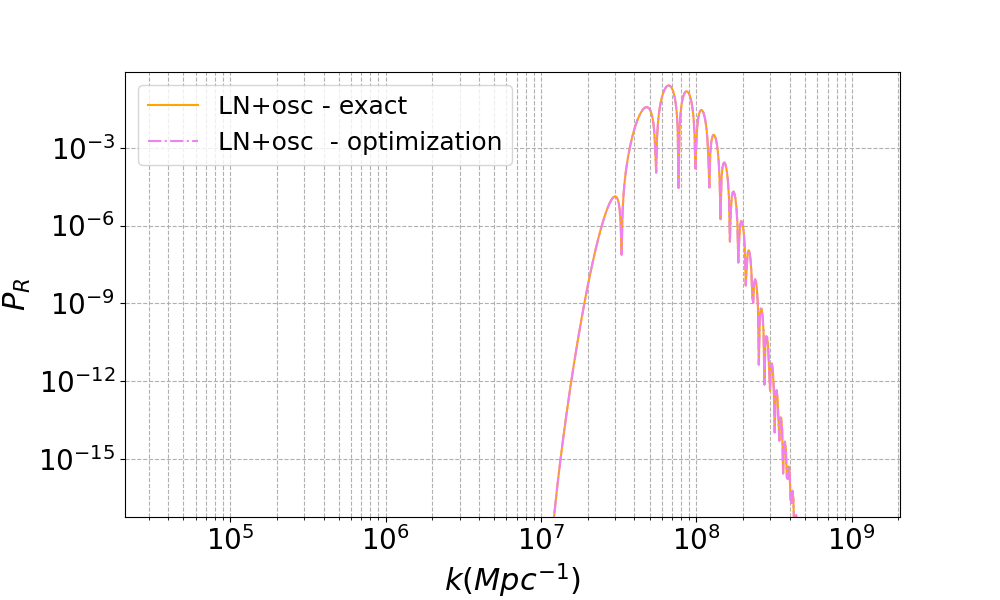}
    \includegraphics[width=0.49\textwidth]{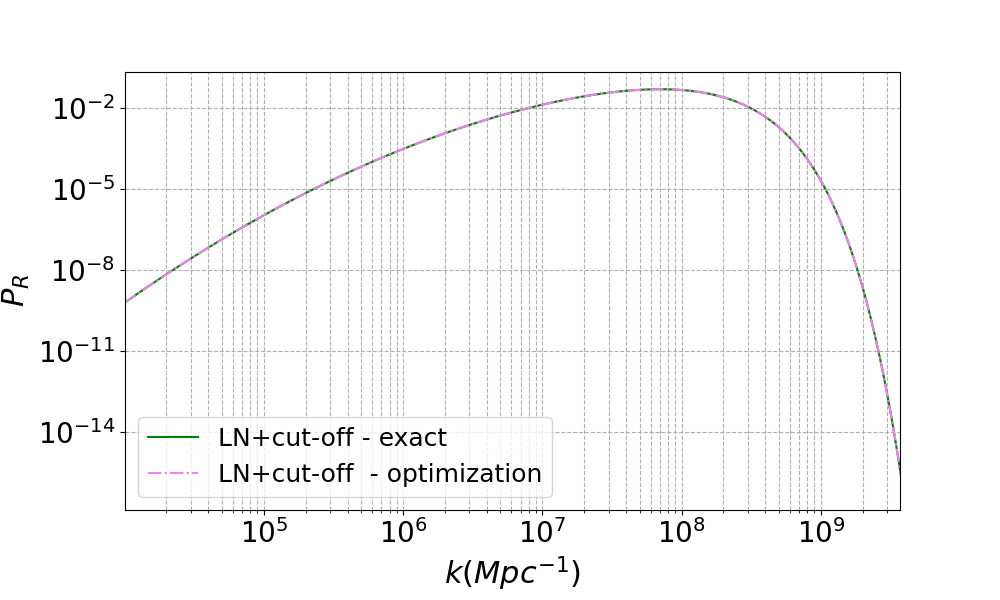}
      \caption{
        Power spectra from a given dataset of gravitational waves:
            broken power law (top-left),
            log-normal (top-right), 
            log-normal with oscillations (bottom-left) and 
            log-normal with a cut-off (bottom-right).
        }
  \label{fig:pr_recon}
\end{figure*}

\begin{figure*}[ht]
    \centering
    \vs{-5mm}
    \includegraphics[width=0.48\textwidth]{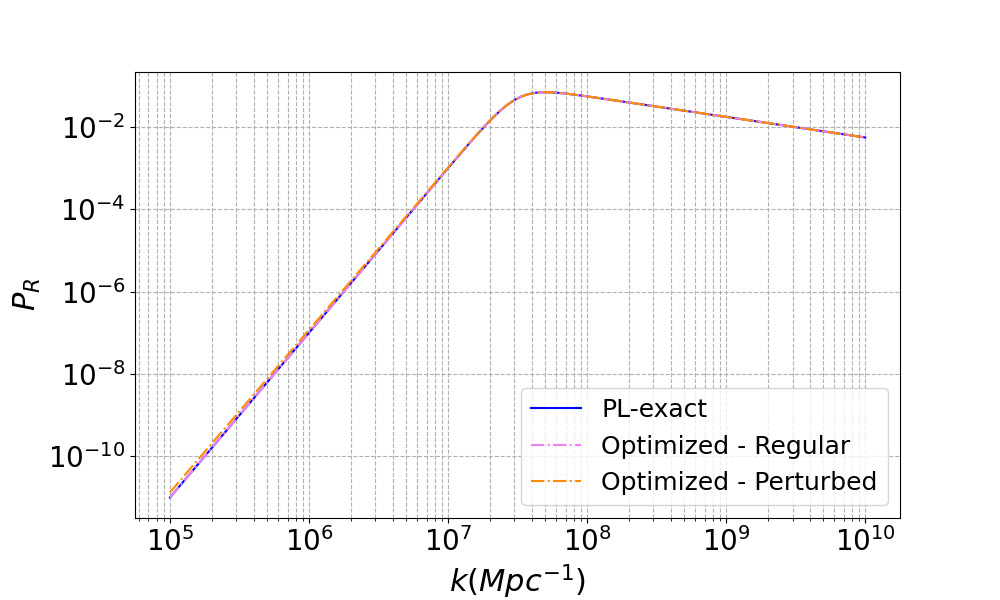}\hspace{3mm}% 
    \includegraphics[width=0.48\textwidth]{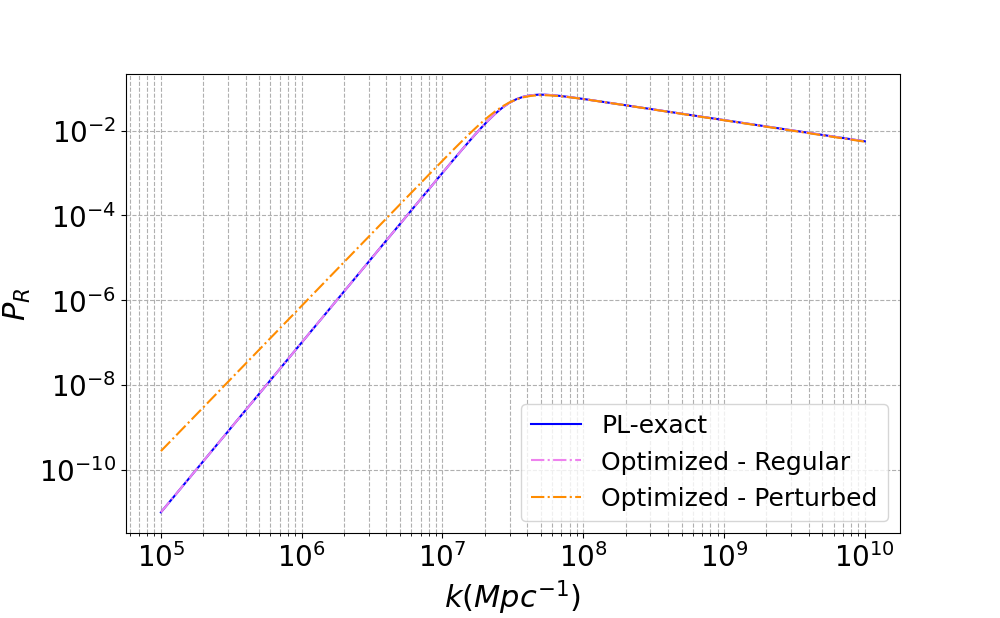}\hspace{3mm}
      \caption{
        Power spectra from a given gravitational-wave dataset with 1\% (left) and 10\% (right) noise.
        }
  \label{fig:pr_pertu}
\end{figure*}

\begin{figure}[ht]
\includegraphics[width=0.49\textwidth]{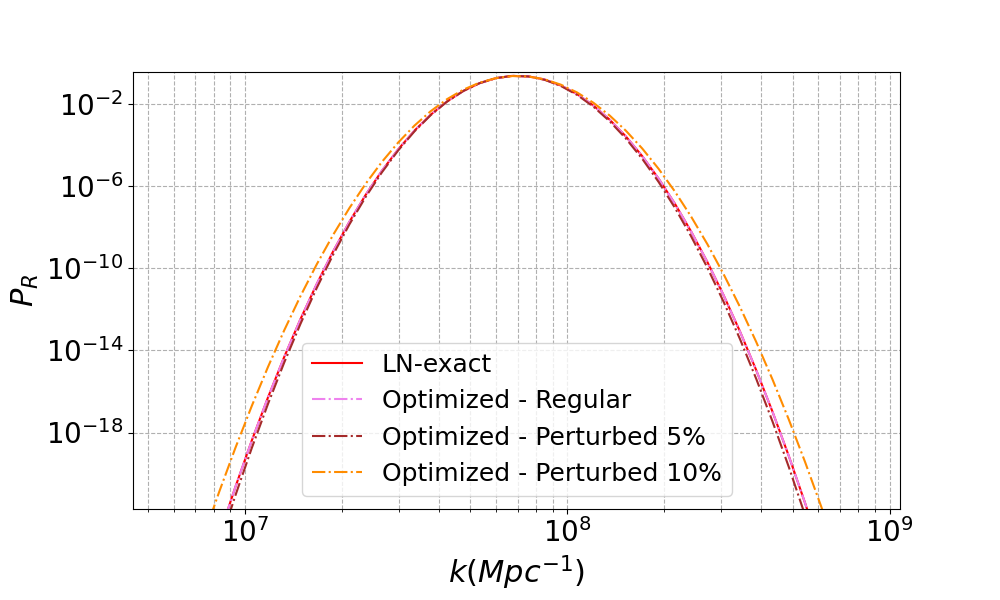}\hspace{3mm}% 
    \caption{
        Power spectra from a given gravitational-wave dataset of with a noise level of 5\% (brown dot-dashed curve) and 10\% (yellow dot-dashed curve) for the case of a log-normal power spectrum.
        }
  \label{fig:pr_pertu2}
\end{figure}

In our analysis, we have selected specific parameters for the broken-power-law model: $\alpha_1 = 4$, $\beta_1 = 0.5$, $A_1 = 0.07$, and $k_{*} = 5 \times 10^{7}\,{\rm Mpc}^{-1}$. This choice is based on their relevance to single-field inflation. Specifically, the hypothesis of adiabatic Gau{\ss}ian scalar fluctuations, characterised by a power spectrum with simple power-law form, stands as a central prediction of standard slow-roll models, which has been tested with unprecedented accuracy~\cite{Planck:2018jri}. We then compute the integral in Eq.~\eqref{eq:omega_gw_1}, generating the dataset as described in Eq.~\eqref{eq:dataset} and depicted in Figs.~\ref{fig:gw_pta} and~\ref{fig:qw_all}, where the blue line represents our results. \textcolor{black}{For our analysis, we utilize the 14 frequency bins from the NANOGrav 15-year dataset (violins) \cite{NANOGrav:2023gor}.} We note that the energy density from the broken-power law \textcolor{black}{ and lognormal with a plateau} align with both the PTA data as well as the detectability range of future observation from LISA and DECIGO (see Fig.~\ref{fig:qw_all}).

Our fitting process is aimed at an accurate reconstruction of the power spectrum. The corresponding results are shown in Fig.~\ref{fig:pr_recon} (top-left panel), with the dashed line indicating the fitted values, and the solid line representing those of the exact power spectrum. Our analysis shows consistency between the theoretical model and observed data.
%\newpage

In order to test the resilience of our methodology against data variability, we conduct an analysis which introduces Gau{\ss}ian noise perturbations into our dataset, simulating observational uncertainties. We apply noise levels of 1$\%$ and 10$\%$ to assess the impact of varying noise intensities on our results. The effects of these perturbations on the reconstructed power spectra are shown in Fig.~\ref{fig:pr_pertu}, with the left panel depicting the 1$\%$-noise scenario and the right panel showing the 10$\%$-noise scenario. We note that order-one noises does not affect the results.
\vs{3mm}

%%%%%%%%%%%%%%%%%%%%%%%%%%%%%%%%%%%%%%%
\subsection{Log-Normal Peak}
\label{sec:Log-Normal-Peak}

A broad class of power spectra, particularly relevant for primordial black hole formation and gravitational-wave production, can be effectively described using a log-normal (LN) function, relevant in diverse inflationary models (see e.g.~Ref.~\cite{Pi:2020otn}). It can be expressed as
\begin{equation}
    \Pcal_{\rm LN}
        =
            \frac{ A_{2} }{ \sqrt{2\mspace{1mu}\pi}\.\Delta }\,
            \exp\!
            \left( 
                - \frac{ \ln^{2}( k / k_{*} ) }
                { 2\.\Delta^{2} }
            \right)
            .
    \label{eq:p_Log_normal}
\end{equation} 
Prior research, particularly in the context of PTA analyses, has identified optimal parameters for this model, these being $\Delta = 0.21$, $A_{2} = 0.12$ and $k_{*}  = 7 \times 10^{7}\,{\rm Mpc}^{-1}$~\cite{Ellis:2023oxs}. The corresponding power spectrum is illustrated in Fig.~\ref{fig:power} by a red line.

We proceed to calculate the gravitational-wave energy density and establish our dataset accordingly. The latter is graphically represented in Figs.~\ref{fig:gw_pta} and \ref{fig:qw_all}, where the red line corresponds to the outcome for the log-normal model. It can be observed that, while our chosen parameters align well with the PTA data, they diverge from the expected results at higher frequencies, as previously noted in the study by Ellis {\it et al.}~\cite{Ellis:2023oxs}. This divergence underscores the model's limitations in capturing signals across a broader frequency range
 
Using our methodology, we successfully reconstruct the power spectrum, the results being showcased in Fig.~\ref{fig:pr_recon} (upper-right panel). The dashed lines in the figure represent our fitting techniques, while the solid lines indicate the exact data. This comparison highlights the efficacy of our approach. Notably, incorporating varying levels of noise into the data does not significantly change the results (cf.~Fig.~\ref{fig:pr_pertu2}).
\vs{4mm}

%%%%%%%%%%%%%%%%%%%%%%%%%%%%%%%%%%%%%%%
\subsection{Log-Normal Peak with Oscillatory Pattern}
\label{Log--Normal-Peak-with-Oscillatory-Pattern}

Extending the investigations of the previous subsection, we introduce an oscillatory pattern to the log-normal distribution. This addition is motivated by multiple factors. First, inflationary potentials featuring step-like characteristics~(see e.g.~Refs.~\cite{Adams:2001vc, Kefala:2020xsx, Cai:2021zsp}) or those within the context of natural inflation, such as multi-axion field models (see e.g.~Refs.~\cite{Ballesteros:2019hus, Mavromatos:2022yql}), can induce oscillations in the power spectra. Second, oscillatory behavior in the power spectrum can manifest as characteristic signal in the energy densities of gravitational waves~\cite{Fumagalli:2021cel}. Additionally, such a pattern allows for a more nuanced fit, resembling a comb structure for individual peaks~\cite{Fumagalli:2021cel}. Finally, as elaborated in the subsequent sections, these oscillations can play a crucial r{\^o}le in avoiding PBH overproduction~\cite{Cook:2022zol}. The power spectrum is given by
\begin{equation}
\begin{split}
    \Pcal_{\rm LN}^{\rm osc}
        =
            \frac{ A_{2} }{ \sqrt{2\mspace{1mu}\pi}\.\Delta}\,
            \exp\!
            \left(
                -\frac{ \ln^{2}( k/k_{*} ) }
                { 2\.\Delta^{2} }
            \right)
            \cos\!
            \left(
                d\.\frac{ k }{ k_{*} }
            \right)^{\!2}
            .
    \label{eq:p_Log_normal_osc}
    \end{split}
\end{equation}
We retain the parameter values from Eq.~\eqref{eq:p_Log_normal} while introducing an additional parameter $d = 10$. This value determines the frequency of oscillations in the power spectrum. It also impacts the computational time required to evaluate the energy densities of the gravitational waves. The selected value corresponds to a minimal choice, ensuring that oscillations are present in the energy densities. The power spectrum of Eq.~\eqref{eq:p_Log_normal} is depicted in Fig.~\ref{fig:power} by an orange line. 

We compute the gravitational-wave energy density to develop our dataset, being presented in Figs.~\ref{fig:gw_pta} and~\ref{fig:qw_all} by orange lines. These reveal distinct oscillatory patterns in the energy density, reflecting the unique characteristic of this model. Figure~\ref{fig:pr_recon} (bottom-left panel) shows the successful reconstruction of the power spectrum using our fitting techniques. The dashed line represents the model after the fitting process. As can be seen, our resulting curve closely aligns with the theoretical model. 
%\newpage

%%%%%%%%%%%%%%%%%%%%%%%%%%%%%%%%%%%%%%%
\subsection{Log-Normal Peak with Exponential UV Cut-Off}
\label{sec:Log-Normal-Peak-with-Exponential-UV-Cut--Off}

A log-normal power spectrum with an exponential UV cut-off has also be studied in Ref.~\cite{Vaskonen:2020lbd}. Here, we assume
\begin{equation}
    \Pcal_{\rm LN}^{\rm cut}
        =
            A_{3}
            \exp\!
            \left\{
                \beta_{3}\!
                \left[
                    1 - \frac{ k }{ k_{*} }
                    + \ln\!
                    \left(
                        \frac{ k }{ k_{*} }
                    \right)
                \right]
                -
                \alpha_{3}\.
                \ln^{2}\!
                \left(
                    \frac{ k }{ k_{*} }
                \right)\!
            \right\}
            .
    \label{eq:p_Log_normal_cut}
\end{equation}
The utilised parameter values correspond to those in Ref.~\cite{Vaskonen:2020lbd}: $\alpha_{3} = 0.17$, $\beta_{3} = 0.62$, $A_{3} = 0.05$, and $k_{*} = 7 \times 10^{7}\,{\rm Mpc}^{-1}$. The resulting spectrum is shown in Fig.~\ref{fig:power} (green line).

As in the previous cases, we obtain our dataset (depicted in Figs.~\ref{fig:gw_pta} and~\ref{fig:qw_all}) for the gravitational-wave energy density. We note that it aligns with the PTA results. Using our fitting techniques, we have reconstructed the power spectrum, as shown in the bottom-right panel of Fig.~\ref{fig:pr_recon}. 
\vs{3mm}

%Therefore, we have successfully reconstructed the power spectrum from a given dataset for all \textcolor{blue}{five} models studied above. We remark that a dataset can be analysed using all of the four different models, allowing to accurately identify the most appropriate model for each case. However, for an arbitrary dataset, one needs to incorporate more models, making it hence imperative to proceed to higher optimisation techniques. 

%%%%%%%%%%%%%%%%%%%%%%%%%%%%%%%%%%%%%%%
\subsection{Log-Normal Peak with Plateau}
\label{sec:Log-Normal-Peak-with-Plateau}

\textcolor{black}{The primordial power spectrum in inflationary models generally features a plateau at CMB scales, which is relevant when studying the power spectra related to PBH and GW production. The presence of such a plateau imposes constraints on the peak's width, as noted in \cite{Frolovsky_2023}. Additionally, the interaction between the plateau and peak modes introduces distinctive features in the induced GW spectrum, especially in cases with sharp peaks \cite{Balaji_2022}. Therefore, it is important to examine power spectra that include both a plateau and a peak.}

\textcolor{black}{A simple fit of this type was proposed in \cite{Frolovsky_2023}, given by the following expression:}
\begin{equation}\label{LN1Plateau}
    \Pcal_{\rm LN}^{\rm pl}
        =
            P_{0} \ln^{2}
            \Bigg(
                \frac{k}{k_{\text{end}}}
            \Bigg)
            + 
            \frac{ A_{2} }{ \sqrt{2\mspace{1mu}\pi}\.\Delta}\,
            \exp\!
            \left(
                -\frac{ \ln^{2}( k/k_{*} ) }
                { 2\.\Delta^{2} }
            \right)
            ,
\end{equation}
\textcolor{black}{where the first term is the power spectrum of T-type $\alpha$-attractor models \cite{Kallosh_2013, Galante_2015} in slow-roll approximation. A power spectrum in the form of a sum of a Log-Normal peak and a plateau is typical for multi-field inflationary models (see Section 2.2 of Ref.~\cite{LISACosmologyWorkingGroup:2023njw}, for a review, and references therein).}

\textcolor{black}{Many inflationary models that predict the formation of PBHs also feature a secondary plateau beyond the peak. For a detailed discussion, see Ref.~\cite{Balaji_2022}.
Here, we extend Eq.~\eqref{LN1Plateau} to account for a second plateau following the peak, which has a different amplitude than the plateau associated with CMB scales:}
\begin{equation}
\label{LNplateau}
\begin{split}
    \Pcal_{\rm LN}^{\rm 2pl}
        &=
            \ln^{2}\Bigg(\frac{k}{k_{\text{end}}}\Bigg)\Bigg(P_{0} + A_{\text{rel}}+A_{\text{rel}}\tanh( k - k_{*} )\Bigg)
            \\
        &+
            \frac{ A_{2} }{ \sqrt{2\mspace{1mu}\pi}\.\Delta}\,
            \exp\!
            \left(
                -\frac{ \ln^{2}( k/k_{*} ) }
                { 2\.\Delta^{2} }
            \right)\,.
    \end{split}
\end{equation}
\textcolor{black}{We maintain the parameter values from Eq.~\eqref{eq:p_Log_normal} and introduce additional parameters. The value $P_{0} = 6.24\times10^{-13}$ fixes the amplitude of the first plateau, $k_{\text{end}} = 7.7 \times 10^{23}\,\text{Mpc}^{-1}$ denotes the mode exiting the horizon at the end of inflation, and $A_{\text{rel}} = A_{2} / 10^{4}$ represents the amplitude of the second plateau.}
\begin{figure}
    \centering
    \includegraphics[width=0.99\linewidth]{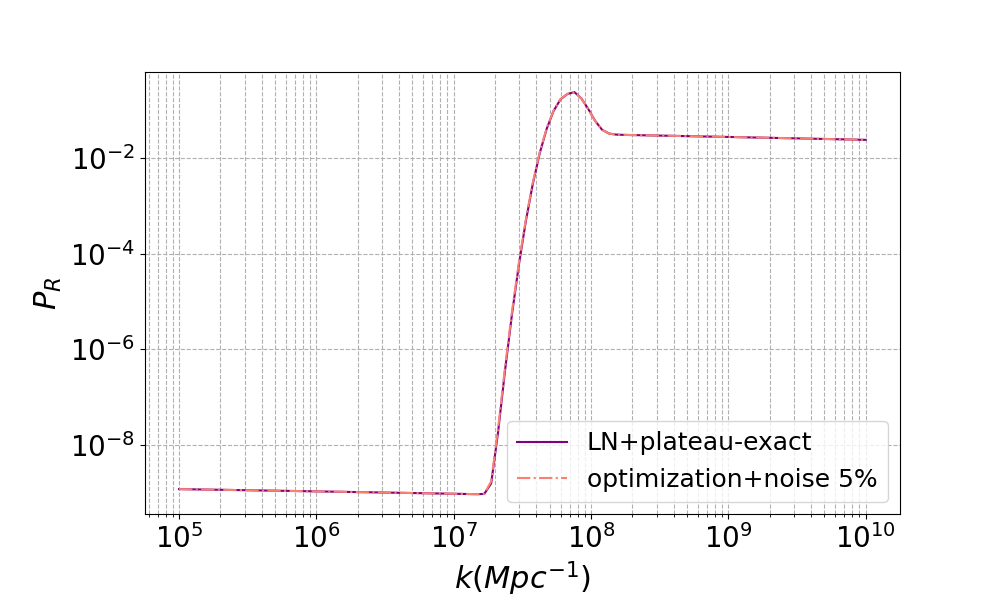}
    \caption{
        Power spectrum for a given gravitational-wave dataset for the case of log-normal power spectrum with a plateau.}
    \label{fig:power_spectra_plateau}
\end{figure}

{As in previous power spectra, we derive our dataset for the gravitational-wave energy density, illustrated in Figs.~\ref{fig:gw_pta} and~\ref{fig:qw_all}, which shows consistency with the PTA observations. By applying our fitting techniques, we successfully reconstruct the power spectrum, presented in the bottom-right panel of Fig.~\ref{fig:power_spectra_plateau}.}

Therefore, we have successfully reconstructed the power spectrum from a given dataset for all {five} models studied above. We remark that a dataset can be analysed using all of the {five} different models, allowing to accurately identify the most appropriate model for each case. However, for an arbitrary dataset, one needs to incorporate more models, making it hence imperative to proceed to higher optimisation techniques.

%%%%%%%%%%%%%%%%%%%%%%%%%%%%%%%%%%%%%%%
\section{Gravitational-Wave Background and Primordial Black Holes}
\label{sec:Gravitational--Wave-Background-and-Primordial-Black-Holes}

In this section, we elaborate on how PBH formation is facilitated by the different power spectra discussed above (see Refs.~\cite{Vaskonen:2020lbd, Inomata:2023zup} for similar investigations). As discussed in the literature (see e.g.~the recent Refs.~\cite{Musco:2018rwt, Musco:2020jjb, Escriva:2020tak, Escriva:2019phb, Stamou:2023vxu}), the shape of the latter has a vital influence on the former. For each of the power spectra (\ref{eq:p_power_Law}--\ref{LN1Plateau}) we carefully take into account the corresponding PBH formation threshold.

\textcolor{black}{ The mass of the PBHs, $M$, can be related to the wavenumber $k$ by }{\cite{Carr:1975qj, Carr:1974nx, Young:2019yug, PhysRevLett.70.9, PhysRevLett.70.9, Evans:1994pj, Koike:1995jm, Frosina:2023nxu} 
\begin{equation}
    M( k )
        =
            \Kcal\.M_{\Hrm}
            \left[
                \left(
                    \delta_{\Lrm} - \frac{1}{4\Phi}\delta_{\Lrm}^{2}
                \right)
                -
                \delta_{\crm}
            \right]^{\gamma}\.,
\end{equation}
{where} 
\begin{equation}
    M_{\Hrm}
        =
            17
            \left(
                \frac{g_{*}}{106.75}
            \right)^{-1/6}
            \left(
                \frac{k}{10^{6}\.{\rm Mpc}^{-1}}
            \right)^{-2}
            M_{\odot}
            \, .
\end{equation}
\textcolor{black}{Here, $\gamma$ depends on the details of gravitational collapse ($\gamma = 0.38$ in our analysis), and $\delta_{\crm}$ is the critical threshold we will discuss later. The parameter $\Phi$ in this equation controls the strength of the quadratic term in $\delta_{\Lrm}$, adjusting how non-linear contributions to $\delta_{\Lrm}$ scale relative to $\zeta$. For a radiation-dominated universe, we adopt the standard value of $\Phi = 2/3$ to align with typical cosmological models. The proportionality constant $\Kcal$ relates the horizon mass $M_{\Hrm}$ to the resulting PBH mass $M$ after collapse, accounting for the efficiency of the collapse process. In a radiation-dominated universe, we take $K = 4.36$  \cite{Frosina:2023nxu}. This factor adjusts the final PBH mass relative to the horizon mass at the time of re-entry, fine-tuning the mass prediction in PBH formation models~\cite{Frosina:2023nxu}.}

\textcolor{black}{In this context, $\delta_{\Lrm}$ denotes the linear component of the density contrast field, quantifying the fractional overdensity relative to the cosmic average as the perturbation re-enters the horizon. Originating from the primordial curvature perturbation $\zeta$, $\delta_{\Lrm}$ incorporates non-Gau{\ss}ianities through quadratic and higher-order terms, influencing the likelihood of exceeding the collapse threshold $\delta_{\text{c}}$ for PBH formation \cite{Young:2019yug, DeLuca:2019qsy, Frosina:2023nxu}. Specifically, $\delta_{\Lrm}$ is expressed as}
\begin{equation}
    \delta_{\Lrm}
        =
            - \frac{4}{3}r_{\mrm}\zeta'(r_{m})
            \, ,
\end{equation}
\textcolor{black}{where $r_{\mrm}$ is the comoving radius at which the compaction function reaches its maximum, and $\zeta'(r_{\mrm})$ is the radial derivative of the curvature perturbation at this scale.}

%\textcolor{red}{The fraction of collapsed horizon patches, denoted as $\beta$, can be estimated using the Press--Schechter (PS) approach~\cite{Press74} or Peak Theory~\cite{Bardeen:1985tr}. Our present study concentrates on the former approach, wherein $\beta$ is determined by the probability that the overdensity $\delta$ surpasses a specific collapse threshold, denoted as $\delta_{\crm}$. Then, the fraction $\beta$ is derived from}
%\vs{2mm}
%\begin{equation}
%    \textcolor{red}{\beta_{\rm PS}( M )
%        =
%            \frac{ 1 }{ 2\.\sqrt{\pi} }\,
%            \Gamma\!
%            \left(
%                \frac{ 1 }{ 2 }, 
%                \frac{ \delta_{\crm}^{2} }
%                { 2\.\sigma^{2} }
%            \right)}
%            ,
%            \label{eq:beta_ps}
%\end{equation}
%\newpage

{The mass fraction of PBHs, denoted as $\beta(M)$, is given by}
\begin{equation}
\begin{split}
    \beta_{\rm PS}( M )
        &=
            \int_{\delta_{\crm}}^{\infty}\d\delta\;
            \frac{M_{\rm PBH}}{M_{\Hrm}}\.P( \delta )
            \\
        &=
            \Mcal \int^{\delta_{\Lrm}^{\text{max}}}_{\delta_{\Lrm}^{\text{min}}}\d\delta\;
            \left(
                \delta_{\Lrm} - \frac{ 1 }{ 4 \rm{\Phi} } {\delta}_{\Lrm} - \delta_{\text{th}}
            \right)^{\!\gamma}
            P_{\Grm}(\delta_{\Lrm})
            \, ,
    \end{split}
\end{equation}
\textcolor{black}{where}
\begin{equation*}
    \delta_{\Lrm}^{\rm max}
        =
            2 \Phi ,\quad 
    \delta_{\Lrm}^{\rm min}
        =
            2\Phi( 1 - \sqrt{1-\delta_{\crm}/\Phi} )
            \, .
\end{equation*}
\textcolor{black}{and for $P_{\Grm}( \delta_{\Lrm} )$ we assume}
\begin{equation}
    P_{\Grm}( \delta_{\Lrm} )
        =
            \frac{1}{\sqrt{2 \sigma^{2}}}\,{\rm Exp}\!
            \left[
                \frac{ -\,\delta_{\Lrm}}
                {2 \sigma^{2}( M_{\Hrm} )}
            \right]\, .
\end{equation}

\noindent \textcolor{black}{Here, $\sigma$ is the coarse-grained density constant, given by}
\begin{equation}
\label{eq:sigma_1}
    \sigma^{2}\big[ M( k ) \big]
        =
            \frac{ 4\.( 1 + w )^{2} }
            { (5 + 3\.w )^{2} }
            \int \frac{ \d k' }{ k' }
            \left(
                \frac{ k' }{ k }
            \right)^{\!4}
            \Pcal_{R}( k' )\,
            {\tilde W}^{2}\!
            \left(
                \frac{ k' }{ k }
            \right)
            ,
\end{equation}
\textcolor{black}{with $\tilde W( k' / k )$ being the Fourier top-hat window function}
\begin{equation}
    W( y )
        =
            3
            \left[
                \frac{ \sin( y ) - y\,\cos( y ) }{ y^{3} }
            \right]
            .
\end{equation}
The choice of the window function introduces corrections in the calculation of the threshold; however, these modifications become less significant in computing the PBH abundance when the same window function is used to evaluate the variations~\cite{Young:2019osy, Tokeshi:2020tjq, Musco:2020jjb}.

As obvious from Eq.~\eqref{eq:beta_ps}, the fraction $\beta$ is extremely sensitive to the value of $\delta_{\crm}$. In order to address the question of whether the power spectra, given in Eqs.~(\ref{eq:p_power_Law}-\ref{LN1Plateau}), can lead to sizeable PBH formation, we carefully evaluate the corresponding thresholds and follow the analytical approach described in Refs.~\cite{Escriva:2020tak, Musco:2020jjb, Escriva:2022duf}. The threshold calculation necessitates an evaluation of the two-point correlation function, which can be expressed as %\textcolor{red}{Change the limit of the intrgral}
\begin{equation}
    g( \hat{r} )
        =
            \frac{ 1 }{ \sigma_{0}^{2} }
            \int^{\infty}_{0}
            \frac{ \d k }{ k }\;
            \frac{ \sin( k\.\hat{r} ) }
            { k\.\hat{r} }\.
            \Pcal_{R}( k )
            \, ,
\label{eq:g_two_point}
\end{equation}
with
\vs{-1mm}
\begin{equation}
    \sigma_{0}^{2}
        =
            \int^{\infty}_{0}
            \frac{ \d k }{ k }\;
            \Pcal_{R}( k )
            \, .
\end{equation}
Equation~\eqref{eq:g_two_point} connects the scalar power spectrum $\Pcal_{R}$ with the two point correlation function.

For calculating $\delta_{\crm}$, it is essential to define the peak value $\hat{r}_{m}$ of the so-called compaction function $C$, which has been introduced in Ref.~\cite{Shibata:1999zs}, and is defied as
\begin{equation}
    C
        \equiv
            2\.\frac{ M( r, t ) - M_{b}( r, t ) }
            { R( r, t ) }
            \, ,
\end{equation}
where $M$ is the Misner--Sharp mass and $M_{b}( r, t ) = 4\pi\rho_{b}\.R^{3}/3$. The compaction function can be approximated as~\cite{Harada:2015yda, Polnarev:2006aa, Polnarev:2012bi, Escriva:2022duf}.
\begin{equation}
    C
        \simeq
            f( w )\.K(r)\.r^{2}
        =
            f( w )\mspace{-2mu}
            \left(
                1
                -
                \big[
                    1 + \hat{r}\.\zeta'( \hat{r} )
                \big]^{2}
            \right)
            ,
\end{equation}
where $f( w ) = 3\.( 1 + w )/( 5 + 3\.w )$ and $w$ is the equation-of-state parameter. Here, $K$ and $\zeta$ are the conserved comoving curvature perturbations defined on the superhorizon scales. The scale $\hat{r}_{m}$ is determined through solving the root equation
\vs{-1mm}
\begin{equation}
    \zeta'( \hat{r}_{m} )
    +
    \hat{r}_{m}\.\zeta''( \hat{r}_{m} )
        =
            0
            \, .
\end{equation}
We introduce the parameter $\mu$ as the amplitude of curvature fluctuation, being given as
\begin{equation}
    \mu
        =
            \zeta( \hat{r} )/g( \hat{r} )
            \, ,
\end{equation}
%\newpage

\noindent or, in terms of the compaction function,
\begin{equation}
    \mu
        =
            \frac{ \pm\sqrt{1 - C( \hat{r}_{m} )/f( w )} - 1 }
            { g'( \hat{r}_{m} )\. \hat{r}_{m} }
            \, .
\end{equation}
The critical amplitude, $\mu_{\crm}$, is obtained at $C( r_{m} ) = \delta_{\crm}$.

In order to obtain the threshold, we define the function
\begin{equation}
    G( \hat{r}_{m} )
        =
            \frac{ g'( \hat{r}_{m} )
            -
            \hat{r}_{m}^{2}\.g'''( \hat{r}_{m} )/2 }
            { g'( \hat{r}_{m} ) }
            \, .
\end{equation}
The threshold $\delta_{\crm}$ can then be derived from
\begin{equation}
    q
        =
            G( r_{m} )\.
            \frac{ 1 }{ \sqrt{1 - \delta_{\crm}( q )/f( w )} }\.
            \frac{ 1 }{ 
            \Big[
                1 + \sqrt{1-\delta_{\crm}( q )/f( w )}
            \Big] }
            \, ,
\label{eq:shapeq}
\end{equation}
where $\delta_{\crm}$ is given by
\begin{equation}
    \delta_{\crm}
        =
            \frac{ 4 }{ 15 }\.
            e^{-1/q}\.
            \frac{ q^{1-5/2\mspace{2mu}q} }
            { \Gamma( 5 / 2\mspace{2mu}q ) - 
            \Gamma( 5 / 2\mspace{2mu}q, 1 / q  ) }
            \, .
            \label{eq:dc}
\end{equation}
The shape parameter $q$ characterises the contour around the peak of the compaction function. Equation~\eqref{eq:dc} provides an analytical method to compute the threshold $\delta_{\crm}$ as a function of the shape parameter $q$ during the radiation era, as discussed in Ref.~\cite{Escriva:2019phb}. Above, $\Gamma$ represents the incomplete gamma function. 

%\begin{figure}[t]
%    \vs{-7mm}
%    \centering
%    \includegraphics[width=0.49\textwidth]{fpbh.png}
%    \caption{
%        Fractional PBH abundances for the different models studied in this work.
%        \vs{5mm}
%        }
%    \label{fig:fbh}
%\end{figure}

The abundances of PBHs can be evaluated from~\cite{Inomata:2017okj}
\vs{2mm}
\begin{align}
    \frac{ \Omega_{\rm PBH}}
    {\Omega_{\rm DM}}\big[ M( k ) \big]
        &=
            1.52 \times 10^{8}
            \left(
                \frac{ \gamma }{ 0.2 }
            \right)^{3/2}
            \left(
                \frac{ g_{*} }{ 106.75 }
            \right)^{-1/4}
            \notag
            \\[2.5mm]
        &\phantom{=\;}\times
            \left(
                \frac{ M( k ) }{ M_{\odot} }
            \right)^{-1/2}
            \beta
            \big[
                M( k )
            \big]
            \, ,
\end{align}
%\newpage

\noindent and the total abundance of PBHs is given by
\begin{equation}
    f_{\rm PBH}
        =
            \int \d\ln M\;
            \frac{ \Omega_{\rm PBH} }{ \Omega_{\rm DM} }
            \, .
\end{equation}

\begin{table}[t]
\vs{1mm}
\centering
\begin{tabular}{ m{13em} m{1.3cm}  } 
    \hline
    \hline
    \q Model$\vphantom{1^{^{^{1}}}}$
        & \;$\delta_{\crm}$
    \\[0.7ex] 
    \hline
    \q Broken Power-Law$\vphantom{1^{^{^{1}}}}$
        & \;$0.506$
         %   & \;$\textcolor{red}{0.1353}$\\ 
         \\
    \q Log-Normal-Peak$\vphantom{1^{^{^{1}}}}$
        & \;$0.507$
       %     & \;$\textcolor{red}{234.01}$\\
       \\
    \q Log-Normal-Peak+osc$\vphantom{1^{^{^{1}}}}$
        & \;$0.495$
           \\
    \q Log-Normal-Peak+cut-off$\vphantom{1^{^{^{1}}}}$
        & \;$0.506$
           \\
            \q Log-Normal-Peak+plateau$\vphantom{1^{^{^{1}}}}$
        & \;$0.464$
             \\[1mm]
    \hline
    \hline
\end{tabular}
\caption{
    Critical thresholds for different  \textcolor{black}{ scenarios}.}
\label{tab:pbh_models}
\end{table}

{Having evaluated the fractional abundances of PBHs using the methodology outlined above, we find that all models consistently encounter a significant issue of PBH overproduction. This overproduction persists across various power spectrum shapes, each attempting to match the gravitational-wave signals observed by PTAs. Despite differing spectral forms, all models predict overproduction of PBH abundances, highlighting a universal challenge in reconciling these models with PTA data.} \textcolor{black}{In Table \ref{tab:pbh_models}, we present the critical threshold values for the five shapes of power spectra that we studied, following the analysis in \cite{Musco:2020jjb,Escriva:2022duf,Escriva:2019phb}. It is important to note that our evaluation does not account for the non-linear effects explicitly analyzed in \cite{Musco:2020jjb}, as this is beyond the scope of this work. However, even if these effects are considered, using the threshold values provided in \cite{Musco:2020jjb}, our results remain unaffected. } 

\textcolor{black}{ Observational constraints in the region of interest further excite the overproduction problem. Specifically, the recent constraints from OGLE~\cite{Mroz:2024mse, Mroz:2024wag} and LVK~\cite{Franciolini:2022tfm, Andres-Carcasona:2024wqk} indicate that PBHs can constitute at most 1\% of the dark matter density in this mass range. These stringent limits underscore the challenge of reconciling PBH abundance predictions with observational data, highlighting the need for additional mechanisms to suppress PBH formation.}

\textcolor{black}{ We need to remark here that the standard process employing Gau{\ss}ian curvature perturbations often predicts an excessive abundance of PBHs. For this reason, recent studies have explored the introduction of non-Gau{\ss}ianities in the curvature perturbation field, which can effectively reduce the likelihood of the large density fluctuations necessary for PBH formation. Particularly, negative non-Gau{\ss}ianities have been shown to substantially diminish the PBH abundance by decreasing the probability of extreme positive fluctuations, thus ensuring consistency with observational limits \cite{Franciolini:2023pbf, Young:2022phe, Ferrante:2022mui, Gow:2022jfb, Ianniccari:2024bkh, Kitajima:2021fpq, Yoo:2022mzl, Perna:2024ehx}. Moreover, these adjustments not only impact the rate of PBH formation but also significantly influence the characteristics of the GW background. Altering the power spectrum's shape and amplitude through non-Gau{\ss}ian effects aligns better with the observed GW signals by PTA, offering a refined method to analyze early-Universe phenomena. Such studies underscore the pivotal role of non-Gau{\ss}ianities in cosmological predictions \cite{Perna:2024ehx}.}

Finally, we remark that the issue of PBH overproduction remains relatively unchanged across a broad range of the critical threshold values, $\delta_{\crm} \in [ 0.45, 0.51 ]$. Specifically, this is true for the broken-power-law model. The mass fraction is evaluated using Press--Schechter approach. Incorporating peaks theory into the calculations affects the results, would lead to larger values for the fractional abundances of PBHs~\cite{Young:2014ana, Wang:2021kbh, Stamou:2021qdk}, these intensifying the problem. We leave a comprehensive discussion to study the effect of incorporating various statistical approaches to future work.

%%%%%%%%%%%%%%%%%%%%%%%%%%%%%%%%%%%%%%%
\section{Conclusion \& Outlook}
\label{sec:Conclusion-and-Outlook}

We have proposed a novel methodology to analyse the relation between the energy density in gravitational waves and primordial power spectra. This methodology differs from conventional Bayesian approaches by offering a direct data interpretation. Our study has the advantage of both computational efficiency and simplifying the parametric-space exploration without the need of prior distributions.

As an application, we investigate several scalar power spectra, focusing on {five} distinct shapes: broken power law, log-normal peak, and its variations incorporating oscillatory patterns, UV cut-offs, \textcolor{black}{ and with a plateau at CMB scales and beyond the peak}. Assuming scalar-induced gravitational waves as a full explanation of the recent PTA measurements, we evaluate the associated energy densities and generate datasets suitable for interpreting the PTA signals. Subsequently, we employ optimisation techniques to reconstruct our models. To demonstrate the adaptability of our analysis, we also introduce noise into our dataset. %\textcolor{red}{We connect our results to the formation of PBHs. Here, the broken-power-law model stands out for its ability to explain not only the PTA gravitational-wave detection but also for offering an all-PBH explanation to the dark matter problem.}

Given the broad applicability of our methodology, we explore the relationship between the energy density of gravitational waves, as described by the various power spectrum shapes, and the abundances of primordial black holes. Knowing that the fractional abundances of primordial black holes strongly depend on the exact value of the threshold at which the fluctuations collapse, we evaluate the threshold for each case. For simplicity, we employ the Press-Schechter approach to evaluate the mass fraction of primordial black holes, using a  \textcolor{black}{top-hat} window function for smoothing. Under these assumptions, we find that
% \textcolor{red}{ the broken power-law shape emerges as the most favourable scenario. It not only predicts the signals of various current and future gravitational-wave detections but also accounts for significant dark matter abundances. On the other hand, the log-normal shape cannot explain a broad range of gravitational-wave signals and has the significant drawback of yielding overproduction of PBHs. However, by introducing oscillations onto the log-normal shape, this problem can be avoided.}
 \textcolor{black}{all the scenarios we considered consistently lead to significant overproduction of PBHs. We should stress, however, that the main point of our article is to introduce a novel methodology; this works regardless of the specific scenario it is applied to.}

Of course, the studied power-spectrum shapes are rather simple, and realistic PBH models are likely to involve more complex structures. The currently most promising and most natural scenario for PBH formation{\,---\,}the {\it Thermal History} scenario~\cite{Carr:2019kxo}{\,---\,}extends over many orders of magnitude in mass and is inherently multimodal. As this can describe around 20 current observations which may be argued to be plausibly attributed to PBHs (see Ref.~\cite{Carr:2023tpt} for the presently most complete overview), we will apply our methodology to investigate the thermal-history power-spectrum shape in a forthcoming study.

Finally, considering the wide applicability of our approach, there are several avenues for application and extension. As regards the former, future work will focus on integrating additional data from forthcoming gravitational-wave observations, including those from LISA and DECIGO. As regards the latter, our approach, suited for extension to other power-spectrum shapes, can well incorporate advantages fitting methods. This specifically includes advanced machine-learning techniques which would enable the exploration of more complex datasets. The training data can be encapsulated in the weights of Eq.~\eqref{eq:omega_num}. Hence, these techniques promise to have potential to identify key features of the signals, thereby deepening our understanding of gravitational waves and their origins{\,---\,}possibly connected to the dark matter in the Universe.

\section*{Acknowledgments}
We thank the anonymous referee for their helpful comments, which led to improvements in this work. DF is supported by the Foundation for Advancement of Theoretical Physics and Mathematics “BASIS”, and by Tomsk State University under the development program Priority-2030.

%%%%%%%%%%%%%%%%%%%%%%%%%%%%%%%%%%%%%%%
%%%%%%%%%%%%%%%%%%%%%%%%%%%%%%%%%%%%%%%
\appendix

%%%%%%%%%%%%%%%%%%%%%%%%%%%%%%%%%%%%%%%
\section{Gravitational Waves}
\label{sec:Gravitational-Waves}

The aim of this section is to provide the basic aspect of the gravitational waves needed in our subsequent analysis. In particular, we show how the energy density of gravitational waves is related to the scalar power spectrum. This section is based on the Refs.~\cite{Kohri:2018awv, Baumann:2007zm, Mollerach:2003nq, Maggiore:1999vm, Espinosa:2018eve}.

%%%%%%%%%%%%%%%%%%%%%%%%%%%%%%%%%%%%%%%
\subsection{Equation of Motion for Gravitational Waves}
\label{sec:Equation-of-Motion-for-Gravitational Waves}

The perturbed metric in conformal Newtonian gauge reads
\begin{equation}
    \d s^{2}
        =
            -\.a^{2}
            \left(
                1 + 2\Phi
            \right)
            \d\eta^{2}
            +
            a^{2}
            \left[
                ( 1 - 2\Psi )\.\delta_{ij} 
                +
                \frac{ 1 }{ 2 }\.h_{ij}
            \right]
            \d x^{i} \d x^{j}
            \, ,
\end{equation}
with $\eta$ being the conformal time, $\Phi$ and $\Psi$ are the Bardeen potential, and $h_{ij}$ is the tensor perturbation, whose equation of motion reads
\begin{equation}
    h''_{ij}
    +
    2\.\Hcal\.h'_{ij}
    -
    \nabla^{2} h_{ij}
        =
            2\.a^{2}\.S_{ij}
            \, .
\label{eq:eoms_tensor}
\end{equation}
Here, $\Hcal$ represents the conformal Hubble parameter and $S_{ij}$ is the source term. For negligible anisotropic stress, $\Phi = \Psi$, the second-order Einstein tensor and energy-momentum tensor contains the quadratic scalar combinations, which act as a source, which is given by
\begin{align}
\begin{split}
    S_{ij}
        &=
            4\.\Phi\.\partial_{i}\partial_{j}\Phi
            +
            2\.\partial_{i}\Phi\.\partial_{j}\Phi
            \\[1mm]
        &\phantom{=\;}
            -
            \frac{ 4 }{ 3\.( 1 + w ) }\,
            \partial_{i}\!
            \left(
                \frac{ \Phi' }{ \Hcal  }+ \Phi
            \right)
            \partial_{j}\!
            \left(
                \frac{ \Phi' }
                { \Hcal  }+ \Phi
            \right)
            .
\end{split}
\end{align}
Upon Fourier transformation,
\begin{equation}
    \Phi_{\kbm}( \eta, \kbm )
        =
            \int \d^{3}x\,
            \Phi( \eta, \bm{x} )\.
            e^{-i\kbm\cdot\bm{x}}
            \, ,
\end{equation}
this becomes 
\begin{align}
    S^{\lambda}_{k}
        &=
            \int\frac{ \d^{3}k }{ ( 2\mspace{1mu}\pi )^{3/2} }\;
            e^{\lambda,ij}( \kbm )\.q_{i}\.q_{j}
            \bigg[
                2\.\Phi_{\qbm}
                \Phi_{\kbm - \qbm}
            \\[2mm]
        &\phantom{=\;}
                +
                \frac{ 4 }{ 3\.( 1 + w ) }
                \big(
                    \Hcal^{-1}
                    \Phi'_{\qbm}
                    +
                    \Phi_{\qbm}
                \big)
                \big(
                    \Hcal^{-1}
                    \Phi'_{\kbm - \qbm}
                    +
                    \Phi_{\kbm - \qbm}
                \big)
            \bigg]
             \, ,
            \notag
\end{align}
where we have used the convolution theorem. With $e_{ij}^{\lambda}$ we denote the polarisation modes $\lambda = ( +, \times )$ (for further details, see Refs.~\cite{Kohri:2018awv, Espinosa:2018eve}). Equation~\eqref{eq:eoms_tensor} describes the evolution of tensor modes sourced by the scalar perturbations $\Phi$, and is solved by
\begin{equation}
    h_{k}^{\lambda}( \eta )
        =
            \frac{ 1 }{ a( \eta ) }
            \int^{\eta}_{0} \d {\tilde{\eta}}\;
            G_{k}( \eta, \tilde{\eta} )\.
            \alpha( \tilde{\eta} )\.
             S^{\lambda}_{k}( \tilde{\eta}, \kbm )
            \, ,
\end{equation}
where the Green's function $G_{k}$ is the solution of
\begin{equation}
    G''_{k}(\eta, \tilde{\eta})
    +
    \left[
        k^{2}
        -
        \frac{ a''( \eta ) }{ a( \eta ) }
    \right]
    G_{k}( \eta, \tilde{\eta} )
        =
            \delta( \eta - \tilde{\eta} )
            \, .
\end{equation}
The Fourier component of the tensor mode is
\begin{equation}
    h_{ij}( \eta, \bm{x} )
        =
            \sum_{\lambda}
            \int\frac{ \d^{3}k }{ ( 2\mspace{1mu}\pi )^{3/2}}
            h_{\lambda}( \eta, \kbm )\.
            e_{ij}^{ \lambda }( k )\.
            e^{i\kbm\cdot\bm{x}}
            \, .
\end{equation}
As opposed to first order of the Einstein equation, at second order, the scalar, vector and tensor components are not independent. There is a second-order contribution to the tensor mode that depends quadratically on the first-order scalar metric perturbation \cite{Baumann:2007zm}.

The power-spectral density of the tensor perturbation is defined as
\begin{equation}
    \big\langle
        h_{\bm q}^{\lambda}( \eta )\.
        h_{\bm k}^{\lambda}( \eta ) 
    \big\rangle
        \equiv
            \frac{ 2\mspace{1mu}\pi^{2} }{ k^{3} }\.
            \Pcal_{h}( \tau, k )\,
            \delta^{(3)}( \qbm + \kbm )
            \, ,
\end{equation}
where $\delta^{(3)}$ is the dimensionless delta distribution. Hence, the power spectrum of the curvature perturbation is 
\begin{equation}
    \overline{\Pcal_{h}( t, k )}
        =
            \int^{\infty}_{0}\d\upsilon
            \int^{1 + u}_{| 1 - u |}\d u\;
            T( u, \upsilon, \eta, k )\.
            \Pcal_{R}( u\mspace{1.5mu}k )\.
            \Pcal_{R}( \upsilon\mspace{1.5mu}k )
            \, ,
\end{equation}
where $T$ is the transfer function, given by
\begin{align}
\begin{split}
    T( u, \upsilon, \eta, k )
        &=
            4
            \left(
                \frac{ 4\.\upsilon^{2} -
                \big(
                    1 + \upsilon^{2} - u^{2}
                \big) }{ 4\.u\mspace{1.5mu}\upsilon }
            \right)^{\!2}
            \\[1mm]
        &\phantom{=\;}
            \times
            \left(
                \frac{ 3 + 3\.w }{ 5 + 3\.w }
            \right)^{2}
            \overline{I^{2}( u, \upsilon, \eta, k )}
            \, ,
\end{split}
\end{align}
and the kernel function is defined as
\begin{equation}
    I( u, \upsilon, \eta, k )
        =
            \int^{x}_{0}\d y\;
            \frac{ a( y/k ) }{ a( \tau ) }\.
            k\,G_{k}( x, y )\.
            f( u, \upsilon, \eta, k )
            \, ,
\end{equation}
with $x \equiv k\.\eta$. The so-called oscillation average can be calculated as
\begin{equation}
    \overline{I^{2}( u, \upsilon, \eta, k )}
        \equiv
            \frac{ 1 }{ 2\mspace{1mu}\pi }
            \int^{x + 2\mspace{1mu}\pi}_{\pi}\d y^{2}\;
            I^{2}( u, \upsilon, \eta, k )
            \, .
\end{equation}

Using the variables $t \equiv u + \upsilon - 1$ and $s \equiv u - \upsilon$, the power spectrum $\Pcal_{h}$ reads~\cite{Kohri:2018awv}
\begin{align}
    \Pcal_{h}( \eta, k )
        &=
            2 \int^{\infty}_{0}\d t\;
            \int^{1}_{-1} \d s\;
            \Bigg[
                \frac{ t\.( t + 2 )\.
                \big( s^{2} - 1 \big) }
                {( 1 - s + t )\.( 1 + s + t )}
            \Bigg]^{2}
            \notag
            \\[2mm]
        &\phantom{=\;2 \int^{\infty}_{0}}
            \times
            I^{2}( u, \upsilon, x )\.
            \Pcal_{R}( k\upsilon )\.
            \Pcal_{R}( ku )
            \, ,
\end{align}
where the function $I$ is given in terms of $u$ and $\upsilon$ (or $t$ and $s$), assuming that the gravitational waves are produced during radiation- or matter-dominated era.

%%%%%%%%%%%%%%%%%%%%%%%%%%%%%%%%%%%%%%%
\subsection{Gravitational-Wave Energy Density}
\label{sec:Gravitational--Wave-Energy-Density}

The fraction of the gravitational-wave energy density per logarithmic wavelength is obtained via
\begin{equation}
\label{eq:omega_gw}
    \Omega_{\rm GW}( \eta, k )
        =
            \frac{ \rho_{\rm GW}( \eta, k ) }
            {\rho_{\rm tot}( k ) }
        =
            \frac{ 1 }{ 24 }
            \left(
                \frac{ k }{ a( \eta )\mspace{1.5mu}H( \eta ) }
            \right)^{\!2}\.
            \overline{\Pcal_{h}( \eta, k ) }
            \, ,
\end{equation}
where
\begin{equation}
    \rho_{\rm GW}( \eta, k )
        =
            \frac{ M_{\rm Pl}^{2} }{ 8 }\.
            \frac{ k^{2} }{ \alpha^{2} }\,
            \overline{\Pcal_{h}( \eta, k )}
            \; .
\end{equation}
We study two cases of radiation- and matter-dominated Universe.

Assuming that the power spectrum of gravitational waves is produced in the radiation-dominated era yields the solution
\begin{align}
    &\overline{I_{\rm RD}^{2}( u, v, x \to \infty )}
        =
            \frac{ 1 }{ 2 }
            \Bigg(
                \frac{ 3\mspace{1.5mu}
                \big( u^{2} + v^{2} - 3 \big) }
                { 4\.u^{3}\.v^{3}\.x }
            \Bigg)^{\!2}
            \displaybreak[1]
            \nonumber
            \\[1mm]
        &\qquad
            \times
            \Bigg[
                \bigg(
                    - 4\.u\mspace{1.5mu}v + ( u^{2} + v^{2} - 3 )\.
                    \log\!
                    \left|
                        \frac{ 3 - ( u + v )^{2} }
                        { 3 - ( u - v )^{2} }
                    \right|
                \bigg)^{\!2}
            \nonumber
            \displaybreak[1]
            \\[1mm]
        &\phantom{\qquad\Bigg[\bigg(}
                +
                \pi^{2}
                \big(
                    u^{2} + v^{2} - 3
                \big)^{2}\,
                \Theta
                \Big(
                    v + u - \sqrt{3}
                \Big)
            \Bigg]
            \, ,
            \label{eq:I_RD_ave}
\end{align}
or, in $t, s$ variables,
\begin{align}
    &\overline{I_{\rm RD}^{2}( t, s, x \to \infty )}
        =
            \frac{ 288\.
            \big(
                s^{2} + t\.( 2 + t ) - 5
            \big)^{2} }
            { x^{2}\.( 1 - s + t )^{6}\.
            ( 1 + s + t )^{6} }
            \displaybreak[1]
            \notag
            \\[1mm]
        &\phantom{=\;}
            \times
            \Bigg\{
                \frac{ \pi^{2} }{ 4 }
                \Big(
                    s^{2} + t\.\big( t + 2 \big) - 5
                \Big)^{\mspace{-2mu}2}\,
                \Theta
                \bigg[
                    t - \Big( \sqrt{3} - 1 \Big)
                \bigg]
            \Bigg\}
            \displaybreak[1]
            \notag
            \\[2mm]
        &\phantom{=\;}
            +
            \Bigg[
                \frac{ 1 }{ 2 }\.
                \big[
                    s^{2} + t\.( t + 2 ) - 5
                \big]\.
                \log
                \bigg|
                    \frac{ ( 2 + t )\.t - 2 }
                    { 3 - s^{2} }
                \bigg|
            \notag
            \\[1mm]
        &\phantom{=\;+\Bigg[}
                + ( s - 1 - t )( t + s + 1 )
            \Bigg]^{2}
            \, .
            \label{eq:I_RD_st} 
\end{align}
For the gravitational-wave energy density, one obtains
\begin{equation}
\begin{split}
    &\Omega_{\rm GW}( k )
        =
            0.387\;\Omega_{\Rrm}\!
            \left(
                \frac{ g_{*,s}^{4}\.g_{*}^{-3} }
                { 106.75 }
            \right)^{\!-1/3}
            \frac{ 1 }{ 6 }
            \frac{ \big( y^{2} - 1 \big)
            \big( x^{2} - 1 \big) }
            { ( x - y )^{2}\.( x + y )^{2} }
            \\[2mm]
        &\times
            \int_{-1}^{1}
            \d x \int_{1}^{\infty}\d y\;
            \Pcal_{R}\!
            \left(
                \frac{ y - x }{ 2 }\.k
            \right)
            \Pcal_{R}\!
            \left(
                \frac{ x + y }{ 2 }\.k
            \right)\,
            \overline{I_{\rm RD}^{2}( x, y )}  
            \, ,
\end{split}
\label{eq:omega_gw_1}
\end{equation}
where $x = s$ and $y = t + 1$. Here, $\Omega_{\Rrm}$ is the radiation abundance ($\Omega_{\Rrm} = 5.38 \times 10^{-5}$), and $g_{*}$ and $g_{*,s}$ are the effective degrees of freedom for the energy and entropy densities, respectively. They can be approximately considered to be equal. 

In a matter-dominated Universe, the function $I( u, v, x )$ is given by
\begin{equation}
    I_{\rm MD}( u, v, x )
        =
            \frac{ 6\.
            \big(
                x^{3} + 3\.x \cos x - 3\.\sin x
            \big) }{ 5\.x^{3} }
            \, ,
            \label{eq:I_mD}
\end{equation}
and for the oscillation average $\overline{I_{\rm MD}^{2}( v, u, x \to \infty )} = 18 / 25$. In this work we focus on the energy density of gravitational waves in the radiation-dominated era.\\[10mm]

%%%%%%%%%%%%%%%%%%%%%%%%%%%%%%%%%%%%%%%
%\setlength{\bibsep}{4pt}
\bibliography{bib}

\begin{thebibliography}{126}%
\makeatletter
\providecommand \@ifxundefined [1]{%
 \@ifx{#1\undefined}
}%
\providecommand \@ifnum [1]{%
 \ifnum #1\expandafter \@firstoftwo
 \else \expandafter \@secondoftwo
 \fi
}%
\providecommand \@ifx [1]{%
 \ifx #1\expandafter \@firstoftwo
 \else \expandafter \@secondoftwo
 \fi
}%
\providecommand \natexlab [1]{#1}%
\providecommand \enquote  [1]{``#1''}%
\providecommand \bibnamefont  [1]{#1}%
\providecommand \bibfnamefont [1]{#1}%
\providecommand \citenamefont [1]{#1}%
\providecommand \href@noop [0]{\@secondoftwo}%
\providecommand \href [0]{\begingroup \@sanitize@url \@href}%
\providecommand \@href[1]{\@@startlink{#1}\@@href}%
\providecommand \@@href[1]{\endgroup#1\@@endlink}%
\providecommand \@sanitize@url [0]{\catcode `\\12\catcode `\$12\catcode
  `\&12\catcode `\#12\catcode `\^12\catcode `\_12\catcode `\%12\relax}%
\providecommand \@@startlink[1]{}%
\providecommand \@@endlink[0]{}%
\providecommand \url  [0]{\begingroup\@sanitize@url \@url }%
\providecommand \@url [1]{\endgroup\@href {#1}{\urlprefix }}%
\providecommand \urlprefix  [0]{URL }%
\providecommand \Eprint [0]{\href }%
\providecommand \doibase [0]{https://doi.org/}%
\providecommand \selectlanguage [0]{\@gobble}%
\providecommand \bibinfo  [0]{\@secondoftwo}%
\providecommand \bibfield  [0]{\@secondoftwo}%
\providecommand \translation [1]{[#1]}%
\providecommand \BibitemOpen [0]{}%
\providecommand \bibitemStop [0]{}%
\providecommand \bibitemNoStop [0]{.\EOS\space}%
\providecommand \EOS [0]{\spacefactor3000\relax}%
\providecommand \BibitemShut  [1]{\csname bibitem#1\endcsname}%
\let\auto@bib@innerbib\@empty
%</preamble>
\bibitem [{\citenamefont {Abbott}\ \emph
  {et~al.}(2016{\natexlab{a}})\citenamefont {Abbott} \emph
  {et~al.}}]{Abbott:2016blz}%
  \BibitemOpen
  \bibfield  {author} {\bibinfo {author} {\bibfnamefont {B.~P.}\ \bibnamefont
  {Abbott}} \emph {et~al.} (\bibinfo {collaboration} {LIGO Scientific,
  Virgo}),\ }\bibfield  {title} {\bibinfo {title} {{Observation of
  Gravitational Waves from a Binary Black Hole Merger}},\ }\href
  {https://doi.org/10.1103/PhysRevLett.116.061102} {\bibfield  {journal}
  {\bibinfo  {journal} {Phys. Rev. Lett.}\ }\textbf {\bibinfo {volume} {116}},\
  \bibinfo {pages} {061102} (\bibinfo {year} {2016}{\natexlab{a}})},\ \Eprint
  {https://arxiv.org/abs/1602.03837} {arXiv:1602.03837 [gr-qc]} \BibitemShut
  {NoStop}%
\bibitem [{\citenamefont {Abbott}\ \emph
  {et~al.}(2017{\natexlab{a}})\citenamefont {Abbott} \emph
  {et~al.}}]{Abbott:2017vtc}%
  \BibitemOpen
  \bibfield  {author} {\bibinfo {author} {\bibfnamefont {B.~P.}\ \bibnamefont
  {Abbott}} \emph {et~al.} (\bibinfo {collaboration} {LIGO Scientific,
  VIRGO}),\ }\bibfield  {title} {\bibinfo {title} {{GW170104: Observation of a
  50-Solar-Mass Binary Black Hole Coalescence at Redshift 0.2}},\ }\href
  {https://doi.org/10.1103/PhysRevLett.118.221101} {\bibfield  {journal}
  {\bibinfo  {journal} {Phys. Rev. Lett.}\ }\textbf {\bibinfo {volume} {118}},\
  \bibinfo {pages} {221101} (\bibinfo {year} {2017}{\natexlab{a}})},\ \bibinfo
  {note} {[Erratum: Phys.Rev.Lett. 121, 129901 (2018)]},\ \Eprint
  {https://arxiv.org/abs/1706.01812} {arXiv:1706.01812 [gr-qc]} \BibitemShut
  {NoStop}%
\bibitem [{\citenamefont {Abbott}\ \emph
  {et~al.}(2017{\natexlab{b}})\citenamefont {Abbott} \emph
  {et~al.}}]{Abbott:2017gyy}%
  \BibitemOpen
  \bibfield  {author} {\bibinfo {author} {\bibfnamefont {B.~P.}\ \bibnamefont
  {Abbott}} \emph {et~al.} (\bibinfo {collaboration} {LIGO Scientific,
  Virgo}),\ }\bibfield  {title} {\bibinfo {title} {{GW170608: Observation of a
  19-solar-mass Binary Black Hole Coalescence}},\ }\href
  {https://doi.org/10.3847/2041-8213/aa9f0c} {\bibfield  {journal} {\bibinfo
  {journal} {Astrophys. J. Lett.}\ }\textbf {\bibinfo {volume} {851}},\
  \bibinfo {pages} {L35} (\bibinfo {year} {2017}{\natexlab{b}})},\ \Eprint
  {https://arxiv.org/abs/1711.05578} {arXiv:1711.05578 [astro-ph.HE]}
  \BibitemShut {NoStop}%
\bibitem [{\citenamefont {Abbott}\ \emph
  {et~al.}(2017{\natexlab{c}})\citenamefont {Abbott} \emph
  {et~al.}}]{Abbott:2017oio}%
  \BibitemOpen
  \bibfield  {author} {\bibinfo {author} {\bibfnamefont {B.~P.}\ \bibnamefont
  {Abbott}} \emph {et~al.} (\bibinfo {collaboration} {LIGO Scientific,
  Virgo}),\ }\bibfield  {title} {\bibinfo {title} {{GW170814: A Three-Detector
  Observation of Gravitational Waves from a Binary Black Hole Coalescence}},\
  }\href {https://doi.org/10.1103/PhysRevLett.119.141101} {\bibfield  {journal}
  {\bibinfo  {journal} {Phys. Rev. Lett.}\ }\textbf {\bibinfo {volume} {119}},\
  \bibinfo {pages} {141101} (\bibinfo {year} {2017}{\natexlab{c}})},\ \Eprint
  {https://arxiv.org/abs/1709.09660} {arXiv:1709.09660 [gr-qc]} \BibitemShut
  {NoStop}%
\bibitem [{\citenamefont {Abbott}\ \emph
  {et~al.}(2016{\natexlab{b}})\citenamefont {Abbott} \emph
  {et~al.}}]{Abbott:2016nmj}%
  \BibitemOpen
  \bibfield  {author} {\bibinfo {author} {\bibfnamefont {B.~P.}\ \bibnamefont
  {Abbott}} \emph {et~al.} (\bibinfo {collaboration} {LIGO Scientific,
  Virgo}),\ }\bibfield  {title} {\bibinfo {title} {{GW151226: Observation of
  Gravitational Waves from a 22-Solar-Mass Binary Black Hole Coalescence}},\
  }\href {https://doi.org/10.1103/PhysRevLett.116.241103} {\bibfield  {journal}
  {\bibinfo  {journal} {Phys. Rev. Lett.}\ }\textbf {\bibinfo {volume} {116}},\
  \bibinfo {pages} {241103} (\bibinfo {year} {2016}{\natexlab{b}})},\ \Eprint
  {https://arxiv.org/abs/1606.04855} {arXiv:1606.04855 [gr-qc]} \BibitemShut
  {NoStop}%
\bibitem [{\citenamefont {Agazie}\ \emph
  {et~al.}(2023{\natexlab{a}})\citenamefont {Agazie} \emph
  {et~al.}}]{NANOGrav:2023gor}%
  \BibitemOpen
  \bibfield  {author} {\bibinfo {author} {\bibfnamefont {G.}~\bibnamefont
  {Agazie}} \emph {et~al.} (\bibinfo {collaboration} {NANOGrav}),\ }\bibfield
  {title} {\bibinfo {title} {{The NANOGrav 15 yr Data Set: Evidence for a
  Gravitational-wave Background}},\ }\href
  {https://doi.org/10.3847/2041-8213/acdac6} {\bibfield  {journal} {\bibinfo
  {journal} {Astrophys. J. Lett.}\ }\textbf {\bibinfo {volume} {951}},\
  \bibinfo {pages} {L8} (\bibinfo {year} {2023}{\natexlab{a}})},\ \Eprint
  {https://arxiv.org/abs/2306.16213} {arXiv:2306.16213 [astro-ph.HE]}
  \BibitemShut {NoStop}%
\bibitem [{\citenamefont {Agazie}\ \emph
  {et~al.}(2023{\natexlab{b}})\citenamefont {Agazie} \emph
  {et~al.}}]{NANOGrav:2023hde}%
  \BibitemOpen
  \bibfield  {author} {\bibinfo {author} {\bibfnamefont {G.}~\bibnamefont
  {Agazie}} \emph {et~al.} (\bibinfo {collaboration} {NANOGrav}),\ }\bibfield
  {title} {\bibinfo {title} {{The NANOGrav 15 yr Data Set: Observations and
  Timing of 68 Millisecond Pulsars}},\ }\href
  {https://doi.org/10.3847/2041-8213/acda9a} {\bibfield  {journal} {\bibinfo
  {journal} {Astrophys. J. Lett.}\ }\textbf {\bibinfo {volume} {951}},\
  \bibinfo {pages} {L9} (\bibinfo {year} {2023}{\natexlab{b}})},\ \Eprint
  {https://arxiv.org/abs/2306.16217} {arXiv:2306.16217 [astro-ph.HE]}
  \BibitemShut {NoStop}%
\bibitem [{\citenamefont {Afzal}\ \emph {et~al.}(2023)\citenamefont {Afzal}
  \emph {et~al.}}]{NANOGrav:2023hvm}%
  \BibitemOpen
  \bibfield  {author} {\bibinfo {author} {\bibfnamefont {A.}~\bibnamefont
  {Afzal}} \emph {et~al.} (\bibinfo {collaboration} {NANOGrav}),\ }\bibfield
  {title} {\bibinfo {title} {{The NANOGrav 15 yr Data Set: Search for Signals
  from New Physics}},\ }\href {https://doi.org/10.3847/2041-8213/acdc91}
  {\bibfield  {journal} {\bibinfo  {journal} {Astrophys. J. Lett.}\ }\textbf
  {\bibinfo {volume} {951}},\ \bibinfo {pages} {L11} (\bibinfo {year}
  {2023})},\ \Eprint {https://arxiv.org/abs/2306.16219} {arXiv:2306.16219
  [astro-ph.HE]} \BibitemShut {NoStop}%
\bibitem [{\citenamefont {Antoniadis}\ \emph
  {et~al.}(2023{\natexlab{a}})\citenamefont {Antoniadis} \emph
  {et~al.}}]{EPTA:2023fyk}%
  \BibitemOpen
  \bibfield  {author} {\bibinfo {author} {\bibfnamefont {J.}~\bibnamefont
  {Antoniadis}} \emph {et~al.} (\bibinfo {collaboration} {EPTA, InPTA:}),\
  }\bibfield  {title} {\bibinfo {title} {{The second data release from the
  European Pulsar Timing Array - III. Search for gravitational wave signals}},\
  }\href {https://doi.org/10.1051/0004-6361/202346844} {\bibfield  {journal}
  {\bibinfo  {journal} {Astron. Astrophys.}\ }\textbf {\bibinfo {volume}
  {678}},\ \bibinfo {pages} {A50} (\bibinfo {year} {2023}{\natexlab{a}})},\
  \Eprint {https://arxiv.org/abs/2306.16214} {arXiv:2306.16214 [astro-ph.HE]}
  \BibitemShut {NoStop}%
\bibitem [{\citenamefont {Antoniadis}\ \emph
  {et~al.}(2023{\natexlab{b}})\citenamefont {Antoniadis} \emph
  {et~al.}}]{EPTA:2023sfo}%
  \BibitemOpen
  \bibfield  {author} {\bibinfo {author} {\bibfnamefont {J.}~\bibnamefont
  {Antoniadis}} \emph {et~al.} (\bibinfo {collaboration} {EPTA}),\ }\bibfield
  {title} {\bibinfo {title} {{The second data release from the European Pulsar
  Timing Array - I. The dataset and timing analysis}},\ }\href
  {https://doi.org/10.1051/0004-6361/202346841} {\bibfield  {journal} {\bibinfo
   {journal} {Astron. Astrophys.}\ }\textbf {\bibinfo {volume} {678}},\
  \bibinfo {pages} {A48} (\bibinfo {year} {2023}{\natexlab{b}})},\ \Eprint
  {https://arxiv.org/abs/2306.16224} {arXiv:2306.16224 [astro-ph.HE]}
  \BibitemShut {NoStop}%
\bibitem [{\citenamefont {Antoniadis}\ \emph
  {et~al.}(2023{\natexlab{c}})\citenamefont {Antoniadis} \emph
  {et~al.}}]{EPTA:2023akd}%
  \BibitemOpen
  \bibfield  {author} {\bibinfo {author} {\bibfnamefont {J.}~\bibnamefont
  {Antoniadis}} \emph {et~al.} (\bibinfo {collaboration} {EPTA, InPTA}),\
  }\bibfield  {title} {\bibinfo {title} {{The second data release from the
  European Pulsar Timing Array - II. Customised pulsar noise models for
  spatially correlated gravitational waves}},\ }\href
  {https://doi.org/10.1051/0004-6361/202346842} {\bibfield  {journal} {\bibinfo
   {journal} {Astron. Astrophys.}\ }\textbf {\bibinfo {volume} {678}},\
  \bibinfo {pages} {A49} (\bibinfo {year} {2023}{\natexlab{c}})},\ \Eprint
  {https://arxiv.org/abs/2306.16225} {arXiv:2306.16225 [astro-ph.HE]}
  \BibitemShut {NoStop}%
\bibitem [{\citenamefont {Antoniadis}\ \emph
  {et~al.}(2023{\natexlab{d}})\citenamefont {Antoniadis} \emph
  {et~al.}}]{EPTA:2023xxk}%
  \BibitemOpen
  \bibfield  {author} {\bibinfo {author} {\bibfnamefont {J.}~\bibnamefont
  {Antoniadis}} \emph {et~al.} (\bibinfo {collaboration} {EPTA}),\ }\bibfield
  {title} {\bibinfo {title} {{The second data release from the European Pulsar
  Timing Array: V. Implications for massive black holes, dark matter and the
  early Universe}},\ }\href@noop {} {\  (\bibinfo {year}
  {2023}{\natexlab{d}})},\ \Eprint {https://arxiv.org/abs/2306.16227}
  {arXiv:2306.16227 [astro-ph.CO]} \BibitemShut {NoStop}%
\bibitem [{\citenamefont {Miles}\ \emph {et~al.}(2025)\citenamefont {Miles}
  \emph {et~al.}}]{Miles:2024rjc}%
  \BibitemOpen
  \bibfield  {author} {\bibinfo {author} {\bibfnamefont {M.~T.}\ \bibnamefont
  {Miles}} \emph {et~al.},\ }\bibfield  {title} {\bibinfo {title} {{The MeerKAT
  Pulsar Timing Array: The $4.5$-year data release and the noise and stochastic
  signals of the millisecond pulsar population}},\ }\href
  {https://doi.org/10.1093/mnras/stae2572} {\bibfield  {journal} {\bibinfo
  {journal} {Mon. Not. Roy. Astron. Soc.}\ }\textbf {\bibinfo {volume} {536}},\
  \bibinfo {pages} {1467} (\bibinfo {year} {2025})},\ \Eprint
  {https://arxiv.org/abs/2412.01148} {arXiv:2412.01148 [astro-ph.HE]}
  \BibitemShut {NoStop}%
\bibitem [{\citenamefont {Amaro-Seoane}\ \emph {et~al.}(2017)\citenamefont
  {Amaro-Seoane} \emph {et~al.}}]{LISA:2017pwj}%
  \BibitemOpen
  \bibfield  {author} {\bibinfo {author} {\bibfnamefont {P.}~\bibnamefont
  {Amaro-Seoane}} \emph {et~al.} (\bibinfo {collaboration} {LISA}),\ }\bibfield
   {title} {\bibinfo {title} {{Laser Interferometer Space Antenna}},\
  }\href@noop {} {\  (\bibinfo {year} {2017})},\ \Eprint
  {https://arxiv.org/abs/1702.00786} {arXiv:1702.00786 [astro-ph.IM]}
  \BibitemShut {NoStop}%
\bibitem [{\citenamefont {Yagi}\ and\ \citenamefont
  {Seto}(2011)}]{Yagi:2011wg}%
  \BibitemOpen
  \bibfield  {author} {\bibinfo {author} {\bibfnamefont {K.}~\bibnamefont
  {Yagi}}\ and\ \bibinfo {author} {\bibfnamefont {N.}~\bibnamefont {Seto}},\
  }\bibfield  {title} {\bibinfo {title} {{Detector configuration of DECIGO/BBO
  and identification of cosmological neutron-star binaries}},\ }\href
  {https://doi.org/10.1103/PhysRevD.83.044011} {\bibfield  {journal} {\bibinfo
  {journal} {Phys. Rev. D}\ }\textbf {\bibinfo {volume} {83}},\ \bibinfo
  {pages} {044011} (\bibinfo {year} {2011})},\ \bibinfo {note} {[Erratum:
  Phys.Rev.D 95, 109901 (2017)]},\ \Eprint {https://arxiv.org/abs/1101.3940}
  {arXiv:1101.3940 [astro-ph.CO]} \BibitemShut {NoStop}%
\bibitem [{\citenamefont {Maggiore}\ \emph {et~al.}(2020)\citenamefont
  {Maggiore} \emph {et~al.}}]{Maggiore:2019uih}%
  \BibitemOpen
  \bibfield  {author} {\bibinfo {author} {\bibfnamefont {M.}~\bibnamefont
  {Maggiore}} \emph {et~al.},\ }\bibfield  {title} {\bibinfo {title} {{Science
  Case for the Einstein Telescope}},\ }\href
  {https://doi.org/10.1088/1475-7516/2020/03/050} {\bibfield  {journal}
  {\bibinfo  {journal} {JCAP}\ }\textbf {\bibinfo {volume} {03}},\ \bibinfo
  {pages} {050}},\ \Eprint {https://arxiv.org/abs/1912.02622} {arXiv:1912.02622
  [astro-ph.CO]} \BibitemShut {NoStop}%
\bibitem [{\citenamefont {Kohri}\ and\ \citenamefont
  {Terada}(2018)}]{Kohri:2018awv}%
  \BibitemOpen
  \bibfield  {author} {\bibinfo {author} {\bibfnamefont {K.}~\bibnamefont
  {Kohri}}\ and\ \bibinfo {author} {\bibfnamefont {T.}~\bibnamefont {Terada}},\
  }\bibfield  {title} {\bibinfo {title} {{Semianalytic calculation of
  gravitational wave spectrum nonlinearly induced from primordial curvature
  perturbations}},\ }\href {https://doi.org/10.1103/PhysRevD.97.123532}
  {\bibfield  {journal} {\bibinfo  {journal} {Phys. Rev. D}\ }\textbf {\bibinfo
  {volume} {97}},\ \bibinfo {pages} {123532} (\bibinfo {year} {2018})},\
  \Eprint {https://arxiv.org/abs/1804.08577} {arXiv:1804.08577 [gr-qc]}
  \BibitemShut {NoStop}%
\bibitem [{\citenamefont {Acquaviva}\ \emph {et~al.}(2003)\citenamefont
  {Acquaviva}, \citenamefont {Bartolo}, \citenamefont {Matarrese},\ and\
  \citenamefont {Riotto}}]{Acquaviva:2002ud}%
  \BibitemOpen
  \bibfield  {author} {\bibinfo {author} {\bibfnamefont {V.}~\bibnamefont
  {Acquaviva}}, \bibinfo {author} {\bibfnamefont {N.}~\bibnamefont {Bartolo}},
  \bibinfo {author} {\bibfnamefont {S.}~\bibnamefont {Matarrese}},\ and\
  \bibinfo {author} {\bibfnamefont {A.}~\bibnamefont {Riotto}},\ }\bibfield
  {title} {\bibinfo {title} {{Second order cosmological perturbations from
  inflation}},\ }\href {https://doi.org/10.1016/S0550-3213(03)00550-9}
  {\bibfield  {journal} {\bibinfo  {journal} {Nucl. Phys. B}\ }\textbf
  {\bibinfo {volume} {667}},\ \bibinfo {pages} {119} (\bibinfo {year}
  {2003})},\ \Eprint {https://arxiv.org/abs/astro-ph/0209156}
  {arXiv:astro-ph/0209156} \BibitemShut {NoStop}%
\bibitem [{\citenamefont {Vagnozzi}(2023)}]{Vagnozzi:2023lwo}%
  \BibitemOpen
  \bibfield  {author} {\bibinfo {author} {\bibfnamefont {S.}~\bibnamefont
  {Vagnozzi}},\ }\bibfield  {title} {\bibinfo {title} {{Inflationary
  interpretation of the stochastic gravitational wave background signal
  detected by pulsar timing array experiments}},\ }\href
  {https://doi.org/10.1016/j.jheap.2023.07.001} {\bibfield  {journal} {\bibinfo
   {journal} {JHEAp}\ }\textbf {\bibinfo {volume} {39}},\ \bibinfo {pages} {81}
  (\bibinfo {year} {2023})},\ \Eprint {https://arxiv.org/abs/2306.16912}
  {arXiv:2306.16912 [astro-ph.CO]} \BibitemShut {NoStop}%
\bibitem [{\citenamefont {Fujikura}\ \emph {et~al.}(2023)\citenamefont
  {Fujikura}, \citenamefont {Girmohanta}, \citenamefont {Nakai},\ and\
  \citenamefont {Suzuki}}]{Fujikura:2023lkn}%
  \BibitemOpen
  \bibfield  {author} {\bibinfo {author} {\bibfnamefont {K.}~\bibnamefont
  {Fujikura}}, \bibinfo {author} {\bibfnamefont {S.}~\bibnamefont
  {Girmohanta}}, \bibinfo {author} {\bibfnamefont {Y.}~\bibnamefont {Nakai}},\
  and\ \bibinfo {author} {\bibfnamefont {M.}~\bibnamefont {Suzuki}},\
  }\bibfield  {title} {\bibinfo {title} {{NANOGrav signal from a dark conformal
  phase transition}},\ }\href {https://doi.org/10.1016/j.physletb.2023.138203}
  {\bibfield  {journal} {\bibinfo  {journal} {Phys. Lett. B}\ }\textbf
  {\bibinfo {volume} {846}},\ \bibinfo {pages} {138203} (\bibinfo {year}
  {2023})},\ \Eprint {https://arxiv.org/abs/2306.17086} {arXiv:2306.17086
  [hep-ph]} \BibitemShut {NoStop}%
\bibitem [{\citenamefont {Addazi}\ \emph {et~al.}(2024)\citenamefont {Addazi},
  \citenamefont {Cai}, \citenamefont {Marciano},\ and\ \citenamefont
  {Visinelli}}]{Addazi:2023jvg}%
  \BibitemOpen
  \bibfield  {author} {\bibinfo {author} {\bibfnamefont {A.}~\bibnamefont
  {Addazi}}, \bibinfo {author} {\bibfnamefont {Y.-F.}\ \bibnamefont {Cai}},
  \bibinfo {author} {\bibfnamefont {A.}~\bibnamefont {Marciano}},\ and\
  \bibinfo {author} {\bibfnamefont {L.}~\bibnamefont {Visinelli}},\ }\bibfield
  {title} {\bibinfo {title} {{Have pulsar timing array methods detected a
  cosmological phase transition?}},\ }\href
  {https://doi.org/10.1103/PhysRevD.109.015028} {\bibfield  {journal} {\bibinfo
   {journal} {Phys. Rev. D}\ }\textbf {\bibinfo {volume} {109}},\ \bibinfo
  {pages} {015028} (\bibinfo {year} {2024})},\ \Eprint
  {https://arxiv.org/abs/2306.17205} {arXiv:2306.17205 [astro-ph.CO]}
  \BibitemShut {NoStop}%
\bibitem [{\citenamefont {Megias}\ \emph {et~al.}(2023)\citenamefont {Megias},
  \citenamefont {Nardini},\ and\ \citenamefont {Quiros}}]{Megias:2023kiy}%
  \BibitemOpen
  \bibfield  {author} {\bibinfo {author} {\bibfnamefont {E.}~\bibnamefont
  {Megias}}, \bibinfo {author} {\bibfnamefont {G.}~\bibnamefont {Nardini}},\
  and\ \bibinfo {author} {\bibfnamefont {M.}~\bibnamefont {Quiros}},\
  }\bibfield  {title} {\bibinfo {title} {{Pulsar timing array stochastic
  background from light Kaluza-Klein resonances}},\ }\href
  {https://doi.org/10.1103/PhysRevD.108.095017} {\bibfield  {journal} {\bibinfo
   {journal} {Phys. Rev. D}\ }\textbf {\bibinfo {volume} {108}},\ \bibinfo
  {pages} {095017} (\bibinfo {year} {2023})},\ \Eprint
  {https://arxiv.org/abs/2306.17071} {arXiv:2306.17071 [hep-ph]} \BibitemShut
  {NoStop}%
\bibitem [{\citenamefont {Zu}\ \emph {et~al.}(2023)\citenamefont {Zu},
  \citenamefont {Zhang}, \citenamefont {Li}, \citenamefont {Gu}, \citenamefont
  {Tsai},\ and\ \citenamefont {Fan}}]{Zu:2023olm}%
  \BibitemOpen
  \bibfield  {author} {\bibinfo {author} {\bibfnamefont {L.}~\bibnamefont
  {Zu}}, \bibinfo {author} {\bibfnamefont {C.}~\bibnamefont {Zhang}}, \bibinfo
  {author} {\bibfnamefont {Y.-Y.}\ \bibnamefont {Li}}, \bibinfo {author}
  {\bibfnamefont {Y.-C.}\ \bibnamefont {Gu}}, \bibinfo {author} {\bibfnamefont
  {Y.-L.~S.}\ \bibnamefont {Tsai}},\ and\ \bibinfo {author} {\bibfnamefont
  {Y.-Z.}\ \bibnamefont {Fan}},\ }\bibfield  {title} {\bibinfo {title} {{Mirror
  QCD phase transition as the origin of the nanohertz Stochastic
  Gravitational-Wave Background}}\ }\href
  {https://doi.org/10.1016/j.scib.2024.01.037} {10.1016/j.scib.2024.01.037}
  (\bibinfo {year} {2023}),\ \Eprint {https://arxiv.org/abs/2306.16769}
  {arXiv:2306.16769 [astro-ph.HE]} \BibitemShut {NoStop}%
\bibitem [{\citenamefont {Winkler}\ and\ \citenamefont
  {Freese}(2024)}]{Winkler:2024olr}%
  \BibitemOpen
  \bibfield  {author} {\bibinfo {author} {\bibfnamefont {M.~W.}\ \bibnamefont
  {Winkler}}\ and\ \bibinfo {author} {\bibfnamefont {K.}~\bibnamefont
  {Freese}},\ }\bibfield  {title} {\bibinfo {title} {{Origin of the Stochastic
  Gravitational Wave Background: First-Order Phase Transition vs. Black Hole
  Mergers}},\ }\href@noop {} {\  (\bibinfo {year} {2024})},\ \Eprint
  {https://arxiv.org/abs/2401.13729} {arXiv:2401.13729 [astro-ph.CO]}
  \BibitemShut {NoStop}%
\bibitem [{\citenamefont {Ellis}\ \emph {et~al.}(2023)\citenamefont {Ellis},
  \citenamefont {Lewicki}, \citenamefont {Lin},\ and\ \citenamefont
  {Vaskonen}}]{Ellis:2023tsl}%
  \BibitemOpen
  \bibfield  {author} {\bibinfo {author} {\bibfnamefont {J.}~\bibnamefont
  {Ellis}}, \bibinfo {author} {\bibfnamefont {M.}~\bibnamefont {Lewicki}},
  \bibinfo {author} {\bibfnamefont {C.}~\bibnamefont {Lin}},\ and\ \bibinfo
  {author} {\bibfnamefont {V.}~\bibnamefont {Vaskonen}},\ }\bibfield  {title}
  {\bibinfo {title} {{Cosmic superstrings revisited in light of NANOGrav
  15-year data}},\ }\href {https://doi.org/10.1103/PhysRevD.108.103511}
  {\bibfield  {journal} {\bibinfo  {journal} {Phys. Rev. D}\ }\textbf {\bibinfo
  {volume} {108}},\ \bibinfo {pages} {103511} (\bibinfo {year} {2023})},\
  \Eprint {https://arxiv.org/abs/2306.17147} {arXiv:2306.17147 [astro-ph.CO]}
  \BibitemShut {NoStop}%
\bibitem [{\citenamefont {Kitajima}\ and\ \citenamefont
  {Nakayama}(2023)}]{Kitajima:2023vre}%
  \BibitemOpen
  \bibfield  {author} {\bibinfo {author} {\bibfnamefont {N.}~\bibnamefont
  {Kitajima}}\ and\ \bibinfo {author} {\bibfnamefont {K.}~\bibnamefont
  {Nakayama}},\ }\bibfield  {title} {\bibinfo {title} {{Nanohertz gravitational
  waves from cosmic strings and dark photon dark matter}},\ }\href
  {https://doi.org/10.1016/j.physletb.2023.138213} {\bibfield  {journal}
  {\bibinfo  {journal} {Phys. Lett. B}\ }\textbf {\bibinfo {volume} {846}},\
  \bibinfo {pages} {138213} (\bibinfo {year} {2023})},\ \Eprint
  {https://arxiv.org/abs/2306.17390} {arXiv:2306.17390 [hep-ph]} \BibitemShut
  {NoStop}%
\bibitem [{\citenamefont {Lazarides}\ \emph {et~al.}(2023)\citenamefont
  {Lazarides}, \citenamefont {Maji},\ and\ \citenamefont
  {Shafi}}]{Lazarides:2023ksx}%
  \BibitemOpen
  \bibfield  {author} {\bibinfo {author} {\bibfnamefont {G.}~\bibnamefont
  {Lazarides}}, \bibinfo {author} {\bibfnamefont {R.}~\bibnamefont {Maji}},\
  and\ \bibinfo {author} {\bibfnamefont {Q.}~\bibnamefont {Shafi}},\ }\bibfield
   {title} {\bibinfo {title} {{Superheavy quasistable strings and walls bounded
  by strings in the light of NANOGrav 15~year data}},\ }\href
  {https://doi.org/10.1103/PhysRevD.108.095041} {\bibfield  {journal} {\bibinfo
   {journal} {Phys. Rev. D}\ }\textbf {\bibinfo {volume} {108}},\ \bibinfo
  {pages} {095041} (\bibinfo {year} {2023})},\ \Eprint
  {https://arxiv.org/abs/2306.17788} {arXiv:2306.17788 [hep-ph]} \BibitemShut
  {NoStop}%
\bibitem [{\citenamefont {Eichhorn}\ \emph {et~al.}(2024)\citenamefont
  {Eichhorn}, \citenamefont {Lino~dos Santos},\ and\ \citenamefont
  {Miqueleto}}]{Eichhorn:2023gat}%
  \BibitemOpen
  \bibfield  {author} {\bibinfo {author} {\bibfnamefont {A.}~\bibnamefont
  {Eichhorn}}, \bibinfo {author} {\bibfnamefont {R.~R.}\ \bibnamefont {Lino~dos
  Santos}},\ and\ \bibinfo {author} {\bibfnamefont {J.~a.~L.}\ \bibnamefont
  {Miqueleto}},\ }\bibfield  {title} {\bibinfo {title} {{From quantum gravity
  to gravitational waves through cosmic strings}},\ }\href
  {https://doi.org/10.1103/PhysRevD.109.026013} {\bibfield  {journal} {\bibinfo
   {journal} {Phys. Rev. D}\ }\textbf {\bibinfo {volume} {109}},\ \bibinfo
  {pages} {026013} (\bibinfo {year} {2024})},\ \Eprint
  {https://arxiv.org/abs/2306.17718} {arXiv:2306.17718 [gr-qc]} \BibitemShut
  {NoStop}%
\bibitem [{\citenamefont {Kitajima}\ \emph {et~al.}(2023)\citenamefont
  {Kitajima}, \citenamefont {Lee}, \citenamefont {Murai}, \citenamefont
  {Takahashi},\ and\ \citenamefont {Yin}}]{Kitajima:2023cek}%
  \BibitemOpen
  \bibfield  {author} {\bibinfo {author} {\bibfnamefont {N.}~\bibnamefont
  {Kitajima}}, \bibinfo {author} {\bibfnamefont {J.}~\bibnamefont {Lee}},
  \bibinfo {author} {\bibfnamefont {K.}~\bibnamefont {Murai}}, \bibinfo
  {author} {\bibfnamefont {F.}~\bibnamefont {Takahashi}},\ and\ \bibinfo
  {author} {\bibfnamefont {W.}~\bibnamefont {Yin}},\ }\bibfield  {title}
  {\bibinfo {title} {{Gravitational Waves from Domain Wall Collapse, and
  Application to Nanohertz Signals with QCD-coupled Axions}},\ }\href@noop {}
  {\  (\bibinfo {year} {2023})},\ \Eprint {https://arxiv.org/abs/2306.17146}
  {arXiv:2306.17146 [hep-ph]} \BibitemShut {NoStop}%
\bibitem [{\citenamefont {Guo}\ \emph {et~al.}(2023)\citenamefont {Guo},
  \citenamefont {Khlopov}, \citenamefont {Liu}, \citenamefont {Wu},
  \citenamefont {Wu},\ and\ \citenamefont {Zhu}}]{Guo:2023hyp}%
  \BibitemOpen
  \bibfield  {author} {\bibinfo {author} {\bibfnamefont {S.-Y.}\ \bibnamefont
  {Guo}}, \bibinfo {author} {\bibfnamefont {M.}~\bibnamefont {Khlopov}},
  \bibinfo {author} {\bibfnamefont {X.}~\bibnamefont {Liu}}, \bibinfo {author}
  {\bibfnamefont {L.}~\bibnamefont {Wu}}, \bibinfo {author} {\bibfnamefont
  {Y.}~\bibnamefont {Wu}},\ and\ \bibinfo {author} {\bibfnamefont
  {B.}~\bibnamefont {Zhu}},\ }\bibfield  {title} {\bibinfo {title} {{Footprints
  of Axion-Like Particle in Pulsar Timing Array Data and JWST Observations}},\
  }\href@noop {} {\  (\bibinfo {year} {2023})},\ \Eprint
  {https://arxiv.org/abs/2306.17022} {arXiv:2306.17022 [hep-ph]} \BibitemShut
  {NoStop}%
\bibitem [{\citenamefont {Blasi}\ \emph {et~al.}(2023)\citenamefont {Blasi},
  \citenamefont {Mariotti}, \citenamefont {Rase},\ and\ \citenamefont
  {Sevrin}}]{Blasi:2023sej}%
  \BibitemOpen
  \bibfield  {author} {\bibinfo {author} {\bibfnamefont {S.}~\bibnamefont
  {Blasi}}, \bibinfo {author} {\bibfnamefont {A.}~\bibnamefont {Mariotti}},
  \bibinfo {author} {\bibfnamefont {A.}~\bibnamefont {Rase}},\ and\ \bibinfo
  {author} {\bibfnamefont {A.}~\bibnamefont {Sevrin}},\ }\bibfield  {title}
  {\bibinfo {title} {{Axionic domain walls at Pulsar Timing Arrays: QCD bias
  and particle friction}},\ }\href {https://doi.org/10.1007/JHEP11(2023)169}
  {\bibfield  {journal} {\bibinfo  {journal} {JHEP}\ }\textbf {\bibinfo
  {volume} {11}},\ \bibinfo {pages} {169}},\ \Eprint
  {https://arxiv.org/abs/2306.17830} {arXiv:2306.17830 [hep-ph]} \BibitemShut
  {NoStop}%
\bibitem [{\citenamefont {Gouttenoire}\ and\ \citenamefont
  {Vitagliano}(2023)}]{Gouttenoire:2023ftk}%
  \BibitemOpen
  \bibfield  {author} {\bibinfo {author} {\bibfnamefont {Y.}~\bibnamefont
  {Gouttenoire}}\ and\ \bibinfo {author} {\bibfnamefont {E.}~\bibnamefont
  {Vitagliano}},\ }\bibfield  {title} {\bibinfo {title} {{Domain wall
  interpretation of the PTA signal confronting black hole overproduction}},\
  }\href@noop {} {\  (\bibinfo {year} {2023})},\ \Eprint
  {https://arxiv.org/abs/2306.17841} {arXiv:2306.17841 [gr-qc]} \BibitemShut
  {NoStop}%
\bibitem [{\citenamefont {Babichev}\ \emph {et~al.}(2023)\citenamefont
  {Babichev}, \citenamefont {Gorbunov}, \citenamefont {Ramazanov},
  \citenamefont {Samanta},\ and\ \citenamefont {Vikman}}]{Babichev_2023}%
  \BibitemOpen
  \bibfield  {author} {\bibinfo {author} {\bibfnamefont {E.}~\bibnamefont
  {Babichev}}, \bibinfo {author} {\bibfnamefont {D.}~\bibnamefont {Gorbunov}},
  \bibinfo {author} {\bibfnamefont {S.}~\bibnamefont {Ramazanov}}, \bibinfo
  {author} {\bibfnamefont {R.}~\bibnamefont {Samanta}},\ and\ \bibinfo {author}
  {\bibfnamefont {A.}~\bibnamefont {Vikman}},\ }\bibfield  {title} {\bibinfo
  {title} {{NANOGrav spectral index \ensuremath{\gamma}=3 from melting domain
  walls}},\ }\href {https://doi.org/10.1103/PhysRevD.108.123529} {\bibfield
  {journal} {\bibinfo  {journal} {Phys. Rev. D}\ }\textbf {\bibinfo {volume}
  {108}},\ \bibinfo {pages} {123529} (\bibinfo {year} {2023})},\ \Eprint
  {https://arxiv.org/abs/2307.04582} {arXiv:2307.04582 [hep-ph]} \BibitemShut
  {NoStop}%
\bibitem [{\citenamefont {Vaskonen}\ and\ \citenamefont
  {Veerm\"ae}(2021)}]{Vaskonen:2020lbd}%
  \BibitemOpen
  \bibfield  {author} {\bibinfo {author} {\bibfnamefont {V.}~\bibnamefont
  {Vaskonen}}\ and\ \bibinfo {author} {\bibfnamefont {H.}~\bibnamefont
  {Veerm\"ae}},\ }\bibfield  {title} {\bibinfo {title} {{Did NANOGrav see a
  signal from primordial black hole formation?}},\ }\href
  {https://doi.org/10.1103/PhysRevLett.126.051303} {\bibfield  {journal}
  {\bibinfo  {journal} {Phys. Rev. Lett.}\ }\textbf {\bibinfo {volume} {126}},\
  \bibinfo {pages} {051303} (\bibinfo {year} {2021})},\ \Eprint
  {https://arxiv.org/abs/2009.07832} {arXiv:2009.07832 [astro-ph.CO]}
  \BibitemShut {NoStop}%
\bibitem [{\citenamefont {Franciolini}\ \emph {et~al.}(2023)\citenamefont
  {Franciolini}, \citenamefont {Iovino}, \citenamefont {Vaskonen},\ and\
  \citenamefont {Veermae}}]{Franciolini:2023pbf}%
  \BibitemOpen
  \bibfield  {author} {\bibinfo {author} {\bibfnamefont {G.}~\bibnamefont
  {Franciolini}}, \bibinfo {author} {\bibfnamefont {A.}~\bibnamefont {Iovino},
  \bibfnamefont {Junior.}}, \bibinfo {author} {\bibfnamefont {V.}~\bibnamefont
  {Vaskonen}},\ and\ \bibinfo {author} {\bibfnamefont {H.}~\bibnamefont
  {Veermae}},\ }\bibfield  {title} {\bibinfo {title} {{Recent Gravitational
  Wave Observation by Pulsar Timing Arrays and Primordial Black Holes: The
  Importance of Non-Gaussianities}},\ }\href
  {https://doi.org/10.1103/PhysRevLett.131.201401} {\bibfield  {journal}
  {\bibinfo  {journal} {Phys. Rev. Lett.}\ }\textbf {\bibinfo {volume} {131}},\
  \bibinfo {pages} {201401} (\bibinfo {year} {2023})},\ \Eprint
  {https://arxiv.org/abs/2306.17149} {arXiv:2306.17149 [astro-ph.CO]}
  \BibitemShut {NoStop}%
\bibitem [{\citenamefont {Inomata}\ \emph {et~al.}(2023)\citenamefont
  {Inomata}, \citenamefont {Kohri},\ and\ \citenamefont
  {Terada}}]{Inomata:2023zup}%
  \BibitemOpen
  \bibfield  {author} {\bibinfo {author} {\bibfnamefont {K.}~\bibnamefont
  {Inomata}}, \bibinfo {author} {\bibfnamefont {K.}~\bibnamefont {Kohri}},\
  and\ \bibinfo {author} {\bibfnamefont {T.}~\bibnamefont {Terada}},\
  }\bibfield  {title} {\bibinfo {title} {{The Detected Stochastic Gravitational
  Waves and Subsolar-Mass Primordial Black Holes}},\ }\href@noop {} {\
  (\bibinfo {year} {2023})},\ \Eprint {https://arxiv.org/abs/2306.17834}
  {arXiv:2306.17834 [astro-ph.CO]} \BibitemShut {NoStop}%
\bibitem [{\citenamefont {Cai}\ \emph {et~al.}(2023)\citenamefont {Cai},
  \citenamefont {He}, \citenamefont {Ma}, \citenamefont {Yan},\ and\
  \citenamefont {Yuan}}]{Cai:2023dls}%
  \BibitemOpen
  \bibfield  {author} {\bibinfo {author} {\bibfnamefont {Y.-F.}\ \bibnamefont
  {Cai}}, \bibinfo {author} {\bibfnamefont {X.-C.}\ \bibnamefont {He}},
  \bibinfo {author} {\bibfnamefont {X.-H.}\ \bibnamefont {Ma}}, \bibinfo
  {author} {\bibfnamefont {S.-F.}\ \bibnamefont {Yan}},\ and\ \bibinfo {author}
  {\bibfnamefont {G.-W.}\ \bibnamefont {Yuan}},\ }\bibfield  {title} {\bibinfo
  {title} {{Limits on scalar-induced gravitational waves from the stochastic
  background by pulsar timing array observations}},\ }\href
  {https://doi.org/10.1016/j.scib.2023.10.027} {\bibfield  {journal} {\bibinfo
  {journal} {Sci. Bull.}\ }\textbf {\bibinfo {volume} {68}},\ \bibinfo {pages}
  {2929} (\bibinfo {year} {2023})},\ \Eprint {https://arxiv.org/abs/2306.17822}
  {arXiv:2306.17822 [gr-qc]} \BibitemShut {NoStop}%
\bibitem [{\citenamefont {Wang}\ \emph {et~al.}(2023)\citenamefont {Wang},
  \citenamefont {Zhao}, \citenamefont {Li},\ and\ \citenamefont
  {Zhu}}]{Wang:2023ost}%
  \BibitemOpen
  \bibfield  {author} {\bibinfo {author} {\bibfnamefont {S.}~\bibnamefont
  {Wang}}, \bibinfo {author} {\bibfnamefont {Z.-C.}\ \bibnamefont {Zhao}},
  \bibinfo {author} {\bibfnamefont {J.-P.}\ \bibnamefont {Li}},\ and\ \bibinfo
  {author} {\bibfnamefont {Q.-H.}\ \bibnamefont {Zhu}},\ }\bibfield  {title}
  {\bibinfo {title} {{Implications of Pulsar Timing Array Data for
  Scalar-Induced Gravitational Waves and Primordial Black Holes: Primordial
  Non-Gaussianity $f_{\mathrm{NL}}$ Considered}},\ }\href@noop {} {\  (\bibinfo
  {year} {2023})},\ \Eprint {https://arxiv.org/abs/2307.00572}
  {arXiv:2307.00572 [astro-ph.CO]} \BibitemShut {NoStop}%
\bibitem [{\citenamefont {Ellis}\ \emph {et~al.}(2024)\citenamefont {Ellis},
  \citenamefont {Fairbairn}, \citenamefont {Franciolini}, \citenamefont
  {H\"utsi}, \citenamefont {Iovino}, \citenamefont {Lewicki}, \citenamefont
  {Raidal}, \citenamefont {Urrutia}, \citenamefont {Vaskonen},\ and\
  \citenamefont {Veerm\"ae}}]{Ellis:2023oxs}%
  \BibitemOpen
  \bibfield  {author} {\bibinfo {author} {\bibfnamefont {J.}~\bibnamefont
  {Ellis}}, \bibinfo {author} {\bibfnamefont {M.}~\bibnamefont {Fairbairn}},
  \bibinfo {author} {\bibfnamefont {G.}~\bibnamefont {Franciolini}}, \bibinfo
  {author} {\bibfnamefont {G.}~\bibnamefont {H\"utsi}}, \bibinfo {author}
  {\bibfnamefont {A.}~\bibnamefont {Iovino}}, \bibinfo {author} {\bibfnamefont
  {M.}~\bibnamefont {Lewicki}}, \bibinfo {author} {\bibfnamefont
  {M.}~\bibnamefont {Raidal}}, \bibinfo {author} {\bibfnamefont
  {J.}~\bibnamefont {Urrutia}}, \bibinfo {author} {\bibfnamefont
  {V.}~\bibnamefont {Vaskonen}},\ and\ \bibinfo {author} {\bibfnamefont
  {H.}~\bibnamefont {Veerm\"ae}},\ }\bibfield  {title} {\bibinfo {title} {{What
  is the source of the PTA GW signal?}},\ }\href
  {https://doi.org/10.1103/PhysRevD.109.023522} {\bibfield  {journal} {\bibinfo
   {journal} {Phys. Rev. D}\ }\textbf {\bibinfo {volume} {109}},\ \bibinfo
  {pages} {023522} (\bibinfo {year} {2024})},\ \Eprint
  {https://arxiv.org/abs/2308.08546} {arXiv:2308.08546 [astro-ph.CO]}
  \BibitemShut {NoStop}%
\bibitem [{\citenamefont {Figueroa}\ \emph {et~al.}(2024)\citenamefont
  {Figueroa}, \citenamefont {Pieroni}, \citenamefont {Ricciardone},\ and\
  \citenamefont {Simakachorn}}]{Figueroa:2023zhu}%
  \BibitemOpen
  \bibfield  {author} {\bibinfo {author} {\bibfnamefont {D.~G.}\ \bibnamefont
  {Figueroa}}, \bibinfo {author} {\bibfnamefont {M.}~\bibnamefont {Pieroni}},
  \bibinfo {author} {\bibfnamefont {A.}~\bibnamefont {Ricciardone}},\ and\
  \bibinfo {author} {\bibfnamefont {P.}~\bibnamefont {Simakachorn}},\
  }\bibfield  {title} {\bibinfo {title} {{Cosmological Background
  Interpretation of Pulsar Timing Array Data}},\ }\href
  {https://doi.org/10.1103/PhysRevLett.132.171002} {\bibfield  {journal}
  {\bibinfo  {journal} {Phys. Rev. Lett.}\ }\textbf {\bibinfo {volume} {132}},\
  \bibinfo {pages} {171002} (\bibinfo {year} {2024})},\ \Eprint
  {https://arxiv.org/abs/2307.02399} {arXiv:2307.02399 [astro-ph.CO]}
  \BibitemShut {NoStop}%
\bibitem [{\citenamefont {Iovino}\ \emph {et~al.}(2024)\citenamefont {Iovino},
  \citenamefont {Perna}, \citenamefont {Riotto},\ and\ \citenamefont
  {Veerm\"ae}}]{Iovino:2024tyg}%
  \BibitemOpen
  \bibfield  {author} {\bibinfo {author} {\bibfnamefont {A.~J.}\ \bibnamefont
  {Iovino}}, \bibinfo {author} {\bibfnamefont {G.}~\bibnamefont {Perna}},
  \bibinfo {author} {\bibfnamefont {A.}~\bibnamefont {Riotto}},\ and\ \bibinfo
  {author} {\bibfnamefont {H.}~\bibnamefont {Veerm\"ae}},\ }\bibfield  {title}
  {\bibinfo {title} {{Curbing PBHs with PTAs}},\ }\href
  {https://doi.org/10.1088/1475-7516/2024/10/050} {\bibfield  {journal}
  {\bibinfo  {journal} {JCAP}\ }\textbf {\bibinfo {volume} {10}},\ \bibinfo
  {pages} {050}},\ \Eprint {https://arxiv.org/abs/2406.20089} {arXiv:2406.20089
  [astro-ph.CO]} \BibitemShut {NoStop}%
\bibitem [{\citenamefont {Carr}\ \emph {et~al.}(2016)\citenamefont {Carr},
  \citenamefont {Kuhnel},\ and\ \citenamefont {Sandstad}}]{Carr:2016drx}%
  \BibitemOpen
  \bibfield  {author} {\bibinfo {author} {\bibfnamefont {B.}~\bibnamefont
  {Carr}}, \bibinfo {author} {\bibfnamefont {F.}~\bibnamefont {Kuhnel}},\ and\
  \bibinfo {author} {\bibfnamefont {M.}~\bibnamefont {Sandstad}},\ }\bibfield
  {title} {\bibinfo {title} {{Primordial Black Holes as Dark Matter}},\ }\href
  {https://doi.org/10.1103/PhysRevD.94.083504} {\bibfield  {journal} {\bibinfo
  {journal} {Phys. Rev. D}\ }\textbf {\bibinfo {volume} {94}},\ \bibinfo
  {pages} {083504} (\bibinfo {year} {2016})},\ \Eprint
  {https://arxiv.org/abs/1607.06077} {arXiv:1607.06077 [astro-ph.CO]}
  \BibitemShut {NoStop}%
\bibitem [{\citenamefont {Carr}\ \emph
  {et~al.}(2021{\natexlab{a}})\citenamefont {Carr}, \citenamefont {Kohri},
  \citenamefont {Sendouda},\ and\ \citenamefont {Yokoyama}}]{Carr:2020gox}%
  \BibitemOpen
  \bibfield  {author} {\bibinfo {author} {\bibfnamefont {B.}~\bibnamefont
  {Carr}}, \bibinfo {author} {\bibfnamefont {K.}~\bibnamefont {Kohri}},
  \bibinfo {author} {\bibfnamefont {Y.}~\bibnamefont {Sendouda}},\ and\
  \bibinfo {author} {\bibfnamefont {J.}~\bibnamefont {Yokoyama}},\ }\bibfield
  {title} {\bibinfo {title} {{Constraints on primordial black holes}},\ }\href
  {https://doi.org/10.1088/1361-6633/ac1e31} {\bibfield  {journal} {\bibinfo
  {journal} {Rept. Prog. Phys.}\ }\textbf {\bibinfo {volume} {84}},\ \bibinfo
  {pages} {116902} (\bibinfo {year} {2021}{\natexlab{a}})},\ \Eprint
  {https://arxiv.org/abs/2002.12778} {arXiv:2002.12778 [astro-ph.CO]}
  \BibitemShut {NoStop}%
\bibitem [{\citenamefont {Carr}\ and\ \citenamefont
  {Kuhnel}(2022)}]{Carr:2021bzv}%
  \BibitemOpen
  \bibfield  {author} {\bibinfo {author} {\bibfnamefont {B.}~\bibnamefont
  {Carr}}\ and\ \bibinfo {author} {\bibfnamefont {F.}~\bibnamefont {Kuhnel}},\
  }\bibfield  {title} {\bibinfo {title} {{Primordial black holes as dark matter
  candidates}},\ }\href {https://doi.org/10.21468/SciPostPhysLectNotes.48}
  {\bibfield  {journal} {\bibinfo  {journal} {SciPost Phys. Lect. Notes}\
  }\textbf {\bibinfo {volume} {48}},\ \bibinfo {pages} {1} (\bibinfo {year}
  {2022})},\ \Eprint {https://arxiv.org/abs/2110.02821} {arXiv:2110.02821
  [astro-ph.CO]} \BibitemShut {NoStop}%
\bibitem [{\citenamefont {Carr}\ \emph {et~al.}(2024)\citenamefont {Carr},
  \citenamefont {Clesse}, \citenamefont {Garcia-Bellido}, \citenamefont
  {Hawkins},\ and\ \citenamefont {Kuhnel}}]{Carr:2023tpt}%
  \BibitemOpen
  \bibfield  {author} {\bibinfo {author} {\bibfnamefont {B.}~\bibnamefont
  {Carr}}, \bibinfo {author} {\bibfnamefont {S.}~\bibnamefont {Clesse}},
  \bibinfo {author} {\bibfnamefont {J.}~\bibnamefont {Garcia-Bellido}},
  \bibinfo {author} {\bibfnamefont {M.}~\bibnamefont {Hawkins}},\ and\ \bibinfo
  {author} {\bibfnamefont {F.}~\bibnamefont {Kuhnel}},\ }\bibfield  {title}
  {\bibinfo {title} {{Observational evidence for primordial black holes: A
  positivist perspective}},\ }\href
  {https://doi.org/10.1016/j.physrep.2023.11.005} {\bibfield  {journal}
  {\bibinfo  {journal} {Phys. Rept.}\ }\textbf {\bibinfo {volume} {1054}},\
  \bibinfo {pages} {1} (\bibinfo {year} {2024})},\ \Eprint
  {https://arxiv.org/abs/2306.03903} {arXiv:2306.03903 [astro-ph.CO]}
  \BibitemShut {NoStop}%
\bibitem [{\citenamefont {Abbott}\ \emph {et~al.}(2022)\citenamefont {Abbott}
  \emph {et~al.}}]{LIGOScientific:2021job}%
  \BibitemOpen
  \bibfield  {author} {\bibinfo {author} {\bibfnamefont {R.}~\bibnamefont
  {Abbott}} \emph {et~al.} (\bibinfo {collaboration} {LIGO Scientific, VIRGO,
  KAGRA}),\ }\bibfield  {title} {\bibinfo {title} {{Search for Subsolar-Mass
  Binaries in the First Half of Advanced LIGO\textquoteright{}s and Advanced
  Virgo\textquoteright{}s Third Observing Run}},\ }\href
  {https://doi.org/10.1103/PhysRevLett.129.061104} {\bibfield  {journal}
  {\bibinfo  {journal} {Phys. Rev. Lett.}\ }\textbf {\bibinfo {volume} {129}},\
  \bibinfo {pages} {061104} (\bibinfo {year} {2022})},\ \Eprint
  {https://arxiv.org/abs/2109.12197} {arXiv:2109.12197 [astro-ph.CO]}
  \BibitemShut {NoStop}%
\bibitem [{\citenamefont {Phukon}\ \emph {et~al.}(2021)\citenamefont {Phukon},
  \citenamefont {Baltus}, \citenamefont {Caudill}, \citenamefont {Clesse},
  \citenamefont {Depasse}, \citenamefont {Fays}, \citenamefont {Fong},
  \citenamefont {Kapadia}, \citenamefont {Magee},\ and\ \citenamefont
  {Tanasijczuk}}]{Phukon:2021cus}%
  \BibitemOpen
  \bibfield  {author} {\bibinfo {author} {\bibfnamefont {K.~S.}\ \bibnamefont
  {Phukon}}, \bibinfo {author} {\bibfnamefont {G.}~\bibnamefont {Baltus}},
  \bibinfo {author} {\bibfnamefont {S.}~\bibnamefont {Caudill}}, \bibinfo
  {author} {\bibfnamefont {S.}~\bibnamefont {Clesse}}, \bibinfo {author}
  {\bibfnamefont {A.}~\bibnamefont {Depasse}}, \bibinfo {author} {\bibfnamefont
  {M.}~\bibnamefont {Fays}}, \bibinfo {author} {\bibfnamefont {H.}~\bibnamefont
  {Fong}}, \bibinfo {author} {\bibfnamefont {S.~J.}\ \bibnamefont {Kapadia}},
  \bibinfo {author} {\bibfnamefont {R.}~\bibnamefont {Magee}},\ and\ \bibinfo
  {author} {\bibfnamefont {A.~J.}\ \bibnamefont {Tanasijczuk}},\ }\bibfield
  {title} {\bibinfo {title} {{The hunt for sub-solar primordial black holes in
  low mass ratio binaries is open}},\ }\href@noop {} {\  (\bibinfo {year}
  {2021})},\ \Eprint {https://arxiv.org/abs/2105.11449} {arXiv:2105.11449
  [astro-ph.CO]} \BibitemShut {NoStop}%
\bibitem [{\citenamefont {Abbott}\ \emph {et~al.}(2023)\citenamefont {Abbott}
  \emph {et~al.}}]{LIGOScientific:2022hai}%
  \BibitemOpen
  \bibfield  {author} {\bibinfo {author} {\bibfnamefont {R.}~\bibnamefont
  {Abbott}} \emph {et~al.} (\bibinfo {collaboration} {LIGO Scientific, VIRGO,
  KAGRA}),\ }\bibfield  {title} {\bibinfo {title} {{Search for subsolar-mass
  black hole binaries in the second part of Advanced LIGO's and Advanced
  Virgo's third observing run}},\ }\href
  {https://doi.org/10.1093/mnras/stad588} {\bibfield  {journal} {\bibinfo
  {journal} {Mon. Not. Roy. Astron. Soc.}\ }\textbf {\bibinfo {volume} {524}},\
  \bibinfo {pages} {5984} (\bibinfo {year} {2023})},\ \bibinfo {note}
  {[Erratum: Mon.Not.Roy.Astron.Soc. 526, 6234 (2023)]},\ \Eprint
  {https://arxiv.org/abs/2212.01477} {arXiv:2212.01477 [astro-ph.HE]}
  \BibitemShut {NoStop}%
\bibitem [{\citenamefont {Morras}\ \emph {et~al.}(2023)\citenamefont {Morras}
  \emph {et~al.}}]{Morras:2023jvb}%
  \BibitemOpen
  \bibfield  {author} {\bibinfo {author} {\bibfnamefont {G.}~\bibnamefont
  {Morras}} \emph {et~al.},\ }\bibfield  {title} {\bibinfo {title} {{Analysis
  of a subsolar-mass compact binary candidate from the second observing run of
  Advanced LIGO}},\ }\href {https://doi.org/10.1016/j.dark.2023.101285}
  {\bibfield  {journal} {\bibinfo  {journal} {Phys. Dark Univ.}\ }\textbf
  {\bibinfo {volume} {42}},\ \bibinfo {pages} {101285} (\bibinfo {year}
  {2023})},\ \Eprint {https://arxiv.org/abs/2301.11619} {arXiv:2301.11619
  [gr-qc]} \BibitemShut {NoStop}%
\bibitem [{\citenamefont {Prunier}\ \emph {et~al.}(2023)\citenamefont
  {Prunier}, \citenamefont {Morr\'as}, \citenamefont {Siles}, \citenamefont
  {Clesse}, \citenamefont {Garc\'\i{}a-Bellido},\ and\ \citenamefont
  {Ruiz~Morales}}]{Prunier:2023cyv}%
  \BibitemOpen
  \bibfield  {author} {\bibinfo {author} {\bibfnamefont {M.}~\bibnamefont
  {Prunier}}, \bibinfo {author} {\bibfnamefont {G.}~\bibnamefont {Morr\'as}},
  \bibinfo {author} {\bibfnamefont {J.~F. N.~n.}\ \bibnamefont {Siles}},
  \bibinfo {author} {\bibfnamefont {S.}~\bibnamefont {Clesse}}, \bibinfo
  {author} {\bibfnamefont {J.}~\bibnamefont {Garc\'\i{}a-Bellido}},\ and\
  \bibinfo {author} {\bibfnamefont {E.}~\bibnamefont {Ruiz~Morales}},\
  }\bibfield  {title} {\bibinfo {title} {{Analysis of the subsolar-mass black
  hole candidate SSM200308 from the second part of the third observing run of
  Advanced LIGO-Virgo}},\ }\href@noop {} {\  (\bibinfo {year} {2023})},\
  \Eprint {https://arxiv.org/abs/2311.16085} {arXiv:2311.16085 [gr-qc]}
  \BibitemShut {NoStop}%
\bibitem [{\citenamefont {Miller}\ \emph {et~al.}(2024)\citenamefont {Miller},
  \citenamefont {Aggarwal}, \citenamefont {Clesse}, \citenamefont {De~Lillo},
  \citenamefont {Sachdev}, \citenamefont {Astone}, \citenamefont {Palomba},
  \citenamefont {Piccinni},\ and\ \citenamefont {Pierini}}]{Miller:2024fpo}%
  \BibitemOpen
  \bibfield  {author} {\bibinfo {author} {\bibfnamefont {A.~L.}\ \bibnamefont
  {Miller}}, \bibinfo {author} {\bibfnamefont {N.}~\bibnamefont {Aggarwal}},
  \bibinfo {author} {\bibfnamefont {S.}~\bibnamefont {Clesse}}, \bibinfo
  {author} {\bibfnamefont {F.}~\bibnamefont {De~Lillo}}, \bibinfo {author}
  {\bibfnamefont {S.}~\bibnamefont {Sachdev}}, \bibinfo {author} {\bibfnamefont
  {P.}~\bibnamefont {Astone}}, \bibinfo {author} {\bibfnamefont
  {C.}~\bibnamefont {Palomba}}, \bibinfo {author} {\bibfnamefont {O.~J.}\
  \bibnamefont {Piccinni}},\ and\ \bibinfo {author} {\bibfnamefont
  {L.}~\bibnamefont {Pierini}},\ }\bibfield  {title} {\bibinfo {title}
  {{Gravitational wave constraints on planetary-mass primordial black holes
  using LIGO O3a data}},\ }\href@noop {} {\  (\bibinfo {year} {2024})},\
  \Eprint {https://arxiv.org/abs/2402.19468} {arXiv:2402.19468 [gr-qc]}
  \BibitemShut {NoStop}%
\bibitem [{\citenamefont {Roper~Pol}\ \emph {et~al.}(2022)\citenamefont
  {Roper~Pol}, \citenamefont {Caprini}, \citenamefont {Neronov},\ and\
  \citenamefont {Semikoz}}]{RoperPol:2022iel}%
  \BibitemOpen
  \bibfield  {author} {\bibinfo {author} {\bibfnamefont {A.}~\bibnamefont
  {Roper~Pol}}, \bibinfo {author} {\bibfnamefont {C.}~\bibnamefont {Caprini}},
  \bibinfo {author} {\bibfnamefont {A.}~\bibnamefont {Neronov}},\ and\ \bibinfo
  {author} {\bibfnamefont {D.}~\bibnamefont {Semikoz}},\ }\bibfield  {title}
  {\bibinfo {title} {{Gravitational wave signal from primordial magnetic fields
  in the Pulsar Timing Array frequency band}},\ }\href
  {https://doi.org/10.1103/PhysRevD.105.123502} {\bibfield  {journal} {\bibinfo
   {journal} {Phys. Rev. D}\ }\textbf {\bibinfo {volume} {105}},\ \bibinfo
  {pages} {123502} (\bibinfo {year} {2022})},\ \Eprint
  {https://arxiv.org/abs/2201.05630} {arXiv:2201.05630 [astro-ph.CO]}
  \BibitemShut {NoStop}%
\bibitem [{\citenamefont {Yi}\ \emph {et~al.}(2023)\citenamefont {Yi},
  \citenamefont {Gao}, \citenamefont {Gong}, \citenamefont {Wang},\ and\
  \citenamefont {Zhang}}]{Yi:2023mbm}%
  \BibitemOpen
  \bibfield  {author} {\bibinfo {author} {\bibfnamefont {Z.}~\bibnamefont
  {Yi}}, \bibinfo {author} {\bibfnamefont {Q.}~\bibnamefont {Gao}}, \bibinfo
  {author} {\bibfnamefont {Y.}~\bibnamefont {Gong}}, \bibinfo {author}
  {\bibfnamefont {Y.}~\bibnamefont {Wang}},\ and\ \bibinfo {author}
  {\bibfnamefont {F.}~\bibnamefont {Zhang}},\ }\bibfield  {title} {\bibinfo
  {title} {{Scalar induced gravitational waves in light of Pulsar Timing Array
  data}},\ }\href {https://doi.org/10.1007/s11433-023-2266-1} {\bibfield
  {journal} {\bibinfo  {journal} {Sci. China Phys. Mech. Astron.}\ }\textbf
  {\bibinfo {volume} {66}},\ \bibinfo {pages} {120404} (\bibinfo {year}
  {2023})},\ \Eprint {https://arxiv.org/abs/2307.02467} {arXiv:2307.02467
  [gr-qc]} \BibitemShut {NoStop}%
\bibitem [{\citenamefont {Dom\`enech}\ \emph {et~al.}(2024)\citenamefont
  {Dom\`enech}, \citenamefont {Pi}, \citenamefont {Wang},\ and\ \citenamefont
  {Wang}}]{Domenech:2024rks}%
  \BibitemOpen
  \bibfield  {author} {\bibinfo {author} {\bibfnamefont {G.}~\bibnamefont
  {Dom\`enech}}, \bibinfo {author} {\bibfnamefont {S.}~\bibnamefont {Pi}},
  \bibinfo {author} {\bibfnamefont {A.}~\bibnamefont {Wang}},\ and\ \bibinfo
  {author} {\bibfnamefont {J.}~\bibnamefont {Wang}},\ }\bibfield  {title}
  {\bibinfo {title} {{Induced Gravitational Wave interpretation of PTA data: a
  complete study for general equation of state}},\ }\href@noop {} {\  (\bibinfo
  {year} {2024})},\ \Eprint {https://arxiv.org/abs/2402.18965}
  {arXiv:2402.18965 [astro-ph.CO]} \BibitemShut {NoStop}%
\bibitem [{\citenamefont {Chen}\ \emph {et~al.}(2024)\citenamefont {Chen},
  \citenamefont {Li}, \citenamefont {Liu},\ and\ \citenamefont
  {Yi}}]{Chen:2024fir}%
  \BibitemOpen
  \bibfield  {author} {\bibinfo {author} {\bibfnamefont {Z.-C.}\ \bibnamefont
  {Chen}}, \bibinfo {author} {\bibfnamefont {J.}~\bibnamefont {Li}}, \bibinfo
  {author} {\bibfnamefont {L.}~\bibnamefont {Liu}},\ and\ \bibinfo {author}
  {\bibfnamefont {Z.}~\bibnamefont {Yi}},\ }\bibfield  {title} {\bibinfo
  {title} {{Probing the speed of scalar-induced gravitational waves with pulsar
  timing arrays}},\ }\href@noop {} {\  (\bibinfo {year} {2024})},\ \Eprint
  {https://arxiv.org/abs/2401.09818} {arXiv:2401.09818 [gr-qc]} \BibitemShut
  {NoStop}%
\bibitem [{\citenamefont {Agazie}\ \emph
  {et~al.}(2023{\natexlab{c}})\citenamefont {Agazie} \emph
  {et~al.}}]{InternationalPulsarTimingArray:2023mzf}%
  \BibitemOpen
  \bibfield  {author} {\bibinfo {author} {\bibfnamefont {G.}~\bibnamefont
  {Agazie}} \emph {et~al.} (\bibinfo {collaboration} {International Pulsar
  Timing Array}),\ }\bibfield  {title} {\bibinfo {title} {{Comparing recent PTA
  results on the nanohertz stochastic gravitational wave background}},\
  }\href@noop {} {\  (\bibinfo {year} {2023}{\natexlab{c}})},\ \Eprint
  {https://arxiv.org/abs/2309.00693} {arXiv:2309.00693 [astro-ph.HE]}
  \BibitemShut {NoStop}%
\bibitem [{\citenamefont {Mitridate}\ \emph {et~al.}(2023)\citenamefont
  {Mitridate}, \citenamefont {Wright}, \citenamefont {von Eckardstein},
  \citenamefont {Schr\"oder}, \citenamefont {Nay}, \citenamefont {Olum},
  \citenamefont {Schmitz},\ and\ \citenamefont {Trickle}}]{Mitridate:2023oar}%
  \BibitemOpen
  \bibfield  {author} {\bibinfo {author} {\bibfnamefont {A.}~\bibnamefont
  {Mitridate}}, \bibinfo {author} {\bibfnamefont {D.}~\bibnamefont {Wright}},
  \bibinfo {author} {\bibfnamefont {R.}~\bibnamefont {von Eckardstein}},
  \bibinfo {author} {\bibfnamefont {T.}~\bibnamefont {Schr\"oder}}, \bibinfo
  {author} {\bibfnamefont {J.}~\bibnamefont {Nay}}, \bibinfo {author}
  {\bibfnamefont {K.}~\bibnamefont {Olum}}, \bibinfo {author} {\bibfnamefont
  {K.}~\bibnamefont {Schmitz}},\ and\ \bibinfo {author} {\bibfnamefont
  {T.}~\bibnamefont {Trickle}},\ }\bibfield  {title} {\bibinfo {title}
  {{PTArcade}},\ }\href@noop {} {\  (\bibinfo {year} {2023})},\ \Eprint
  {https://arxiv.org/abs/2306.16377} {arXiv:2306.16377 [hep-ph]} \BibitemShut
  {NoStop}%
\bibitem [{\citenamefont {You}\ \emph {et~al.}(2023)\citenamefont {You},
  \citenamefont {Yi},\ and\ \citenamefont {Wu}}]{You:2023rmn}%
  \BibitemOpen
  \bibfield  {author} {\bibinfo {author} {\bibfnamefont {Z.-Q.}\ \bibnamefont
  {You}}, \bibinfo {author} {\bibfnamefont {Z.}~\bibnamefont {Yi}},\ and\
  \bibinfo {author} {\bibfnamefont {Y.}~\bibnamefont {Wu}},\ }\bibfield
  {title} {\bibinfo {title} {{Constraints on primordial curvature power
  spectrum with pulsar timing arrays}},\ }\href
  {https://doi.org/10.1088/1475-7516/2023/11/065} {\bibfield  {journal}
  {\bibinfo  {journal} {JCAP}\ }\textbf {\bibinfo {volume} {11}},\ \bibinfo
  {pages} {065}},\ \Eprint {https://arxiv.org/abs/2307.04419} {arXiv:2307.04419
  [gr-qc]} \BibitemShut {NoStop}%
\bibitem [{\citenamefont {Handley}\ and\ \citenamefont
  {Lemos}(2021)}]{Handley:2020hdp}%
  \BibitemOpen
  \bibfield  {author} {\bibinfo {author} {\bibfnamefont {W.}~\bibnamefont
  {Handley}}\ and\ \bibinfo {author} {\bibfnamefont {P.}~\bibnamefont
  {Lemos}},\ }\bibfield  {title} {\bibinfo {title} {{Quantifying the global
  parameter tensions between ACT, SPT and Planck}},\ }\href
  {https://doi.org/10.1103/PhysRevD.103.063529} {\bibfield  {journal} {\bibinfo
   {journal} {Phys. Rev. D}\ }\textbf {\bibinfo {volume} {103}},\ \bibinfo
  {pages} {063529} (\bibinfo {year} {2021})},\ \Eprint
  {https://arxiv.org/abs/2007.08496} {arXiv:2007.08496 [astro-ph.CO]}
  \BibitemShut {NoStop}%
\bibitem [{\citenamefont {Ashok}\ \emph {et~al.}(2024)\citenamefont {Ashok},
  \citenamefont {Covas}, \citenamefont {Prix},\ and\ \citenamefont
  {Papa}}]{Ashok:2024fts}%
  \BibitemOpen
  \bibfield  {author} {\bibinfo {author} {\bibfnamefont {A.}~\bibnamefont
  {Ashok}}, \bibinfo {author} {\bibfnamefont {P.~B.}\ \bibnamefont {Covas}},
  \bibinfo {author} {\bibfnamefont {R.}~\bibnamefont {Prix}},\ and\ \bibinfo
  {author} {\bibfnamefont {M.~A.}\ \bibnamefont {Papa}},\ }\bibfield  {title}
  {\bibinfo {title} {{Bayesian $\mathcal{F}$-statistic-based parameter
  estimation of continuous gravitational waves from known pulsars}},\
  }\href@noop {} {\  (\bibinfo {year} {2024})},\ \Eprint
  {https://arxiv.org/abs/2401.17025} {arXiv:2401.17025 [gr-qc]} \BibitemShut
  {NoStop}%
\bibitem [{\citenamefont {Breschi}\ \emph {et~al.}(2024)\citenamefont
  {Breschi}, \citenamefont {Gamba}, \citenamefont {Carullo}, \citenamefont
  {Godzieba}, \citenamefont {Bernuzzi}, \citenamefont {Perego},\ and\
  \citenamefont {Radice}}]{Breschi:2024qlc}%
  \BibitemOpen
  \bibfield  {author} {\bibinfo {author} {\bibfnamefont {M.}~\bibnamefont
  {Breschi}}, \bibinfo {author} {\bibfnamefont {R.}~\bibnamefont {Gamba}},
  \bibinfo {author} {\bibfnamefont {G.}~\bibnamefont {Carullo}}, \bibinfo
  {author} {\bibfnamefont {D.}~\bibnamefont {Godzieba}}, \bibinfo {author}
  {\bibfnamefont {S.}~\bibnamefont {Bernuzzi}}, \bibinfo {author}
  {\bibfnamefont {A.}~\bibnamefont {Perego}},\ and\ \bibinfo {author}
  {\bibfnamefont {D.}~\bibnamefont {Radice}},\ }\bibfield  {title} {\bibinfo
  {title} {{Bayesian inference of multimessenger astrophysical data: Joint and
  coherent inference of gravitational waves and kilonovae}},\ }\href@noop {} {\
   (\bibinfo {year} {2024})},\ \Eprint {https://arxiv.org/abs/2401.03750}
  {arXiv:2401.03750 [astro-ph.HE]} \BibitemShut {NoStop}%
\bibitem [{\citenamefont {Aurrekoetxea}\ \emph {et~al.}(2023)\citenamefont
  {Aurrekoetxea}, \citenamefont {Hoy},\ and\ \citenamefont
  {Hannam}}]{Aurrekoetxea:2023vtp}%
  \BibitemOpen
  \bibfield  {author} {\bibinfo {author} {\bibfnamefont {J.~C.}\ \bibnamefont
  {Aurrekoetxea}}, \bibinfo {author} {\bibfnamefont {C.}~\bibnamefont {Hoy}},\
  and\ \bibinfo {author} {\bibfnamefont {M.}~\bibnamefont {Hannam}},\
  }\bibfield  {title} {\bibinfo {title} {{Revisiting the cosmic string origin
  of GW190521}},\ }\href@noop {} {\  (\bibinfo {year} {2023})},\ \Eprint
  {https://arxiv.org/abs/2312.03860} {arXiv:2312.03860 [gr-qc]} \BibitemShut
  {NoStop}%
\bibitem [{\citenamefont {Gupta}\ and\ \citenamefont
  {Cornish}(2024)}]{Gupta:2023jrn}%
  \BibitemOpen
  \bibfield  {author} {\bibinfo {author} {\bibfnamefont {T.}~\bibnamefont
  {Gupta}}\ and\ \bibinfo {author} {\bibfnamefont {N.~J.}\ \bibnamefont
  {Cornish}},\ }\bibfield  {title} {\bibinfo {title} {{Bayesian power spectral
  estimation of gravitational wave detector noise revisited}},\ }\href
  {https://doi.org/10.1103/PhysRevD.109.064040} {\bibfield  {journal} {\bibinfo
   {journal} {Phys. Rev. D}\ }\textbf {\bibinfo {volume} {109}},\ \bibinfo
  {pages} {064040} (\bibinfo {year} {2024})},\ \Eprint
  {https://arxiv.org/abs/2312.11808} {arXiv:2312.11808 [gr-qc]} \BibitemShut
  {NoStop}%
\bibitem [{\citenamefont {H\"ubner}\ \emph {et~al.}(2020)\citenamefont
  {H\"ubner}, \citenamefont {Talbot}, \citenamefont {Lasky},\ and\
  \citenamefont {Thrane}}]{Hubner:2019sly}%
  \BibitemOpen
  \bibfield  {author} {\bibinfo {author} {\bibfnamefont {M.}~\bibnamefont
  {H\"ubner}}, \bibinfo {author} {\bibfnamefont {C.}~\bibnamefont {Talbot}},
  \bibinfo {author} {\bibfnamefont {P.~D.}\ \bibnamefont {Lasky}},\ and\
  \bibinfo {author} {\bibfnamefont {E.}~\bibnamefont {Thrane}},\ }\bibfield
  {title} {\bibinfo {title} {{Measuring gravitational-wave memory in the first
  LIGO/Virgo gravitational-wave transient catalog}},\ }\href
  {https://doi.org/10.1103/PhysRevD.101.023011} {\bibfield  {journal} {\bibinfo
   {journal} {Phys. Rev. D}\ }\textbf {\bibinfo {volume} {101}},\ \bibinfo
  {pages} {023011} (\bibinfo {year} {2020})},\ \Eprint
  {https://arxiv.org/abs/1911.12496} {arXiv:1911.12496 [astro-ph.HE]}
  \BibitemShut {NoStop}%
\bibitem [{\citenamefont {Matas}\ and\ \citenamefont
  {Romano}(2021)}]{Matas:2020roi}%
  \BibitemOpen
  \bibfield  {author} {\bibinfo {author} {\bibfnamefont {A.}~\bibnamefont
  {Matas}}\ and\ \bibinfo {author} {\bibfnamefont {J.~D.}\ \bibnamefont
  {Romano}},\ }\bibfield  {title} {\bibinfo {title} {{Frequentist versus
  Bayesian analyses: Cross-correlation as an approximate sufficient statistic
  for LIGO-Virgo stochastic background searches}},\ }\href
  {https://doi.org/10.1103/PhysRevD.103.062003} {\bibfield  {journal} {\bibinfo
   {journal} {Phys. Rev. D}\ }\textbf {\bibinfo {volume} {103}},\ \bibinfo
  {pages} {062003} (\bibinfo {year} {2021})},\ \Eprint
  {https://arxiv.org/abs/2012.00907} {arXiv:2012.00907 [gr-qc]} \BibitemShut
  {NoStop}%
\bibitem [{\citenamefont {Benetti}\ \emph {et~al.}(2022)\citenamefont
  {Benetti}, \citenamefont {Graef},\ and\ \citenamefont
  {Vagnozzi}}]{Benetti:2021uea}%
  \BibitemOpen
  \bibfield  {author} {\bibinfo {author} {\bibfnamefont {M.}~\bibnamefont
  {Benetti}}, \bibinfo {author} {\bibfnamefont {L.~L.}\ \bibnamefont {Graef}},\
  and\ \bibinfo {author} {\bibfnamefont {S.}~\bibnamefont {Vagnozzi}},\
  }\bibfield  {title} {\bibinfo {title} {{Primordial gravitational waves from
  NANOGrav: A broken power-law approach}},\ }\href
  {https://doi.org/10.1103/PhysRevD.105.043520} {\bibfield  {journal} {\bibinfo
   {journal} {Phys. Rev. D}\ }\textbf {\bibinfo {volume} {105}},\ \bibinfo
  {pages} {043520} (\bibinfo {year} {2022})},\ \Eprint
  {https://arxiv.org/abs/2111.04758} {arXiv:2111.04758 [astro-ph.CO]}
  \BibitemShut {NoStop}%
\bibitem [{\citenamefont {Lamb}\ \emph {et~al.}(2023)\citenamefont {Lamb},
  \citenamefont {Taylor},\ and\ \citenamefont {van Haasteren}}]{Lamb:2023jls}%
  \BibitemOpen
  \bibfield  {author} {\bibinfo {author} {\bibfnamefont {W.~G.}\ \bibnamefont
  {Lamb}}, \bibinfo {author} {\bibfnamefont {S.~R.}\ \bibnamefont {Taylor}},\
  and\ \bibinfo {author} {\bibfnamefont {R.}~\bibnamefont {van Haasteren}},\
  }\bibfield  {title} {\bibinfo {title} {{Rapid refitting techniques for
  Bayesian spectral characterization of the gravitational wave background using
  pulsar timing arrays}},\ }\href {https://doi.org/10.1103/PhysRevD.108.103019}
  {\bibfield  {journal} {\bibinfo  {journal} {Phys. Rev. D}\ }\textbf {\bibinfo
  {volume} {108}},\ \bibinfo {pages} {103019} (\bibinfo {year} {2023})},\
  \Eprint {https://arxiv.org/abs/2303.15442} {arXiv:2303.15442 [astro-ph.HE]}
  \BibitemShut {NoStop}%
\bibitem [{\citenamefont {Piessens}\ \emph {et~al.}(1983)\citenamefont
  {Piessens}, \citenamefont {Doncker-Kapenga}, \citenamefont {Uberhuber},\ and\
  \citenamefont {Kahaner}}]{Piessens1983-wl}%
  \BibitemOpen
  \bibfield  {author} {\bibinfo {author} {\bibfnamefont {R.}~\bibnamefont
  {Piessens}}, \bibinfo {author} {\bibfnamefont {E.~D.}\ \bibnamefont
  {Doncker-Kapenga}}, \bibinfo {author} {\bibfnamefont {C.}~\bibnamefont
  {Uberhuber}},\ and\ \bibinfo {author} {\bibfnamefont {D.~K.}\ \bibnamefont
  {Kahaner}},\ }\bibfield  {title} {{\bibinfo {title}
  {Quadpack}},\ }\href@noop {} {\ \bibinfo {series} {Springer series in
  computational mathematics} (\bibinfo {year} {1983})}\BibitemShut {NoStop}%
\bibitem [{\citenamefont {Press}\ \emph {et~al.}(1996)\citenamefont {Press},
  \citenamefont {Teukolsky}, \citenamefont {Vetterling},\ and\ \citenamefont
  {Flannery}}]{10.5555/232468}%
  \BibitemOpen
  \bibfield  {author} {\bibinfo {author} {\bibfnamefont {W.~H.}\ \bibnamefont
  {Press}}, \bibinfo {author} {\bibfnamefont {S.~A.}\ \bibnamefont
  {Teukolsky}}, \bibinfo {author} {\bibfnamefont {W.~T.}\ \bibnamefont
  {Vetterling}},\ and\ \bibinfo {author} {\bibfnamefont {B.~P.}\ \bibnamefont
  {Flannery}},\ }\bibfield  {title} {\bibinfo {title} {Numerical recipes in
  fortran 90 (2nd ed.): the art of parallel scientific computing},\ }\href@noop
  {} {\  (\bibinfo {year} {1996})}\BibitemShut {NoStop}%
\bibitem [{\citenamefont {Nelder}\ and\ \citenamefont
  {Mead}(1965)}]{NeldMead65}%
  \BibitemOpen
  \bibfield  {author} {\bibinfo {author} {\bibfnamefont {J.~A.}\ \bibnamefont
  {Nelder}}\ and\ \bibinfo {author} {\bibfnamefont {R.}~\bibnamefont {Mead}},\
  }\bibfield  {title} {\bibinfo {title} {A simplex method for function
  minimization},\ }\href@noop {} {\bibfield  {journal} {\bibinfo  {journal}
  {Computer Journal}\ }\textbf {\bibinfo {volume} {7}},\ \bibinfo {pages} {308}
  (\bibinfo {year} {1965})}\BibitemShut {NoStop}%
\bibitem [{\citenamefont {Desesquelles}\ \emph {et~al.}(2009)\citenamefont
  {Desesquelles}, \citenamefont {Ha}, \citenamefont {Korichi}, \citenamefont
  {Le~Blanc},\ and\ \citenamefont {Petrache}}]{Desesquelles:2009kw}%
  \BibitemOpen
  \bibfield  {author} {\bibinfo {author} {\bibfnamefont {P.}~\bibnamefont
  {Desesquelles}}, \bibinfo {author} {\bibfnamefont {T.~M.~H.}\ \bibnamefont
  {Ha}}, \bibinfo {author} {\bibfnamefont {A.}~\bibnamefont {Korichi}},
  \bibinfo {author} {\bibfnamefont {F.}~\bibnamefont {Le~Blanc}},\ and\
  \bibinfo {author} {\bibfnamefont {C.~M.}\ \bibnamefont {Petrache}},\
  }\bibfield  {title} {\bibinfo {title} {{NNLC: Non-Negative Least Chi-square
  minimization and application to HPGe detectors}},\ }\href
  {https://doi.org/10.1088/0954-3899/36/3/037001} {\bibfield  {journal}
  {\bibinfo  {journal} {J. Phys. G}\ }\textbf {\bibinfo {volume} {36}},\
  \bibinfo {pages} {037001} (\bibinfo {year} {2009})},\ \Eprint
  {https://arxiv.org/abs/0902.3814} {arXiv:0902.3814 [nucl-ex]} \BibitemShut
  {NoStop}%
\bibitem [{\citenamefont {Byrnes}\ \emph {et~al.}(2019)\citenamefont {Byrnes},
  \citenamefont {Cole},\ and\ \citenamefont {Patil}}]{Byrnes:2018txb}%
  \BibitemOpen
  \bibfield  {author} {\bibinfo {author} {\bibfnamefont {C.~T.}\ \bibnamefont
  {Byrnes}}, \bibinfo {author} {\bibfnamefont {P.~S.}\ \bibnamefont {Cole}},\
  and\ \bibinfo {author} {\bibfnamefont {S.~P.}\ \bibnamefont {Patil}},\
  }\bibfield  {title} {\bibinfo {title} {{Steepest growth of the power spectrum
  and primordial black holes}},\ }\href
  {https://doi.org/10.1088/1475-7516/2019/06/028} {\bibfield  {journal}
  {\bibinfo  {journal} {JCAP}\ }\textbf {\bibinfo {volume} {06}},\ \bibinfo
  {pages} {028}},\ \Eprint {https://arxiv.org/abs/1811.11158} {arXiv:1811.11158
  [astro-ph.CO]} \BibitemShut {NoStop}%
\bibitem [{\citenamefont {Fumagalli}\ \emph {et~al.}(2021)\citenamefont
  {Fumagalli}, \citenamefont {Renaux-Petel},\ and\ \citenamefont
  {Witkowski}}]{Fumagalli:2021cel}%
  \BibitemOpen
  \bibfield  {author} {\bibinfo {author} {\bibfnamefont {J.}~\bibnamefont
  {Fumagalli}}, \bibinfo {author} {\bibfnamefont {S.~e.}\ \bibnamefont
  {Renaux-Petel}},\ and\ \bibinfo {author} {\bibfnamefont {L.~T.}\ \bibnamefont
  {Witkowski}},\ }\bibfield  {title} {\bibinfo {title} {{Resonant features in
  the stochastic gravitational wave background}},\ }\href
  {https://doi.org/10.1088/1475-7516/2021/08/059} {\bibfield  {journal}
  {\bibinfo  {journal} {JCAP}\ }\textbf {\bibinfo {volume} {08}},\ \bibinfo
  {pages} {059}},\ \Eprint {https://arxiv.org/abs/2105.06481} {arXiv:2105.06481
  [astro-ph.CO]} \BibitemShut {NoStop}%
\bibitem [{\citenamefont {Balaji}\ \emph {et~al.}(2022)\citenamefont {Balaji},
  \citenamefont {Domènech},\ and\ \citenamefont {Silk}}]{Balaji_2022}%
  \BibitemOpen
  \bibfield  {author} {\bibinfo {author} {\bibfnamefont {S.}~\bibnamefont
  {Balaji}}, \bibinfo {author} {\bibfnamefont {G.}~\bibnamefont {Domènech}},\
  and\ \bibinfo {author} {\bibfnamefont {J.}~\bibnamefont {Silk}},\ }\bibfield
  {title} {\bibinfo {title} {Induced gravitational waves from slow-roll
  inflation after an enhancing phase},\ }\href
  {https://doi.org/10.1088/1475-7516/2022/09/016} {\bibfield  {journal}
  {\bibinfo  {journal} {Journal of Cosmology and Astroparticle Physics}\
  }\textbf {\bibinfo {volume} {2022}}\bibinfo  {number} { (09)},\ \bibinfo
  {pages} {016}}\BibitemShut {NoStop}%
\bibitem [{\citenamefont {Frolovsky}\ and\ \citenamefont
  {Ketov}(2023)}]{Frolovsky_2023}%
  \BibitemOpen
\bibfield  {number} {  }\bibfield  {author} {\bibinfo {author} {\bibfnamefont
  {D.}~\bibnamefont {Frolovsky}}\ and\ \bibinfo {author} {\bibfnamefont
  {S.~V.}\ \bibnamefont {Ketov}},\ }\bibfield  {title} {\bibinfo {title}
  {Fitting power spectrum of scalar perturbations for primordial black hole
  production during inflation},\ }\href
  {https://doi.org/10.3390/astronomy2010005} {\bibfield  {journal} {\bibinfo
  {journal} {Astronomy}\ }\textbf {\bibinfo {volume} {2}},\ \bibinfo {pages}
  {47–57} (\bibinfo {year} {2023})}\BibitemShut {NoStop}%
\bibitem [{\citenamefont {Akrami}\ \emph {et~al.}(2020)\citenamefont {Akrami}
  \emph {et~al.}}]{Planck:2018jri}%
  \BibitemOpen
  \bibfield  {author} {\bibinfo {author} {\bibfnamefont {Y.}~\bibnamefont
  {Akrami}} \emph {et~al.} (\bibinfo {collaboration} {Planck}),\ }\bibfield
  {title} {\bibinfo {title} {{Planck 2018 results. X. Constraints on
  inflation}},\ }\href {https://doi.org/10.1051/0004-6361/201833887} {\bibfield
   {journal} {\bibinfo  {journal} {Astron. Astrophys.}\ }\textbf {\bibinfo
  {volume} {641}},\ \bibinfo {pages} {A10} (\bibinfo {year} {2020})},\ \Eprint
  {https://arxiv.org/abs/1807.06211} {arXiv:1807.06211 [astro-ph.CO]}
  \BibitemShut {NoStop}%
\bibitem [{\citenamefont {Pi}\ and\ \citenamefont {Sasaki}(2020)}]{Pi:2020otn}%
  \BibitemOpen
  \bibfield  {author} {\bibinfo {author} {\bibfnamefont {S.}~\bibnamefont
  {Pi}}\ and\ \bibinfo {author} {\bibfnamefont {M.}~\bibnamefont {Sasaki}},\
  }\bibfield  {title} {\bibinfo {title} {{Gravitational Waves Induced by Scalar
  Perturbations with a Lognormal Peak}},\ }\href
  {https://doi.org/10.1088/1475-7516/2020/09/037} {\bibfield  {journal}
  {\bibinfo  {journal} {JCAP}\ }\textbf {\bibinfo {volume} {09}},\ \bibinfo
  {pages} {037}},\ \Eprint {https://arxiv.org/abs/2005.12306} {arXiv:2005.12306
  [gr-qc]} \BibitemShut {NoStop}%
\bibitem [{\citenamefont {Adams}\ \emph {et~al.}(2001)\citenamefont {Adams},
  \citenamefont {Cresswell},\ and\ \citenamefont {Easther}}]{Adams:2001vc}%
  \BibitemOpen
  \bibfield  {author} {\bibinfo {author} {\bibfnamefont {J.~A.}\ \bibnamefont
  {Adams}}, \bibinfo {author} {\bibfnamefont {B.}~\bibnamefont {Cresswell}},\
  and\ \bibinfo {author} {\bibfnamefont {R.}~\bibnamefont {Easther}},\
  }\bibfield  {title} {\bibinfo {title} {{Inflationary perturbations from a
  potential with a step}},\ }\href {https://doi.org/10.1103/PhysRevD.64.123514}
  {\bibfield  {journal} {\bibinfo  {journal} {Phys. Rev. D}\ }\textbf {\bibinfo
  {volume} {64}},\ \bibinfo {pages} {123514} (\bibinfo {year} {2001})},\
  \Eprint {https://arxiv.org/abs/astro-ph/0102236} {arXiv:astro-ph/0102236}
  \BibitemShut {NoStop}%
\bibitem [{\citenamefont {Kefala}\ \emph {et~al.}(2021)\citenamefont {Kefala},
  \citenamefont {Kodaxis}, \citenamefont {Stamou},\ and\ \citenamefont
  {Tetradis}}]{Kefala:2020xsx}%
  \BibitemOpen
  \bibfield  {author} {\bibinfo {author} {\bibfnamefont {K.}~\bibnamefont
  {Kefala}}, \bibinfo {author} {\bibfnamefont {G.~P.}\ \bibnamefont {Kodaxis}},
  \bibinfo {author} {\bibfnamefont {I.~D.}\ \bibnamefont {Stamou}},\ and\
  \bibinfo {author} {\bibfnamefont {N.}~\bibnamefont {Tetradis}},\ }\bibfield
  {title} {\bibinfo {title} {{Features of the inflaton potential and the power
  spectrum of cosmological perturbations}},\ }\href
  {https://doi.org/10.1103/PhysRevD.104.023506} {\bibfield  {journal} {\bibinfo
   {journal} {Phys. Rev. D}\ }\textbf {\bibinfo {volume} {104}},\ \bibinfo
  {pages} {023506} (\bibinfo {year} {2021})},\ \Eprint
  {https://arxiv.org/abs/2010.12483} {arXiv:2010.12483 [astro-ph.CO]}
  \BibitemShut {NoStop}%
\bibitem [{\citenamefont {Cai}\ \emph {et~al.}(2022)\citenamefont {Cai},
  \citenamefont {Ma}, \citenamefont {Sasaki}, \citenamefont {Wang},\ and\
  \citenamefont {Zhou}}]{Cai:2021zsp}%
  \BibitemOpen
  \bibfield  {author} {\bibinfo {author} {\bibfnamefont {Y.-F.}\ \bibnamefont
  {Cai}}, \bibinfo {author} {\bibfnamefont {X.-H.}\ \bibnamefont {Ma}},
  \bibinfo {author} {\bibfnamefont {M.}~\bibnamefont {Sasaki}}, \bibinfo
  {author} {\bibfnamefont {D.-G.}\ \bibnamefont {Wang}},\ and\ \bibinfo
  {author} {\bibfnamefont {Z.}~\bibnamefont {Zhou}},\ }\bibfield  {title}
  {\bibinfo {title} {{One small step for an inflaton, one giant leap for
  inflation: A novel non-Gaussian tail and primordial black holes}},\ }\href
  {https://doi.org/10.1016/j.physletb.2022.137461} {\bibfield  {journal}
  {\bibinfo  {journal} {Phys. Lett. B}\ }\textbf {\bibinfo {volume} {834}},\
  \bibinfo {pages} {137461} (\bibinfo {year} {2022})},\ \Eprint
  {https://arxiv.org/abs/2112.13836} {arXiv:2112.13836 [astro-ph.CO]}
  \BibitemShut {NoStop}%
\bibitem [{\citenamefont {Ballesteros}\ \emph {et~al.}(2020)\citenamefont
  {Ballesteros}, \citenamefont {Rey},\ and\ \citenamefont
  {Rompineve}}]{Ballesteros:2019hus}%
  \BibitemOpen
  \bibfield  {author} {\bibinfo {author} {\bibfnamefont {G.}~\bibnamefont
  {Ballesteros}}, \bibinfo {author} {\bibfnamefont {J.}~\bibnamefont {Rey}},\
  and\ \bibinfo {author} {\bibfnamefont {F.}~\bibnamefont {Rompineve}},\
  }\bibfield  {title} {\bibinfo {title} {{Detuning primordial black hole dark
  matter with early matter domination and axion monodromy}},\ }\href
  {https://doi.org/10.1088/1475-7516/2020/06/014} {\bibfield  {journal}
  {\bibinfo  {journal} {JCAP}\ }\textbf {\bibinfo {volume} {06}},\ \bibinfo
  {pages} {014}},\ \Eprint {https://arxiv.org/abs/1912.01638} {arXiv:1912.01638
  [astro-ph.CO]} \BibitemShut {NoStop}%
\bibitem [{\citenamefont {Mavromatos}\ \emph {et~al.}(2022)\citenamefont
  {Mavromatos}, \citenamefont {Spanos},\ and\ \citenamefont
  {Stamou}}]{Mavromatos:2022yql}%
  \BibitemOpen
  \bibfield  {author} {\bibinfo {author} {\bibfnamefont {N.~E.}\ \bibnamefont
  {Mavromatos}}, \bibinfo {author} {\bibfnamefont {V.~C.}\ \bibnamefont
  {Spanos}},\ and\ \bibinfo {author} {\bibfnamefont {I.~D.}\ \bibnamefont
  {Stamou}},\ }\bibfield  {title} {\bibinfo {title} {{Primordial black holes
  and gravitational waves in multiaxion-Chern-Simons inflation}},\ }\href
  {https://doi.org/10.1103/PhysRevD.106.063532} {\bibfield  {journal} {\bibinfo
   {journal} {Phys. Rev. D}\ }\textbf {\bibinfo {volume} {106}},\ \bibinfo
  {pages} {063532} (\bibinfo {year} {2022})},\ \Eprint
  {https://arxiv.org/abs/2206.07963} {arXiv:2206.07963 [hep-th]} \BibitemShut
  {NoStop}%
\bibitem [{\citenamefont {Cook}(2023)}]{Cook:2022zol}%
  \BibitemOpen
  \bibfield  {author} {\bibinfo {author} {\bibfnamefont {J.~L.}\ \bibnamefont
  {Cook}},\ }\bibfield  {title} {\bibinfo {title} {{Primordial black hole
  production in natural and hilltop inflation}},\ }\href
  {https://doi.org/10.1088/1475-7516/2023/07/031} {\bibfield  {journal}
  {\bibinfo  {journal} {JCAP}\ }\textbf {\bibinfo {volume} {07}},\ \bibinfo
  {pages} {031}},\ \Eprint {https://arxiv.org/abs/2209.05674} {arXiv:2209.05674
  [astro-ph.CO]} \BibitemShut {NoStop}%
\bibitem [{\citenamefont {Kallosh}\ and\ \citenamefont
  {Linde}(2013)}]{Kallosh_2013}%
  \BibitemOpen
  \bibfield  {author} {\bibinfo {author} {\bibfnamefont {R.}~\bibnamefont
  {Kallosh}}\ and\ \bibinfo {author} {\bibfnamefont {A.}~\bibnamefont
  {Linde}},\ }\bibfield  {title} {\bibinfo {title} {Universality class in
  conformal inflation},\ }\href {https://doi.org/10.1088/1475-7516/2013/07/002}
  {\bibfield  {journal} {\bibinfo  {journal} {Journal of Cosmology and
  Astroparticle Physics}\ }\textbf {\bibinfo {volume} {2013}}\bibinfo  {number}
  { (07)},\ \bibinfo {pages} {002–002}}\BibitemShut {NoStop}%
\bibitem [{\citenamefont {Galante}\ \emph {et~al.}(2015)\citenamefont
  {Galante}, \citenamefont {Kallosh}, \citenamefont {Linde},\ and\
  \citenamefont {Roest}}]{Galante_2015}%
  \BibitemOpen
\bibfield  {number} {  }\bibfield  {author} {\bibinfo {author} {\bibfnamefont
  {M.}~\bibnamefont {Galante}}, \bibinfo {author} {\bibfnamefont
  {R.}~\bibnamefont {Kallosh}}, \bibinfo {author} {\bibfnamefont
  {A.}~\bibnamefont {Linde}},\ and\ \bibinfo {author} {\bibfnamefont
  {D.}~\bibnamefont {Roest}},\ }\bibfield  {title} {\bibinfo {title} {Unity of
  cosmological inflation attractors},\ }\bibfield  {journal} {\bibinfo
  {journal} {Physical Review Letters}\ }\textbf {\bibinfo {volume} {114}},\
  \href {https://doi.org/10.1103/physrevlett.114.141302}
  {10.1103/physrevlett.114.141302} (\bibinfo {year} {2015})\BibitemShut
  {NoStop}%
\bibitem [{\citenamefont {Bagui}\ \emph {et~al.}(2023)\citenamefont {Bagui}
  \emph {et~al.}}]{LISACosmologyWorkingGroup:2023njw}%
  \BibitemOpen
  \bibfield  {author} {\bibinfo {author} {\bibfnamefont {E.}~\bibnamefont
  {Bagui}} \emph {et~al.} (\bibinfo {collaboration} {LISA Cosmology Working
  Group}),\ }\bibfield  {title} {\bibinfo {title} {{Primordial black holes and
  their gravitational-wave signatures}},\ }\href@noop {} {\  (\bibinfo {year}
  {2023})},\ \Eprint {https://arxiv.org/abs/2310.19857} {arXiv:2310.19857
  [astro-ph.CO]} \BibitemShut {NoStop}%
\bibitem [{\citenamefont {Musco}(2019)}]{Musco:2018rwt}%
  \BibitemOpen
  \bibfield  {author} {\bibinfo {author} {\bibfnamefont {I.}~\bibnamefont
  {Musco}},\ }\bibfield  {title} {\bibinfo {title} {{Threshold for primordial
  black holes: Dependence on the shape of the cosmological perturbations}},\
  }\href {https://doi.org/10.1103/PhysRevD.100.123524} {\bibfield  {journal}
  {\bibinfo  {journal} {Phys. Rev. D}\ }\textbf {\bibinfo {volume} {100}},\
  \bibinfo {pages} {123524} (\bibinfo {year} {2019})},\ \Eprint
  {https://arxiv.org/abs/1809.02127} {arXiv:1809.02127 [gr-qc]} \BibitemShut
  {NoStop}%
\bibitem [{\citenamefont {Musco}\ \emph {et~al.}(2021)\citenamefont {Musco},
  \citenamefont {De~Luca}, \citenamefont {Franciolini},\ and\ \citenamefont
  {Riotto}}]{Musco:2020jjb}%
  \BibitemOpen
  \bibfield  {author} {\bibinfo {author} {\bibfnamefont {I.}~\bibnamefont
  {Musco}}, \bibinfo {author} {\bibfnamefont {V.}~\bibnamefont {De~Luca}},
  \bibinfo {author} {\bibfnamefont {G.}~\bibnamefont {Franciolini}},\ and\
  \bibinfo {author} {\bibfnamefont {A.}~\bibnamefont {Riotto}},\ }\bibfield
  {title} {\bibinfo {title} {{Threshold for primordial black holes. II. A
  simple analytic prescription}},\ }\href
  {https://doi.org/10.1103/PhysRevD.103.063538} {\bibfield  {journal} {\bibinfo
   {journal} {Phys. Rev. D}\ }\textbf {\bibinfo {volume} {103}},\ \bibinfo
  {pages} {063538} (\bibinfo {year} {2021})},\ \Eprint
  {https://arxiv.org/abs/2011.03014} {arXiv:2011.03014 [astro-ph.CO]}
  \BibitemShut {NoStop}%
\bibitem [{\citenamefont {Escriv\`a}\ \emph {et~al.}(2021)\citenamefont
  {Escriv\`a}, \citenamefont {Germani},\ and\ \citenamefont
  {Sheth}}]{Escriva:2020tak}%
  \BibitemOpen
  \bibfield  {author} {\bibinfo {author} {\bibfnamefont {A.}~\bibnamefont
  {Escriv\`a}}, \bibinfo {author} {\bibfnamefont {C.}~\bibnamefont {Germani}},\
  and\ \bibinfo {author} {\bibfnamefont {R.~K.}\ \bibnamefont {Sheth}},\
  }\bibfield  {title} {\bibinfo {title} {{Analytical thresholds for black hole
  formation in general cosmological backgrounds}},\ }\href
  {https://doi.org/10.1088/1475-7516/2021/01/030} {\bibfield  {journal}
  {\bibinfo  {journal} {JCAP}\ }\textbf {\bibinfo {volume} {01}},\ \bibinfo
  {pages} {030}},\ \Eprint {https://arxiv.org/abs/2007.05564} {arXiv:2007.05564
  [gr-qc]} \BibitemShut {NoStop}%
\bibitem [{\citenamefont {Escriv\`a}\ \emph {et~al.}(2020)\citenamefont
  {Escriv\`a}, \citenamefont {Germani},\ and\ \citenamefont
  {Sheth}}]{Escriva:2019phb}%
  \BibitemOpen
  \bibfield  {author} {\bibinfo {author} {\bibfnamefont {A.}~\bibnamefont
  {Escriv\`a}}, \bibinfo {author} {\bibfnamefont {C.}~\bibnamefont {Germani}},\
  and\ \bibinfo {author} {\bibfnamefont {R.~K.}\ \bibnamefont {Sheth}},\
  }\bibfield  {title} {\bibinfo {title} {{Universal threshold for primordial
  black hole formation}},\ }\href {https://doi.org/10.1103/PhysRevD.101.044022}
  {\bibfield  {journal} {\bibinfo  {journal} {Phys. Rev. D}\ }\textbf {\bibinfo
  {volume} {101}},\ \bibinfo {pages} {044022} (\bibinfo {year} {2020})},\
  \Eprint {https://arxiv.org/abs/1907.13311} {arXiv:1907.13311 [gr-qc]}
  \BibitemShut {NoStop}%
\bibitem [{\citenamefont {Stamou}(2023)}]{Stamou:2023vxu}%
  \BibitemOpen
  \bibfield  {author} {\bibinfo {author} {\bibfnamefont {I.~D.}\ \bibnamefont
  {Stamou}},\ }\bibfield  {title} {\bibinfo {title} {{Exploring critical
  overdensity thresholds in inflationary models of primordial black holes
  formation}},\ }\href {https://doi.org/10.1103/PhysRevD.108.063515} {\bibfield
   {journal} {\bibinfo  {journal} {Phys. Rev. D}\ }\textbf {\bibinfo {volume}
  {108}},\ \bibinfo {pages} {063515} (\bibinfo {year} {2023})},\ \Eprint
  {https://arxiv.org/abs/2306.02758} {arXiv:2306.02758 [astro-ph.CO]}
  \BibitemShut {NoStop}%
\bibitem [{\citenamefont {Carr}(1975)}]{Carr:1975qj}%
  \BibitemOpen
  \bibfield  {author} {\bibinfo {author} {\bibfnamefont {B.~J.}\ \bibnamefont
  {Carr}},\ }\bibfield  {title} {\bibinfo {title} {{The Primordial black hole
  mass spectrum}},\ }\href {https://doi.org/10.1086/153853} {\bibfield
  {journal} {\bibinfo  {journal} {Astrophys. J.}\ }\textbf {\bibinfo {volume}
  {201}},\ \bibinfo {pages} {1} (\bibinfo {year} {1975})}\BibitemShut {NoStop}%
\bibitem [{\citenamefont {Carr}\ and\ \citenamefont
  {Hawking}(1974)}]{Carr:1974nx}%
  \BibitemOpen
  \bibfield  {author} {\bibinfo {author} {\bibfnamefont {B.~J.}\ \bibnamefont
  {Carr}}\ and\ \bibinfo {author} {\bibfnamefont {S.~W.}\ \bibnamefont
  {Hawking}},\ }\bibfield  {title} {\bibinfo {title} {{Black holes in the early
  Universe}},\ }\href {https://doi.org/10.1093/mnras/168.2.399} {\bibfield
  {journal} {\bibinfo  {journal} {Mon. Not. Roy. Astron. Soc.}\ }\textbf
  {\bibinfo {volume} {168}},\ \bibinfo {pages} {399} (\bibinfo {year}
  {1974})}\BibitemShut {NoStop}%
\bibitem [{\citenamefont {Young}\ \emph {et~al.}(2019)\citenamefont {Young},
  \citenamefont {Musco},\ and\ \citenamefont {Byrnes}}]{Young:2019yug}%
  \BibitemOpen
  \bibfield  {author} {\bibinfo {author} {\bibfnamefont {S.}~\bibnamefont
  {Young}}, \bibinfo {author} {\bibfnamefont {I.}~\bibnamefont {Musco}},\ and\
  \bibinfo {author} {\bibfnamefont {C.~T.}\ \bibnamefont {Byrnes}},\ }\bibfield
   {title} {\bibinfo {title} {{Primordial black hole formation and abundance:
  contribution from the non-linear relation between the density and curvature
  perturbation}},\ }\href {https://doi.org/10.1088/1475-7516/2019/11/012}
  {\bibfield  {journal} {\bibinfo  {journal} {JCAP}\ }\textbf {\bibinfo
  {volume} {11}},\ \bibinfo {pages} {012}},\ \Eprint
  {https://arxiv.org/abs/1904.00984} {arXiv:1904.00984 [astro-ph.CO]}
  \BibitemShut {NoStop}%
\bibitem [{\citenamefont {Choptuik}(1993)}]{PhysRevLett.70.9}%
  \BibitemOpen
  \bibfield  {author} {\bibinfo {author} {\bibfnamefont {M.~W.}\ \bibnamefont
  {Choptuik}},\ }\bibfield  {title} {\bibinfo {title} {Universality and scaling
  in gravitational collapse of a massless scalar field},\ }\href
  {https://doi.org/10.1103/PhysRevLett.70.9} {\bibfield  {journal} {\bibinfo
  {journal} {Phys. Rev. Lett.}\ }\textbf {\bibinfo {volume} {70}},\ \bibinfo
  {pages} {9} (\bibinfo {year} {1993})}\BibitemShut {NoStop}%
\bibitem [{\citenamefont {Evans}\ and\ \citenamefont
  {Coleman}(1994)}]{Evans:1994pj}%
  \BibitemOpen
  \bibfield  {author} {\bibinfo {author} {\bibfnamefont {C.~R.}\ \bibnamefont
  {Evans}}\ and\ \bibinfo {author} {\bibfnamefont {J.~S.}\ \bibnamefont
  {Coleman}},\ }\bibfield  {title} {\bibinfo {title} {{Observation of critical
  phenomena and selfsimilarity in the gravitational collapse of radiation
  fluid}},\ }\href {https://doi.org/10.1103/PhysRevLett.72.1782} {\bibfield
  {journal} {\bibinfo  {journal} {Phys. Rev. Lett.}\ }\textbf {\bibinfo
  {volume} {72}},\ \bibinfo {pages} {1782} (\bibinfo {year} {1994})},\ \Eprint
  {https://arxiv.org/abs/gr-qc/9402041} {arXiv:gr-qc/9402041} \BibitemShut
  {NoStop}%
\bibitem [{\citenamefont {Koike}\ \emph {et~al.}(1995)\citenamefont {Koike},
  \citenamefont {Hara},\ and\ \citenamefont {Adachi}}]{Koike:1995jm}%
  \BibitemOpen
  \bibfield  {author} {\bibinfo {author} {\bibfnamefont {T.}~\bibnamefont
  {Koike}}, \bibinfo {author} {\bibfnamefont {T.}~\bibnamefont {Hara}},\ and\
  \bibinfo {author} {\bibfnamefont {S.}~\bibnamefont {Adachi}},\ }\bibfield
  {title} {\bibinfo {title} {{Critical behavior in gravitational collapse of
  radiation fluid: A Renormalization group (linear perturbation) analysis}},\
  }\href {https://doi.org/10.1103/PhysRevLett.74.5170} {\bibfield  {journal}
  {\bibinfo  {journal} {Phys. Rev. Lett.}\ }\textbf {\bibinfo {volume} {74}},\
  \bibinfo {pages} {5170} (\bibinfo {year} {1995})},\ \Eprint
  {https://arxiv.org/abs/gr-qc/9503007} {arXiv:gr-qc/9503007} \BibitemShut
  {NoStop}%
\bibitem [{\citenamefont {Frosina}\ and\ \citenamefont
  {Urbano}(2023)}]{Frosina:2023nxu}%
  \BibitemOpen
  \bibfield  {author} {\bibinfo {author} {\bibfnamefont {L.}~\bibnamefont
  {Frosina}}\ and\ \bibinfo {author} {\bibfnamefont {A.}~\bibnamefont
  {Urbano}},\ }\bibfield  {title} {\bibinfo {title} {{Inflationary
  interpretation of the nHz gravitational-wave background}},\ }\href
  {https://doi.org/10.1103/PhysRevD.108.103544} {\bibfield  {journal} {\bibinfo
   {journal} {Phys. Rev. D}\ }\textbf {\bibinfo {volume} {108}},\ \bibinfo
  {pages} {103544} (\bibinfo {year} {2023})},\ \Eprint
  {https://arxiv.org/abs/2308.06915} {arXiv:2308.06915 [astro-ph.CO]}
  \BibitemShut {NoStop}%
\bibitem [{\citenamefont {De~Luca}\ \emph {et~al.}(2019)\citenamefont
  {De~Luca}, \citenamefont {Franciolini}, \citenamefont {Kehagias},
  \citenamefont {Peloso}, \citenamefont {Riotto},\ and\ \citenamefont
  {\"Unal}}]{DeLuca:2019qsy}%
  \BibitemOpen
  \bibfield  {author} {\bibinfo {author} {\bibfnamefont {V.}~\bibnamefont
  {De~Luca}}, \bibinfo {author} {\bibfnamefont {G.}~\bibnamefont
  {Franciolini}}, \bibinfo {author} {\bibfnamefont {A.}~\bibnamefont
  {Kehagias}}, \bibinfo {author} {\bibfnamefont {M.}~\bibnamefont {Peloso}},
  \bibinfo {author} {\bibfnamefont {A.}~\bibnamefont {Riotto}},\ and\ \bibinfo
  {author} {\bibfnamefont {C.}~\bibnamefont {\"Unal}},\ }\bibfield  {title}
  {\bibinfo {title} {{The Ineludible non-Gaussianity of the Primordial Black
  Hole Abundance}},\ }\href {https://doi.org/10.1088/1475-7516/2019/07/048}
  {\bibfield  {journal} {\bibinfo  {journal} {JCAP}\ }\textbf {\bibinfo
  {volume} {07}},\ \bibinfo {pages} {048}},\ \Eprint
  {https://arxiv.org/abs/1904.00970} {arXiv:1904.00970 [astro-ph.CO]}
  \BibitemShut {NoStop}%
\bibitem [{\citenamefont {Young}(2019)}]{Young:2019osy}%
  \BibitemOpen
  \bibfield  {author} {\bibinfo {author} {\bibfnamefont {S.}~\bibnamefont
  {Young}},\ }\bibfield  {title} {\bibinfo {title} {{The primordial black hole
  formation criterion re-examined: Parametrisation, timing and the choice of
  window function}},\ }\href {https://doi.org/10.1142/S0218271820300025}
  {\bibfield  {journal} {\bibinfo  {journal} {Int. J. Mod. Phys. D}\ }\textbf
  {\bibinfo {volume} {29}},\ \bibinfo {pages} {2030002} (\bibinfo {year}
  {2019})},\ \Eprint {https://arxiv.org/abs/1905.01230} {arXiv:1905.01230
  [astro-ph.CO]} \BibitemShut {NoStop}%
\bibitem [{\citenamefont {Tokeshi}\ \emph {et~al.}(2020)\citenamefont
  {Tokeshi}, \citenamefont {Inomata},\ and\ \citenamefont
  {Yokoyama}}]{Tokeshi:2020tjq}%
  \BibitemOpen
  \bibfield  {author} {\bibinfo {author} {\bibfnamefont {K.}~\bibnamefont
  {Tokeshi}}, \bibinfo {author} {\bibfnamefont {K.}~\bibnamefont {Inomata}},\
  and\ \bibinfo {author} {\bibfnamefont {J.}~\bibnamefont {Yokoyama}},\
  }\bibfield  {title} {\bibinfo {title} {{Window function dependence of the
  novel mass function of primordial black holes}},\ }\href
  {https://doi.org/10.1088/1475-7516/2020/12/038} {\bibfield  {journal}
  {\bibinfo  {journal} {JCAP}\ }\textbf {\bibinfo {volume} {12}},\ \bibinfo
  {pages} {038}},\ \Eprint {https://arxiv.org/abs/2005.07153} {arXiv:2005.07153
  [astro-ph.CO]} \BibitemShut {NoStop}%
\bibitem [{\citenamefont {Escriv\`a}\ \emph {et~al.}(2022)\citenamefont
  {Escriv\`a}, \citenamefont {Kuhnel},\ and\ \citenamefont
  {Tada}}]{Escriva:2022duf}%
  \BibitemOpen
  \bibfield  {author} {\bibinfo {author} {\bibfnamefont {A.}~\bibnamefont
  {Escriv\`a}}, \bibinfo {author} {\bibfnamefont {F.}~\bibnamefont {Kuhnel}},\
  and\ \bibinfo {author} {\bibfnamefont {Y.}~\bibnamefont {Tada}},\ }\bibfield
  {title} {\bibinfo {title} {{Primordial Black Holes}},\ }\href@noop {} {\
  (\bibinfo {year} {2022})},\ \Eprint {https://arxiv.org/abs/2211.05767}
  {arXiv:2211.05767 [astro-ph.CO]} \BibitemShut {NoStop}%
\bibitem [{\citenamefont {Shibata}\ and\ \citenamefont
  {Sasaki}(1999)}]{Shibata:1999zs}%
  \BibitemOpen
  \bibfield  {author} {\bibinfo {author} {\bibfnamefont {M.}~\bibnamefont
  {Shibata}}\ and\ \bibinfo {author} {\bibfnamefont {M.}~\bibnamefont
  {Sasaki}},\ }\bibfield  {title} {\bibinfo {title} {{Black hole formation in
  the Friedmann universe: Formulation and computation in numerical
  relativity}},\ }\href {https://doi.org/10.1103/PhysRevD.60.084002} {\bibfield
   {journal} {\bibinfo  {journal} {Phys. Rev. D}\ }\textbf {\bibinfo {volume}
  {60}},\ \bibinfo {pages} {084002} (\bibinfo {year} {1999})},\ \Eprint
  {https://arxiv.org/abs/gr-qc/9905064} {arXiv:gr-qc/9905064} \BibitemShut
  {NoStop}%
\bibitem [{\citenamefont {Harada}\ \emph {et~al.}(2015)\citenamefont {Harada},
  \citenamefont {Yoo}, \citenamefont {Nakama},\ and\ \citenamefont
  {Koga}}]{Harada:2015yda}%
  \BibitemOpen
  \bibfield  {author} {\bibinfo {author} {\bibfnamefont {T.}~\bibnamefont
  {Harada}}, \bibinfo {author} {\bibfnamefont {C.-M.}\ \bibnamefont {Yoo}},
  \bibinfo {author} {\bibfnamefont {T.}~\bibnamefont {Nakama}},\ and\ \bibinfo
  {author} {\bibfnamefont {Y.}~\bibnamefont {Koga}},\ }\bibfield  {title}
  {\bibinfo {title} {{Cosmological long-wavelength solutions and primordial
  black hole formation}},\ }\href {https://doi.org/10.1103/PhysRevD.91.084057}
  {\bibfield  {journal} {\bibinfo  {journal} {Phys. Rev. D}\ }\textbf {\bibinfo
  {volume} {91}},\ \bibinfo {pages} {084057} (\bibinfo {year} {2015})},\
  \Eprint {https://arxiv.org/abs/1503.03934} {arXiv:1503.03934 [gr-qc]}
  \BibitemShut {NoStop}%
\bibitem [{\citenamefont {Polnarev}\ and\ \citenamefont
  {Musco}(2007)}]{Polnarev:2006aa}%
  \BibitemOpen
  \bibfield  {author} {\bibinfo {author} {\bibfnamefont {A.~G.}\ \bibnamefont
  {Polnarev}}\ and\ \bibinfo {author} {\bibfnamefont {I.}~\bibnamefont
  {Musco}},\ }\bibfield  {title} {\bibinfo {title} {{Curvature profiles as
  initial conditions for primordial black hole formation}},\ }\href
  {https://doi.org/10.1088/0264-9381/24/6/003} {\bibfield  {journal} {\bibinfo
  {journal} {Class. Quant. Grav.}\ }\textbf {\bibinfo {volume} {24}},\ \bibinfo
  {pages} {1405} (\bibinfo {year} {2007})},\ \Eprint
  {https://arxiv.org/abs/gr-qc/0605122} {arXiv:gr-qc/0605122} \BibitemShut
  {NoStop}%
\bibitem [{\citenamefont {Polnarev}\ \emph {et~al.}(2012)\citenamefont
  {Polnarev}, \citenamefont {Nakama},\ and\ \citenamefont
  {Yokoyama}}]{Polnarev:2012bi}%
  \BibitemOpen
  \bibfield  {author} {\bibinfo {author} {\bibfnamefont {A.~G.}\ \bibnamefont
  {Polnarev}}, \bibinfo {author} {\bibfnamefont {T.}~\bibnamefont {Nakama}},\
  and\ \bibinfo {author} {\bibfnamefont {J.}~\bibnamefont {Yokoyama}},\
  }\bibfield  {title} {\bibinfo {title} {{Self-consistent initial conditions
  for primordial black hole formation}},\ }\href
  {https://doi.org/10.1088/1475-7516/2012/09/027} {\bibfield  {journal}
  {\bibinfo  {journal} {JCAP}\ }\textbf {\bibinfo {volume} {09}},\ \bibinfo
  {pages} {027}},\ \Eprint {https://arxiv.org/abs/1204.6601} {arXiv:1204.6601
  [gr-qc]} \BibitemShut {NoStop}%
\bibitem [{\citenamefont {Inomata}\ \emph {et~al.}(2017)\citenamefont
  {Inomata}, \citenamefont {Kawasaki}, \citenamefont {Mukaida}, \citenamefont
  {Tada},\ and\ \citenamefont {Yanagida}}]{Inomata:2017okj}%
  \BibitemOpen
  \bibfield  {author} {\bibinfo {author} {\bibfnamefont {K.}~\bibnamefont
  {Inomata}}, \bibinfo {author} {\bibfnamefont {M.}~\bibnamefont {Kawasaki}},
  \bibinfo {author} {\bibfnamefont {K.}~\bibnamefont {Mukaida}}, \bibinfo
  {author} {\bibfnamefont {Y.}~\bibnamefont {Tada}},\ and\ \bibinfo {author}
  {\bibfnamefont {T.~T.}\ \bibnamefont {Yanagida}},\ }\bibfield  {title}
  {\bibinfo {title} {{Inflationary Primordial Black Holes as All Dark
  Matter}},\ }\href {https://doi.org/10.1103/PhysRevD.96.043504} {\bibfield
  {journal} {\bibinfo  {journal} {Phys. Rev. D}\ }\textbf {\bibinfo {volume}
  {96}},\ \bibinfo {pages} {043504} (\bibinfo {year} {2017})},\ \Eprint
  {https://arxiv.org/abs/1701.02544} {arXiv:1701.02544 [astro-ph.CO]}
  \BibitemShut {NoStop}%
\bibitem [{\citenamefont {Mr\'oz}\ \emph
  {et~al.}(2024{\natexlab{a}})\citenamefont {Mr\'oz} \emph
  {et~al.}}]{Mroz:2024mse}%
  \BibitemOpen
  \bibfield  {author} {\bibinfo {author} {\bibfnamefont {P.}~\bibnamefont
  {Mr\'oz}} \emph {et~al.},\ }\bibfield  {title} {\bibinfo {title} {{No massive
  black holes in the Milky Way halo}},\ }\href
  {https://doi.org/10.1038/s41586-024-07704-6} {\bibfield  {journal} {\bibinfo
  {journal} {Nature}\ }\textbf {\bibinfo {volume} {632}},\ \bibinfo {pages}
  {749} (\bibinfo {year} {2024}{\natexlab{a}})},\ \Eprint
  {https://arxiv.org/abs/2403.02386} {arXiv:2403.02386 [astro-ph.GA]}
  \BibitemShut {NoStop}%
\bibitem [{\citenamefont {Mr\'oz}\ \emph
  {et~al.}(2024{\natexlab{b}})\citenamefont {Mr\'oz} \emph
  {et~al.}}]{Mroz:2024wag}%
  \BibitemOpen
  \bibfield  {author} {\bibinfo {author} {\bibfnamefont {P.}~\bibnamefont
  {Mr\'oz}} \emph {et~al.},\ }\bibfield  {title} {\bibinfo {title}
  {{Microlensing Optical Depth and Event Rate toward the Large Magellanic Cloud
  Based on 20 yr of OGLE Observations}},\ }\href
  {https://doi.org/10.3847/1538-4365/ad452e} {\bibfield  {journal} {\bibinfo
  {journal} {Astrophys. J. Suppl.}\ }\textbf {\bibinfo {volume} {273}},\
  \bibinfo {pages} {4} (\bibinfo {year} {2024}{\natexlab{b}})},\ \Eprint
  {https://arxiv.org/abs/2403.02398} {arXiv:2403.02398 [astro-ph.GA]}
  \BibitemShut {NoStop}%
\bibitem [{\citenamefont {Franciolini}\ \emph {et~al.}(2022)\citenamefont
  {Franciolini}, \citenamefont {Musco}, \citenamefont {Pani},\ and\
  \citenamefont {Urbano}}]{Franciolini:2022tfm}%
  \BibitemOpen
  \bibfield  {author} {\bibinfo {author} {\bibfnamefont {G.}~\bibnamefont
  {Franciolini}}, \bibinfo {author} {\bibfnamefont {I.}~\bibnamefont {Musco}},
  \bibinfo {author} {\bibfnamefont {P.}~\bibnamefont {Pani}},\ and\ \bibinfo
  {author} {\bibfnamefont {A.}~\bibnamefont {Urbano}},\ }\bibfield  {title}
  {\bibinfo {title} {{From inflation to black hole mergers and back again:
  Gravitational-wave data-driven constraints on inflationary scenarios with a
  first-principle model of primordial black holes across the QCD epoch}},\
  }\href {https://doi.org/10.1103/PhysRevD.106.123526} {\bibfield  {journal}
  {\bibinfo  {journal} {Phys. Rev. D}\ }\textbf {\bibinfo {volume} {106}},\
  \bibinfo {pages} {123526} (\bibinfo {year} {2022})},\ \Eprint
  {https://arxiv.org/abs/2209.05959} {arXiv:2209.05959 [astro-ph.CO]}
  \BibitemShut {NoStop}%
\bibitem [{\citenamefont {Andr\'es-Carcasona}\ \emph
  {et~al.}(2024)\citenamefont {Andr\'es-Carcasona}, \citenamefont {Iovino},
  \citenamefont {Vaskonen}, \citenamefont {Veerm\"ae}, \citenamefont
  {Mart\'\i{}nez}, \citenamefont {Pujol\`as},\ and\ \citenamefont
  {Mir}}]{Andres-Carcasona:2024wqk}%
  \BibitemOpen
  \bibfield  {author} {\bibinfo {author} {\bibfnamefont {M.}~\bibnamefont
  {Andr\'es-Carcasona}}, \bibinfo {author} {\bibfnamefont {A.~J.}\ \bibnamefont
  {Iovino}}, \bibinfo {author} {\bibfnamefont {V.}~\bibnamefont {Vaskonen}},
  \bibinfo {author} {\bibfnamefont {H.}~\bibnamefont {Veerm\"ae}}, \bibinfo
  {author} {\bibfnamefont {M.}~\bibnamefont {Mart\'\i{}nez}}, \bibinfo {author}
  {\bibfnamefont {O.}~\bibnamefont {Pujol\`as}},\ and\ \bibinfo {author}
  {\bibfnamefont {L.~M.}\ \bibnamefont {Mir}},\ }\bibfield  {title} {\bibinfo
  {title} {{Constraints on primordial black holes from LIGO-Virgo-KAGRA O3
  events}},\ }\href {https://doi.org/10.1103/PhysRevD.110.023040} {\bibfield
  {journal} {\bibinfo  {journal} {Phys. Rev. D}\ }\textbf {\bibinfo {volume}
  {110}},\ \bibinfo {pages} {023040} (\bibinfo {year} {2024})},\ \Eprint
  {https://arxiv.org/abs/2405.05732} {arXiv:2405.05732 [astro-ph.CO]}
  \BibitemShut {NoStop}%
\bibitem [{\citenamefont {Young}(2022)}]{Young:2022phe}%
  \BibitemOpen
  \bibfield  {author} {\bibinfo {author} {\bibfnamefont {S.}~\bibnamefont
  {Young}},\ }\bibfield  {title} {\bibinfo {title} {{Peaks and primordial black
  holes: the~effect of non-Gaussianity}},\ }\href
  {https://doi.org/10.1088/1475-7516/2022/05/037} {\bibfield  {journal}
  {\bibinfo  {journal} {JCAP}\ }\textbf {\bibinfo {volume} {05}}\bibfield
  {number} {\bibinfo  {number} { (05)},\ \bibinfo {pages} {037}},\ }\Eprint
  {https://arxiv.org/abs/2201.13345} {arXiv:2201.13345 [astro-ph.CO]}
  \BibitemShut {NoStop}%
\bibitem [{\citenamefont {Ferrante}\ \emph {et~al.}(2023)\citenamefont
  {Ferrante}, \citenamefont {Franciolini}, \citenamefont {Iovino},\ and\
  \citenamefont {Urbano}}]{Ferrante:2022mui}%
  \BibitemOpen
  \bibfield  {author} {\bibinfo {author} {\bibfnamefont {G.}~\bibnamefont
  {Ferrante}}, \bibinfo {author} {\bibfnamefont {G.}~\bibnamefont
  {Franciolini}}, \bibinfo {author} {\bibfnamefont {A.}~\bibnamefont {Iovino},
  \bibfnamefont {Junior.}},\ and\ \bibinfo {author} {\bibfnamefont
  {A.}~\bibnamefont {Urbano}},\ }\bibfield  {title} {\bibinfo {title}
  {{Primordial non-Gaussianity up to all orders: Theoretical aspects and
  implications for primordial black hole models}},\ }\href
  {https://doi.org/10.1103/PhysRevD.107.043520} {\bibfield  {journal} {\bibinfo
   {journal} {Phys. Rev. D}\ }\textbf {\bibinfo {volume} {107}},\ \bibinfo
  {pages} {043520} (\bibinfo {year} {2023})},\ \Eprint
  {https://arxiv.org/abs/2211.01728} {arXiv:2211.01728 [astro-ph.CO]}
  \BibitemShut {NoStop}%
\bibitem [{\citenamefont {Gow}\ \emph {et~al.}(2023)\citenamefont {Gow},
  \citenamefont {Assadullahi}, \citenamefont {Jackson}, \citenamefont {Koyama},
  \citenamefont {Vennin},\ and\ \citenamefont {Wands}}]{Gow:2022jfb}%
  \BibitemOpen
  \bibfield  {author} {\bibinfo {author} {\bibfnamefont {A.~D.}\ \bibnamefont
  {Gow}}, \bibinfo {author} {\bibfnamefont {H.}~\bibnamefont {Assadullahi}},
  \bibinfo {author} {\bibfnamefont {J.~H.~P.}\ \bibnamefont {Jackson}},
  \bibinfo {author} {\bibfnamefont {K.}~\bibnamefont {Koyama}}, \bibinfo
  {author} {\bibfnamefont {V.}~\bibnamefont {Vennin}},\ and\ \bibinfo {author}
  {\bibfnamefont {D.}~\bibnamefont {Wands}},\ }\bibfield  {title} {\bibinfo
  {title} {{Non-perturbative non-Gaussianity and primordial black holes}},\
  }\href {https://doi.org/10.1209/0295-5075/acd417} {\bibfield  {journal}
  {\bibinfo  {journal} {EPL}\ }\textbf {\bibinfo {volume} {142}},\ \bibinfo
  {pages} {49001} (\bibinfo {year} {2023})},\ \Eprint
  {https://arxiv.org/abs/2211.08348} {arXiv:2211.08348 [astro-ph.CO]}
  \BibitemShut {NoStop}%
\bibitem [{\citenamefont {Ianniccari}\ \emph {et~al.}(2024)\citenamefont
  {Ianniccari}, \citenamefont {Iovino}, \citenamefont {Kehagias}, \citenamefont
  {Perrone},\ and\ \citenamefont {Riotto}}]{Ianniccari:2024bkh}%
  \BibitemOpen
  \bibfield  {author} {\bibinfo {author} {\bibfnamefont {A.}~\bibnamefont
  {Ianniccari}}, \bibinfo {author} {\bibfnamefont {A.~J.}\ \bibnamefont
  {Iovino}}, \bibinfo {author} {\bibfnamefont {A.}~\bibnamefont {Kehagias}},
  \bibinfo {author} {\bibfnamefont {D.}~\bibnamefont {Perrone}},\ and\ \bibinfo
  {author} {\bibfnamefont {A.}~\bibnamefont {Riotto}},\ }\bibfield  {title}
  {\bibinfo {title} {{Primordial black hole abundance: The importance of
  broadness}},\ }\href {https://doi.org/10.1103/PhysRevD.109.123549} {\bibfield
   {journal} {\bibinfo  {journal} {Phys. Rev. D}\ }\textbf {\bibinfo {volume}
  {109}},\ \bibinfo {pages} {123549} (\bibinfo {year} {2024})},\ \Eprint
  {https://arxiv.org/abs/2402.11033} {arXiv:2402.11033 [astro-ph.CO]}
  \BibitemShut {NoStop}%
\bibitem [{\citenamefont {Kitajima}\ \emph {et~al.}(2021)\citenamefont
  {Kitajima}, \citenamefont {Tada}, \citenamefont {Yokoyama},\ and\
  \citenamefont {Yoo}}]{Kitajima:2021fpq}%
  \BibitemOpen
  \bibfield  {author} {\bibinfo {author} {\bibfnamefont {N.}~\bibnamefont
  {Kitajima}}, \bibinfo {author} {\bibfnamefont {Y.}~\bibnamefont {Tada}},
  \bibinfo {author} {\bibfnamefont {S.}~\bibnamefont {Yokoyama}},\ and\
  \bibinfo {author} {\bibfnamefont {C.-M.}\ \bibnamefont {Yoo}},\ }\bibfield
  {title} {\bibinfo {title} {{Primordial black holes in peak theory with a
  non-Gaussian tail}},\ }\href {https://doi.org/10.1088/1475-7516/2021/10/053}
  {\bibfield  {journal} {\bibinfo  {journal} {JCAP}\ }\textbf {\bibinfo
  {volume} {10}},\ \bibinfo {pages} {053}},\ \Eprint
  {https://arxiv.org/abs/2109.00791} {arXiv:2109.00791 [astro-ph.CO]}
  \BibitemShut {NoStop}%
\bibitem [{\citenamefont {Yoo}(2022)}]{Yoo:2022mzl}%
  \BibitemOpen
  \bibfield  {author} {\bibinfo {author} {\bibfnamefont {C.-M.}\ \bibnamefont
  {Yoo}},\ }\bibfield  {title} {\bibinfo {title} {{The Basics of Primordial
  Black Hole Formation and Abundance Estimation}},\ }\href
  {https://doi.org/10.3390/galaxies10060112} {\bibfield  {journal} {\bibinfo
  {journal} {Galaxies}\ }\textbf {\bibinfo {volume} {10}},\ \bibinfo {pages}
  {112} (\bibinfo {year} {2022})},\ \Eprint {https://arxiv.org/abs/2211.13512}
  {arXiv:2211.13512 [astro-ph.CO]} \BibitemShut {NoStop}%
\bibitem [{\citenamefont {Perna}\ \emph {et~al.}(2024)\citenamefont {Perna},
  \citenamefont {Testini}, \citenamefont {Ricciardone},\ and\ \citenamefont
  {Matarrese}}]{Perna:2024ehx}%
  \BibitemOpen
  \bibfield  {author} {\bibinfo {author} {\bibfnamefont {G.}~\bibnamefont
  {Perna}}, \bibinfo {author} {\bibfnamefont {C.}~\bibnamefont {Testini}},
  \bibinfo {author} {\bibfnamefont {A.}~\bibnamefont {Ricciardone}},\ and\
  \bibinfo {author} {\bibfnamefont {S.}~\bibnamefont {Matarrese}},\ }\bibfield
  {title} {\bibinfo {title} {{Fully non-Gaussian Scalar-Induced Gravitational
  Waves}},\ }\href {https://doi.org/10.1088/1475-7516/2024/05/086} {\bibfield
  {journal} {\bibinfo  {journal} {JCAP}\ }\textbf {\bibinfo {volume} {05}},\
  \bibinfo {pages} {086}},\ \Eprint {https://arxiv.org/abs/2403.06962}
  {arXiv:2403.06962 [astro-ph.CO]} \BibitemShut {NoStop}%
\bibitem [{\citenamefont {Young}\ \emph {et~al.}(2014)\citenamefont {Young},
  \citenamefont {Byrnes},\ and\ \citenamefont {Sasaki}}]{Young:2014ana}%
  \BibitemOpen
  \bibfield  {author} {\bibinfo {author} {\bibfnamefont {S.}~\bibnamefont
  {Young}}, \bibinfo {author} {\bibfnamefont {C.~T.}\ \bibnamefont {Byrnes}},\
  and\ \bibinfo {author} {\bibfnamefont {M.}~\bibnamefont {Sasaki}},\
  }\bibfield  {title} {\bibinfo {title} {{Calculating the mass fraction of
  primordial black holes}},\ }\href
  {https://doi.org/10.1088/1475-7516/2014/07/045} {\bibfield  {journal}
  {\bibinfo  {journal} {JCAP}\ }\textbf {\bibinfo {volume} {07}},\ \bibinfo
  {pages} {045}},\ \Eprint {https://arxiv.org/abs/1405.7023} {arXiv:1405.7023
  [gr-qc]} \BibitemShut {NoStop}%
\bibitem [{\citenamefont {Wang}\ \emph {et~al.}(2021)\citenamefont {Wang},
  \citenamefont {Liu}, \citenamefont {Su},\ and\ \citenamefont
  {Li}}]{Wang:2021kbh}%
  \BibitemOpen
  \bibfield  {author} {\bibinfo {author} {\bibfnamefont {Q.}~\bibnamefont
  {Wang}}, \bibinfo {author} {\bibfnamefont {Y.-C.}\ \bibnamefont {Liu}},
  \bibinfo {author} {\bibfnamefont {B.-Y.}\ \bibnamefont {Su}},\ and\ \bibinfo
  {author} {\bibfnamefont {N.}~\bibnamefont {Li}},\ }\bibfield  {title}
  {\bibinfo {title} {{Primordial black holes from the perturbations in the
  inflaton potential in peak theory}},\ }\href
  {https://doi.org/10.1103/PhysRevD.104.083546} {\bibfield  {journal} {\bibinfo
   {journal} {Phys. Rev. D}\ }\textbf {\bibinfo {volume} {104}},\ \bibinfo
  {pages} {083546} (\bibinfo {year} {2021})},\ \Eprint
  {https://arxiv.org/abs/2111.10028} {arXiv:2111.10028 [astro-ph.CO]}
  \BibitemShut {NoStop}%
\bibitem [{\citenamefont {Stamou}(2021)}]{Stamou:2021qdk}%
  \BibitemOpen
  \bibfield  {author} {\bibinfo {author} {\bibfnamefont {I.~D.}\ \bibnamefont
  {Stamou}},\ }\bibfield  {title} {\bibinfo {title} {{Mechanisms of producing
  primordial black holes by breaking the $SU(2, 1)/SU(2)\times U(1)$
  symmetry}},\ }\href {https://doi.org/10.1103/PhysRevD.103.083512} {\bibfield
  {journal} {\bibinfo  {journal} {Phys. Rev. D}\ }\textbf {\bibinfo {volume}
  {103}},\ \bibinfo {pages} {083512} (\bibinfo {year} {2021})},\ \Eprint
  {https://arxiv.org/abs/2104.08654} {arXiv:2104.08654 [hep-ph]} \BibitemShut
  {NoStop}%
\bibitem [{\citenamefont {Carr}\ \emph
  {et~al.}(2021{\natexlab{b}})\citenamefont {Carr}, \citenamefont {Clesse},
  \citenamefont {Garc\'\i{}a-Bellido},\ and\ \citenamefont
  {K\"uhnel}}]{Carr:2019kxo}%
  \BibitemOpen
  \bibfield  {author} {\bibinfo {author} {\bibfnamefont {B.}~\bibnamefont
  {Carr}}, \bibinfo {author} {\bibfnamefont {S.}~\bibnamefont {Clesse}},
  \bibinfo {author} {\bibfnamefont {J.}~\bibnamefont {Garc\'\i{}a-Bellido}},\
  and\ \bibinfo {author} {\bibfnamefont {F.}~\bibnamefont {K\"uhnel}},\
  }\bibfield  {title} {\bibinfo {title} {{Cosmic conundra explained by thermal
  history and primordial black holes}},\ }\href
  {https://doi.org/10.1016/j.dark.2020.100755} {\bibfield  {journal} {\bibinfo
  {journal} {Phys. Dark Univ.}\ }\textbf {\bibinfo {volume} {31}},\ \bibinfo
  {pages} {100755} (\bibinfo {year} {2021}{\natexlab{b}})},\ \Eprint
  {https://arxiv.org/abs/1906.08217} {arXiv:1906.08217 [astro-ph.CO]}
  \BibitemShut {NoStop}%
\bibitem [{\citenamefont {Baumann}\ \emph {et~al.}(2007)\citenamefont
  {Baumann}, \citenamefont {Steinhardt}, \citenamefont {Takahashi},\ and\
  \citenamefont {Ichiki}}]{Baumann:2007zm}%
  \BibitemOpen
  \bibfield  {author} {\bibinfo {author} {\bibfnamefont {D.}~\bibnamefont
  {Baumann}}, \bibinfo {author} {\bibfnamefont {P.~J.}\ \bibnamefont
  {Steinhardt}}, \bibinfo {author} {\bibfnamefont {K.}~\bibnamefont
  {Takahashi}},\ and\ \bibinfo {author} {\bibfnamefont {K.}~\bibnamefont
  {Ichiki}},\ }\bibfield  {title} {\bibinfo {title} {{Gravitational Wave
  Spectrum Induced by Primordial Scalar Perturbations}},\ }\href
  {https://doi.org/10.1103/PhysRevD.76.084019} {\bibfield  {journal} {\bibinfo
  {journal} {Phys. Rev. D}\ }\textbf {\bibinfo {volume} {76}},\ \bibinfo
  {pages} {084019} (\bibinfo {year} {2007})},\ \Eprint
  {https://arxiv.org/abs/hep-th/0703290} {arXiv:hep-th/0703290} \BibitemShut
  {NoStop}%
\bibitem [{\citenamefont {Mollerach}\ \emph {et~al.}(2004)\citenamefont
  {Mollerach}, \citenamefont {Harari},\ and\ \citenamefont
  {Matarrese}}]{Mollerach:2003nq}%
  \BibitemOpen
  \bibfield  {author} {\bibinfo {author} {\bibfnamefont {S.}~\bibnamefont
  {Mollerach}}, \bibinfo {author} {\bibfnamefont {D.}~\bibnamefont {Harari}},\
  and\ \bibinfo {author} {\bibfnamefont {S.}~\bibnamefont {Matarrese}},\
  }\bibfield  {title} {\bibinfo {title} {{CMB polarization from secondary
  vector and tensor modes}},\ }\href
  {https://doi.org/10.1103/PhysRevD.69.063002} {\bibfield  {journal} {\bibinfo
  {journal} {Phys. Rev. D}\ }\textbf {\bibinfo {volume} {69}},\ \bibinfo
  {pages} {063002} (\bibinfo {year} {2004})},\ \Eprint
  {https://arxiv.org/abs/astro-ph/0310711} {arXiv:astro-ph/0310711}
  \BibitemShut {NoStop}%
\bibitem [{\citenamefont {Maggiore}(2000)}]{Maggiore:1999vm}%
  \BibitemOpen
  \bibfield  {author} {\bibinfo {author} {\bibfnamefont {M.}~\bibnamefont
  {Maggiore}},\ }\bibfield  {title} {\bibinfo {title} {{Gravitational wave
  experiments and early universe cosmology}},\ }\href
  {https://doi.org/10.1016/S0370-1573(99)00102-7} {\bibfield  {journal}
  {\bibinfo  {journal} {Phys. Rept.}\ }\textbf {\bibinfo {volume} {331}},\
  \bibinfo {pages} {283} (\bibinfo {year} {2000})},\ \Eprint
  {https://arxiv.org/abs/gr-qc/9909001} {arXiv:gr-qc/9909001} \BibitemShut
  {NoStop}%
\bibitem [{\citenamefont {Espinosa}\ \emph {et~al.}(2018)\citenamefont
  {Espinosa}, \citenamefont {Racco},\ and\ \citenamefont
  {Riotto}}]{Espinosa:2018eve}%
  \BibitemOpen
  \bibfield  {author} {\bibinfo {author} {\bibfnamefont {J.~R.}\ \bibnamefont
  {Espinosa}}, \bibinfo {author} {\bibfnamefont {D.}~\bibnamefont {Racco}},\
  and\ \bibinfo {author} {\bibfnamefont {A.}~\bibnamefont {Riotto}},\
  }\bibfield  {title} {\bibinfo {title} {{A Cosmological Signature of the SM
  Higgs Instability: Gravitational Waves}},\ }\href
  {https://doi.org/10.1088/1475-7516/2018/09/012} {\bibfield  {journal}
  {\bibinfo  {journal} {JCAP}\ }\textbf {\bibinfo {volume} {09}},\ \bibinfo
  {pages} {012}},\ \Eprint {https://arxiv.org/abs/1804.07732} {arXiv:1804.07732
  [hep-ph]} \BibitemShut {NoStop}%
\end{thebibliography}%

\end{document}